\documentclass[opre,nonblindrev]{informs3} 

\OneAndAHalfSpacedXI 

\usepackage{natbib}
 \bibpunct[, ]{(}{)}{,}{a}{}{,}%
 \def\bibfont{\small}%
 \def\bibsep{\smallskipamount}%
 \def\bibhang{24pt}%
\TheoremsNumberedThrough     
\ECRepeatTheorems

\EquationsNumberedThrough    

\usepackage{amsfonts,latexsym,xcolor}
\usepackage{url}
\usepackage{framed}
\usepackage{bbm}
\usepackage{subfigure}
\usepackage{enumerate}

\newcommand{\bOne}{\mathbf 1}
\newcommand{\Indicator}{\mathbbm{1}}
\newcommand{\Reals}{\mathbb R}
\newcommand{\Integers}{\mathbb Z}
\newcommand{\IntegersP}{\Integers_+}
\newcommand{\RealsP}{\Reals_+}
\newcommand{\Prob}{\mathbb{P}}
\newcommand{\E}{\mathbb{E}}
\newcommand{\pagerank}{PageRank}
\newcommand{\tmix}{t_{\text{\normalfont mix}} }

\newtheorem{fact}{Fact}

\begin{document}


\RUNAUTHOR{Lee, Ozdaglar, and Shah}

\RUNTITLE{Estimating Stationary Probability of Single State}

\TITLE{Approximating the Stationary Probability \\of a Single State in a Markov chain}

\ARTICLEAUTHORS{%
\AUTHOR{Christina E. Lee, Asuman Ozdaglar, Devavrat Shah}
\AFF{Laboratory for Information and Decision Systems, Massachusetts Institute of Technology, Cambridge, MA 02139, \\ \EMAIL{celee@mit.edu}, \EMAIL{asuman@mit.edu}, \EMAIL{devavrat@mit.edu}} 
} 

\ABSTRACT{
In this paper, we present a novel iterative Monte Carlo method for approximating the stationary probability of a single state of a positive recurrent Markov chain. We utilize the characterization that the stationary probability of a state $i$ is inversely proportional to the expected return time of a random walk beginning at $i$. Our method obtains an $\epsilon$-multiplicative close estimate with probability greater than $1 - \alpha$ using at most $\tilde{O}\left(\tmix \ln(1/\alpha) / \pi_i \epsilon^2 \right)$ simulated random walk steps on the Markov chain across all iterations, where $\tmix$ is the standard mixing time and $\pi_i$ is the stationary probability. In addition, the estimate at each iteration is guaranteed to be an upper bound with high probability, and is decreasing in expectation with the iteration count, allowing us to monitor the progress of the algorithm and design effective termination criteria. We propose a termination criteria which guarantees a $\epsilon (1 + 4 \ln(2) \tmix)$ multiplicative error performance for states with stationary probability larger than $\Delta$, while providing an additive error for states with stationary probability less than $\Delta \in (0,1)$. The algorithm along with this termination criteria uses at most $\tilde{O}\left(\frac{\ln(1/\alpha)}{\epsilon^2} \min\left(\frac{\tmix}{\pi_i}, \frac{1}{\epsilon \Delta}\right)\right)$ simulated random walk steps, which is bounded by a constant with respect to the Markov Chain. We provide a tight analysis of our algorithm based on a locally weighted variant of the mixing time. Our results naturally extend for countably infinite state space Markov chains via Lyapunov function analysis.}



\maketitle

\section{Introduction}

Given a discrete-time, irreducible, positive-recurrent Markov chain $\{X_t\}_{t \geq 0}$ on a countable state space $\Sigma$ with transition probability matrix $P$, we consider the problem of approximating the stationary probability of a chosen state $i \in \Sigma$. This is equivalent to computing the $i^{th}$ component of the largest eigenvector of $P$. The classical approach aims to estimate the entire stationary distribution by computing the largest eigenvector of matrix $P$ using either algebraic, graph theoretic, or simulation based techniques, which often involve computations with the full matrix. In this paper, we focus on computing the stationary probability of a particular state $i$, specifically in settings when $P$ is sparse and the dimension is large. Due to the large scale of the system, it becomes useful to have an algorithm which can approximate only a few components of the solution without the full cost of computing the entire stationary distribution.

Computing the stationary distribution of a Markov chain with a large state space (finite, or countably infinite) has become a basic building block to many algorithms and applications across disciplines. For example, the Markov Chain Monte Carlo (MCMC) method is widely used in statistical inference for approximating or generating samples from distributions that are difficult to specifically compute.
We are particularly motivated by the application of computing stationary distributions of Markov chains for network analysis.
Many decision problems over networks rely on information about the importance of different nodes as quantified by network centrality measures. Network centrality measures are functions assigning ``importance'' values to each node in the network. A few examples of network centrality measures that can be formulated as the stationary distribution of a specific random walk on the underlying network include \pagerank, which is commonly used in Internet search algorithms \citep{Page99}, Bonacich centrality and eigencentrality measures, encountered in the analysis of social networks \citep{Candogan12, Chasparis10}, rumor centrality, utilized for finding influential individuals in social media like Twitter \citep{Shah11}, and rank centrality, used to find a ranking over items within a network of pairwise comparisons \citep{Negahban12}.

There are many natural contexts in which one may be interested in computing the network centrality of a specific agent, or a subset of agents in the network. For example, a particular business owner may be interested in computing the PageRank of his webpage and that of his nearby competitors within the webgraph, without incurring the cost of estimating the full PageRank vector. These settings call for an algorithm which estimates the stationary probability of a given state of a Markov chain using only information adjacent to the state within some local neighborhood as described by the graph induced by matrix $P$, in which the edge $(i,j)$ has weight $P_{ij}$.

\subsection{Contributions}

We provide a novel Monte Carlo algorithm
which is designed based on the characterization of the stationary probability of state $i$ given by 
\[\pi_i = \frac{1}{\E_i[T_i]},\]
where $T_i \triangleq \inf\{t \geq 1 ~:~ X_t = i\}$ and $\E_i[\cdot] \triangleq \E[\cdot | X_0 = i]$. Standard MCMC methods estimate the full stationary distribution by using the property that the distribution of the current state of a random walk over the Markov chain will converge to the stationary distribution as time goes to infinity. Therefore, such methods generate approximate samples by simulating a long random walk until the distribution of the terminal state is close to the stationary distribution. Our key insight is that when we focus on solving for the stationary probability of a specific state $i$, we can center the computation and the samples around the state of interest by sampling return walks that begin at state $i$ and return to state $i$. This method is suitable when we are specifically interested in the estimates for states with high stationary probability, since the corresponding expected return time $\E_i[T_i]$ will be short due to it being inversely proportional to $\pi_i$. In order to keep the computation within a local neighborhood and limit the cost, we truncate the sample random walks at a threshold. To determine the appropriate truncation threshold and sufficient number of samples, we iteratively increase the truncation threshold and the number of samples to obtain successively closer estimates. Thus the method systematically increases the size of the local neighborhood that it computes over, iteratively refining the estimates in a way that exploits the local structure. The estimates along the computation path are always upper bounds with high probability, allowing us to observe the progress of the algorithm as the estimates converge from above.

Given an oracle for transitions of a Markov chain defined on state space $\Sigma$, state $i \in \Sigma$, and scalars $\alpha, \epsilon \in (0,1)$, our method obtains an $\epsilon$-multiplicative close estimate with probability greater than $1 - \alpha$ using at most $\tilde{O}\left(\tmix \ln(1/\alpha) / \pi_i \epsilon^2 \right)$ oracle calls (i.e. number of steps of the Markov chain), where $\tmix$ is the mixing time defined by $\tmix \triangleq \min\{t : \max_{x \in \Sigma} \| P^t(x,\cdot) - \pi\|_{TV} \leq 1/4\}.$\footnote{We denote $\tilde{O}(f(a) g(b)) = O(f(a) \text{polylog} f(a))O(g(b) \text{polylog} g(b))$. $P^t(x,\cdot)$ denotes the $x^{\text{th}}$ row of matrix $P^t$, and $\|\cdot\|_{TV}$ denotes the total variation distance.}
Thus the number of simulated random walk steps used by of our method scales with the same order as standard MCMC approaches up to polylogarithmic factors. Our method has the added benefit that the estimates along the computation path are always upper bounds with high probability, so we can monitor the progress as our estimate converges to the true stationary probability. This allows for easier design of a verifiable termination criterion, i.e., a procedure for determining how many random walks to sample and at what length to truncate the paths.

When the objective is to estimate the stationary probability of states with values larger than some $\Delta \in (0,1)$, by allowing coarser estimates for states with small stationary probability, we are able to provide a verifiable termination criterion such that the computation cost (i.e. number of simulated random walk steps) is bounded by a constant with respect to the mixing properties of the Markov chain. The termination criterion guarantees a $O(\epsilon \tmix)$ multiplicative error performance for states with stationary probability larger than $\Delta$, while providing an additive error for states with stationary probability less than $\Delta \in (0,1)$. More precisely, using our suggested termination criteria, the algorithm outputs an estimate $\hat{\pi}_i$ which satisfies either
\begin{enumerate}[~~~~(a)]
\item $\hat{\pi}_i < \Delta/(1 + \epsilon) \implies \pi_i < \Delta$ with high probability, or
\item $\hat{\pi}_i \geq \Delta/(1 + \epsilon) \implies \left(1 - \epsilon (1 + 4 \ln(2) \tmix) \right) \hat{\pi}_i \leq \pi_i \leq (1 + \epsilon) \hat{\pi}_i$ with high probability.
\end{enumerate}
With probability greater than $1 - \alpha$, the total number of oracle calls, i.e., number of steps of the Markov chain, that the algorithm uses before satisfying the termination criteria is bounded by
\[\tilde{O}\left( \frac{\ln(\frac{1}{\alpha})}{\epsilon^2} \min\left(\frac{\tmix}{\pi_i}, \frac{1}{\epsilon \Delta}\right)\right) = \tilde{O}\left( \frac{\ln(\frac{1}{\alpha})}{\epsilon^3 \Delta}\right).\]
The termination criteria that we propose does not require any knowledge of the properties of the Markov chain, but only depends on the parameters $\Delta, \epsilon$, and the intermediate vectors obtained through the computation. Therefore, a feature of our algorithm is that its cost and performance adapts to the mixing properties of the Markov chain without requiring prior knowledge of the Markov chain. Specifically, the cost of the computation naturally reduces for Markov chains that mix quickly. Standard MCMC methods in contrast often involve choosing the parameters such as number of samples heuristically a priori based on given assumptions about the mixing properties. We show that our analysis of the error is tight for a family of Markov chains, and the precise characterization of the error depends on a locally weighted variant of the classic mixing time. In scenarios where these ``local mixing times'' for different states differ from the global mixing time, then our algorithm which utilizes local random walk samples may provide tighter results than standard MCMC methods. This also suggests an important mathematical inquiry of how to characterize local mixing times of Markov chains, and in what settings they are homogenous as opposed to heterogenous.

We utilize the exponential concentration of return times in Markov chains to establish theoretical guarantees for the algorithm. Our analysis extends to countably infinite state space Markov chains, suggesting an equivalent notion of mixing time for analyzing countable state space Markov chains. We provide an exponential concentration bound on tail of the distribution of return times to a given state, utilizing an bound by \citet{Hajek82} on the concentration of certain types of hitting times in a countably infinite state space Markov chain. For Markov chains that mix quickly, the distribution over return times concentrates more sharply around its mean, resulting in tighter performance guarantees. Our analysis in the countably infinite state space setting lends insights towards understanding the key properties of large scale finite state space Markov chains.

Due to the truncation used within the original algorithm, the estimates obtained are biased. Therefore we also provide a bias correction for the estimates, at no additional computation cost, which we show performs surprisingly well in simulations. Whereas the original algorithm gave coarser estimates for states with low stationary probability, the bias corrected algorithm outputted close estimates for all states in simulations, even the states with stationary probability less than parameter $\Delta$. We provide theoretical analysis that sheds insight into the class of Markov chainns for which the bias correction is effective. In addition we present a modification of our algorithm which reuses the same simulated random walks to obtain estimates of the stationary probabilities of other states in the neighborhood of state $i$, based on the frequency of visits to other s. Again this modification does not require any extra computation cost in terms of simulated random walk steps, and yet provides estimates for the full stationary distribution. We provide theoretical bounds in addition to simulations that show its effectiveness.


\subsection{Related Work}

We provide a brief overview of the standard methods used for computing stationary distributions.

\subsubsection{Monte Carlo Markov Chain}

Monte Carlo Markov chain (MCMC) methods involve simulating long random walks over a carefully designed Markov chain in order to obtain samples from a target stationary distribution \citep{Metropolis53, MH}. In equilibrium, i.e. as time tends to infinity, the distribution of the random walk over the state space approaches the stationary distribution. These algorithms also leverage the ergodic property of Markov chains, which states that the Markov chain exhibits the same distribution when averaged over time and over space. In other words, as $t$ tends to infinity, the distribution over states visited by the Markov chain from time 0 to $t$ will converge to the distribution of the state of the Markov chain at time $t$. Therefore, MCMC methods approximate the stationary distribution by simulating the Markov chain for a sufficiently long time, and then either averaging over the states visited along the Markov chain, or using the last visited state as an approximate sample from the stationary distribution. This process is repeated many times to collect independent samples from $\pi_i$.

When applying MCMC methods in practice, it is often difficult to determine confidently when the Markov chain has been simulated for a sufficiently long time. Therefore, many heuristics are used for the termination criteria. The majority of work following the initial introduction of the MCMC method involves analyzing the convergence rate of the random walk for Markov chains under different conditions \citep{Aldous02, Peres}. Techniques involve spectral analysis (i.e. bounding the convergence rate as a function of the spectral gap of $P$) or coupling arguments. Graph properties such as conductance provide ways to characterize the spectrum of the graph. Most results are specific to reversible finite state space Markov chains, which are equivalent to random walks on weighted undirected graphs. A detailed summary of the major developments and analysis techniques for MCMC methods can be found in articles by \citet{DSC} and \citet{diaconis}.

Our algorithm also falls within the class of MCMC methods, as it is based upon simulating random walks over the Markov chain, and using concentration results to prove guarantees on the estimate. Its major distinction is due to the use of a different characterization of $\pi_i = 1/\E[T_i]$, which naturally lends itself to a component centered approximation. By sampling returning random walks which terminate when the initial state is revisited, we are able to design an intuitive termination criterion that is able to provide a one-sided guarantee.

\subsubsection{Power Iteration}

The power-iteration method is an equally old and well-established method for computing leading eigenvectors of matrices \citep{Golub96, Stewart94, Koury84}. Given a matrix $A$ and an initial vector $x_0$, it recursively compute iterates $x_{t+1} = A x_{t} / \|A x_t\|$. If matrix A has a single eigenvalue that is strictly greater in magnitude than all other eigenvalues, and if $x_0$ is not orthogonal to the eigenvector associated with the dominant eigenvalue, then a subsequence of $\{x_1, x_2, x_3, \dots\}$ converges to the eigenvector associated with the dominant eigenvalue. Recursive multiplications involving large matrices can become expensive very fast as the matrix grows. When the matrix is sparse, computation can be saved by implementing it through `message-passing' techniques; however it still requires computation to take place at every state in the state space. The convergence rate is governed by the {\em spectral gap}, or the difference between the two largest eigenvalues of $A$. Techniques used for analyzing the spectral gap and mixing times as discussed above are also used in analyzing the convergence of power iteration, thus many results again only pertain to reversible Markov chains. For large Markov chains, the mixing properties may scale poorly with the size, making it difficult to obtain good estimates in a reasonable amount of time. This is particularly ill suited for very large or countably infinite state space.

In the setting of computing \pagerank for nodes in a network, there have been efforts to modify the algorithm to execute power iteration over local subsets of the graph and combine the results to obtain estimates for the global \pagerank. These methods rely upon key assumptions on the underlying graph, which are difficult to verify. Kamvar et. al. observed that there may be obvious ways to partition the web graph (i.e. by domain names) such that power iteration can be used to estimate the local \pagerank~within these partitions \citep{Kamvar03}. They use heuristics to estimate the relative weights of these partitions, and combine the local \pagerank~within each partition according to the weights to obtain an initial estimate for \pagerank. This initial estimate is used to initialize the power iteration method over the global Markov chain, with the hope that this initialization may speed up convergence. Chen et. al. proposed a method for estimating the \pagerank~of a subset of nodes given only the local neighborhood of this subset \citep{Chen04}. Their method uses heuristics such as weighted in-degree as estimates for the \pagerank~values of nodes on the boundary of the given neighborhood. After fixing the boundary estimates, standard power iteration is used to obtain estimates for nodes within the local neighborhood. The error in this method depends on how close the true \pagerank~of nodes on the boundary correspond to the heuristic guesses such as weighted in-degree. Unfortunately, we rarely have enough information to make accurate heuristic guesses of these boundary nodes.

\subsubsection{Computing \pagerank~Locally}
There has been much recent effort to develop local algorithms for computing
\pagerank~for the web graph. Given a directed graph of $n$ nodes with an $n\times n$ adjacency matrix $A$ (i.e., $A_{ij} = 1$ if $(i,j) \in E$ and 0 otherwise), the \pagerank~vector $\pi$ is given by the stationary distribution of a Markov chain over $n$ states, whose transition matrix $P$ is given by 
\begin{align}\label{eq:decompose}
P & = (1 - \beta) D^{-1} A + \beta \bOne \cdot r^T.
\end{align}
$D$ denotes the diagonal matrix whose diagonal entries are the out-degrees of the nodes; $\beta \in (0,1)$ is a fixed scalar; and $r$ is a fixed probability vector over the $n$ nodes\footnote{$\bOne$ denotes the all ones vector.}. In each step the random walk with probability $(1 - \beta)$ chooses one of the neighbors of the current node equally likely, and with probability $\beta$ chooses any of the nodes in the graph according to $r$. Thus, the \pagerank~vector $\pi$ satisfies
\begin{align} \label{eq:pagerank}
\pi^T = \pi^T P = (1 - \beta) \pi^T D^{-1} A + \beta r^T
\end{align}
where $\pi^T \cdot \bOne = 1$. This definition of \pagerank~is also known as personalized \pagerank, because $r$ can be tailored to the personal preferences of a particular web surfer. When $r = \frac{1}{n} \cdot \bOne$, then $\pi$ equals the standard global \pagerank~vector. If $r = e_i$, then  $\pi$ describes the personalized \pagerank~that jumps back to node $i$ with probability $\beta$ in every step\footnote{$e_i$ denotes the standard basis vector having value one in coordinate $i$ and zero for all other coordinates.}.

Computationally, the design of local algorithms for computing the personalized \pagerank~has been of interest since its discovery. Most of the algorithms and analyses crucially rely on the specific structure of the random walk describing \pagerank: $P$ decomposes into a natural random walk matrix $D^{-1} A$, and a rank-1 matrix $\bOne \cdot r^T$, with strictly positive weights $(1-\beta)$ and $\beta$ respectively, cf. \eqref{eq:decompose}. \citet{JehWidom03} and \citet{Haveliwala03} observed a key linearity relation -- the global \pagerank~vector is the average of the $n$ personalized \pagerank~vectors corresponding to those obtained by setting $r = e_i$ for $1\leq i\leq n$. That is, these $n$ personalized \pagerank~vectors centered at each node form a basis for all personalized \pagerank~vectors, including the global \pagerank. Therefore, the problem boils down to computing the personalized \pagerank~for a given node. \citet{Fogaras05} used the fact that for the personalized \pagerank~centered at a given node $i$ (i.e., $r = e_i$), the associated random walk has probability $\beta$ at every step to jump back to node $i$, ``resetting'' the random walk. The distribution over the last node visited before a ``reset'' is equivalent to the personalized \pagerank~vector corresponding to node $i$. Therefore, they propose an algorithm which samples from the personalized \pagerank~vector by simulating short geometric-length random walks beginning from node $i$, and recording the last visited node of each sample walk. The performance of the estimate can be established using standard concentration results.

Subsequent to the key observations mentioned above, \citet{Avrachenkov07} surveyed variants to Fogaras' random walk algorithm, such as computing the frequency of visits to nodes across the sample path rather than only the end node. \citet{Bahmani10} addressed how to incrementally update the \pagerank~vector for dynamically evolving graphs, or graphs where the edges arrive in a streaming manner. Das Sarma {\em et al.} extended the algorithm to streaming graph models \citep{DasSarma11}, and distributed computing models \citep{DasSarma12PR}, ``stitching'' together short random walks to obtain longer samples, and thus reducing communication overhead. More recently, building on the same sets of observation, \citet{Borgs12} provided a sublinear time algorithm for estimating global \pagerank~using multi-scale matrix sampling. They use geometric-length random walk samples, but do not require samples for all $n$ personalized \pagerank~vectors. The algorithm returns a set of ``important'' nodes such that the set contains all nodes with \pagerank~greater than a given threshold, $\Delta$, and does not contain any node with \pagerank~less than $\Delta/c$ with probability $1 - o(1)$, for a given $c > 1$. The algorithm runs in time $\tilde{O}\left(n/\Delta\right)$.

\citet{Andersen07} designed a backward variant of these algorithms. Previously, to compute the global \pagerank~of a specific node $j$, we would average over all personalized \pagerank~vectors. The algorithm proposed by Andersen et al. estimates the global \pagerank~of a node $j$ by approximating the ``contribution vector'', i.e. estimating for the $j^{th}$ coordinates of the personalized \pagerank~vectors that contribute the most to $\pi_j$.


All of these algorithms rely on the crucial property that the random walk has renewal time that is distributed geometrically with constant parameter $\beta > 0$ that does not scale with graph size $n$. This is because the transition matrix $P$ decomposes according to \eqref{eq:decompose}, with a fixed $\beta$. In general, the transition matrix of any irreducible, positive-recurrent Markov chain will not have such a decomposition property (and hence known renewal time), making the above algorithms inapplicable in general. Our work can be seen as extending this approach of local approximation via short random walks beyond the restricted class of personalized PageRank to all Markov Chains. We utilize the fundamental invariant that stationary distribution is inversely proportional to average return times, which leads to a natural sampling scheme in which repeated visits to the state of interest behaves as the ``renewal events'' of the stochastic process, whereas the PageRank algorithm used the teleportation steps as the renewal events.

\subsection{Outline of Paper}
For the remainder of the paper, we will formalize the problem, present the main theorem results, provide intuition and proof sketches for the results, and demonstrate the algorithm through basic simulations. Section 2 includes the definition of the problem statement and a background review of key properties of Markov chains. We also provide an example to show that if the mixing properties of the Markov chain could be arbitrarily poor, then any Monte Carlo algorithm which samples only random walks within a local neighborhood of state $i$ cannot distinguish between a family of Markov chains which look similar locally, but very different globally. Section 3 presents the main theorem results of the analysis of convergence time and approximation error for the algorithms which we develop. Sections 4 to 7 present the proof sketch and intuition behind the analysis, including the simple proofs, but leaving the more complex details of the proofs to the appendices. Section 4 shows that the random variables used in the algorithm for the estimates and termination conditions indeeed concentrate around their means with high probability. Section 5 shows that for any positive recurrent Markov chain, the distribution over the return time of a random walk decays exponentially in the length, where the rate of decay is a function of the mixing properties of the Markov chain. This section provides the foundation for proving that our results extend to countably infinite state space Markov chains. 
Section 6 provides the proof sketch for proving bounds on the approximation error of the estimates in each iteration.
Section 7 provides the proof sketch for proving bounds on the convergence and computation time of the algorithm.
Section 8 presents the results from basic simulations in which we implemented and executed our algorithms on simple Markov chains.

\section{Setup}

In this section, we present our problem setup, and review the key definitions and properties of Markov chains which are useful to understanding our algorithm and analysis.

\subsection{Problem Statement}

Consider a discrete time, irreducible, positive recurrent Markov chain $\{X_t\}_{t \geq 0}$ on a countable state space $\Sigma$ with transition probability matrix $P: \Sigma \times \Sigma \to [0, 1]$. Given state $i \in \Sigma$, our goal is to estimate the stationary probability of state $i$, denoted by $\pi_i$. We consider the regime where the state space is large, thus it becomes critical to have an algorithm that scales well with the size of the state space. We limit ourselves to crawl operations originating from state $i$, simulating a limited access setting that occurs when the algorithm is run by a third-party user of the network who does not own or have full access to the network. We also focus on the setting when we are particularly interested in states with large stationary probability, specifically, when there is some threshold $\Delta$ such that we only consider a state significant if it has stationary probability larger than $\Delta$. Thus, for states with stationary probability less than $\Delta$, we are satisfied with a rough estimate; however, for states with stationary probability larger than $\Delta$, we would like an estimate which is has bounded multiplicative error.

\subsection{Basic Definitions and Notation}

Given the transition probability matrix $P$, let $P_{xy}^n$ denote the value of entry $(x,y)$ in the matrix $P^n$. If the state space is countably infinite, then $P: \Sigma \times \Sigma \rightarrow [0,1]$ is a function such that for all $x,y \in \Sigma$,
\[P_{xy} = \Prob(X_{t+1} = y | X_t = x).\]
Similarly, $P_{xy}^n$ is defined for all $x,y \in \Sigma$ to be
\[P_{xy}^n \triangleq \Prob(X_n = y | X_0 = x).\]
The {\em stationary distribution} is the largest eigenvector of $P$, also described by a function $\pi: \Sigma \rightarrow [0,1]$ such that $\sum_{i \in \Sigma} \pi_i = 1$ and $\pi_i = \sum_{j \in \Sigma} \pi_j P_{ji}$ for all $i \in \Sigma$.

The Markov chain can be visualized as a random walk over a weighted directed graph $G = (\Sigma, E, P)$, where $\Sigma$ is the set of states, $E = \{(i,j) \in \Sigma \times \Sigma: P_{ij} > 0\}$ is the set of edges, and $P$ describes the weights of the edges. We refer to G as the {\em Markov chain graph}. Throughout the paper, {\em Markov chain} and {\em random walk} on a graph are used interchangeably; similarly {\em nodes} and {\em states} are used interchangeably. If the state space $\Sigma$ is finite, let $n = |\Sigma|$ denote the number of states in the graph. We assume throughout this paper that the Markov chain $\{X_t\}$ is irreducible and positive recurrent.\footnote{A Markov chain is irreducible if and only if the corresponding Markov chain graph is strongly connected, i.e. for all $x,y \in \Sigma$, there exists a path from $x$ to $y$. A Markov chain is positive recurrent if the expected time for a random walk beginning at state $i$ to return to state $i$ is finite. This means that the random walk cannot ``drift to infinity''. This is true for all irreducible finite state space Markov chains.} This guarantees that there exists a unique stationary distribution.

Our algorithm involves generating sample sequences of the Markov chain by simulating a random walk on the graph. These sample sequences allow us to observe return times $T_i$ and visit frequencies $F_j$ to different states, where $T_i$ and $F_j$ are defined as:
\begin{align} \label{eq:def_ret_time}
T_i \triangleq \inf\{t \geq 1 ~:~ X_t = i\},
\end{align}
and
\[F_j \triangleq \sum_{t=1}^{\infty} \Indicator\{X_t = j\} \Indicator\{t \leq T_i\} = \sum_{t=1}^{T_i} \Indicator\{X_t = j\}.\]

Throughout this paper, we denote $\E_i[\cdot] \triangleq \E[\cdot | X_0 = i]$, and $\Prob_i(\cdot) \triangleq \Prob(\cdot | X_0 = i)$. The following characterization of the stationary distribution of a Markov chain is the theorem upon which our algorithm and analysis stands. Given samples of $T_i$ and $F_j$, we use this theorem to construct estimates for the stationary probabilities.

\begin{lemma} [{\normalfont {\em c.f. Meyn and Tweedie 1993}}] \label{eq:property} 
An irreducible positive recurrent Markov chain has a unique stationary distribution $\pi$ with the following form:

~~{\bf(a) }For any fixed $i \in \Sigma$,
\begin{align*} 
\pi_j & = \frac{\E_i[F_j]}{\E_i[T_i]}, ~~~~ j \in \Sigma.
\end{align*}

~~{\bf(b) }An equivalent expression for this distribution is
\begin{align*}
\pi_i & = \frac{1}{\E_i[T_i]}.
\end{align*}
\end{lemma}

Lemma \ref{eq:property}(b) states that the stationary probability of a state $i$ is inversely proportional to the expected return time of a random walk beginning at state $i$ and ending at its first return to state $i$. The basic algorithm we propose is based upon this characterization of stationary probability. Lemma \ref{eq:property}(a) states that the stationary probability of state $j$ is equivalent to the fraction of expected visits to state $j$ out of the total number of steps taken along a random walk beginning at state $i$ and returning to state $i$. This characterization is used to show that given returning random walks from state $i$ to state $i$, we can also obtain estimates of the stationary probability of other states $j$ by observing the frequency of visits to state $j$ along those sample paths.

\subsection{Mixing Properties} \label{sec:mixing}

The mixing properties of the Markov chain affect the ease to which our algorithm approximates the stationary probabilities. Our analysis and bounds will be a function of a few related quantities which we will proceed to define and discuss. In the finite state space setting, the error bounds on the estimate produced for the stationary probability of state $i$ will be given as a function of the maximal hitting time $H_i$, and the fundamental matrix $Z$. This measures how well connected the graph is globally. The {\em maximal hitting time} to a state $i$ in a finite state space Markov chain is defined as
\begin{align} \label{eq:hitting_time}
H_i \triangleq \max_{j \in \Sigma} \E_j[T_i].
\end{align}
The {\em fundamental matrix} $Z$ of a finite state space Markov chain is
\[Z \triangleq \sum_{t=0}^{\infty} \left(P^t - \bOne \pi^T\right) = \left(I-P+\bOne \pi^T\right)^{-1},\]
i.e., the entries of the fundamental matrix $Z$ are defined by
\[Z_{jk} \triangleq \sum_{t=0}^{\infty} \left(P_{jk}^t - \pi_k\right).\]
Since $P_{jk}^t$ denotes the probability that a random walk beginning at state $j$ is at state $k$ after $t$ steps, $Z_{jk}$ represents how quickly the probability mass at state $k$ from a random walk beginning at state $j$ converges to $\pi_k$. 
We will use the following property, stated by \citet{Aldous02} in Chapter 2 Section 2.2 Lemma 12, to relate entries in the fundamental matrix to expected return times.
\begin{lemma} \label{lemma:aldous_1}
For $j \neq k$,
\[\E_j[T_k] = \frac{Z_{kk} - Z_{jk}}{\pi_k}.\]
\end{lemma}
We define $Z_{\max}(i) \triangleq \max_{k \in \Sigma} |Z_{ki}|$. The relationship between $Z_{\max}(i)$ and $H_i$ is described by
\[Z_{\max}(i) \leq \pi_i H_i \leq 2 Z_{\max}(i).\]

A standard definition of mixing time is the amount of time until the total variation distance between the distribution of a random walk and the stationary distribution is below $1/4$. We formalize the definition and review some well known and useful properties below. For further details, read chapter 4 of \citet{Peres}.
\[\|\mu-\nu\|_{TV} = \frac{1}{2} \sum_{x \in \Sigma} |\mu(x) - \nu(x)| = \sum_{x \in \Sigma, \mu(x) \geq \nu(x)} |\mu(x) - \nu(x)|.\]
\[d(t) \triangleq \max_{x \in \Sigma} \| P^t(x, \cdot) - \pi \|_{TV} = \sup_{\mu \in \mathcal{P}} \| \mu P^t - \pi \|_{TV}.\]
\[\bar{d}(t) \triangleq \max_{x, y \in \Sigma} \| P^t(x, \cdot) - P^t(y, \cdot) \|_{TV} = \sup_{\mu, \nu \in \mathcal{P}} \| \mu P^t - \nu P^t \|_{TV}.\]
\[\tmix(\epsilon) \triangleq \min\{t: d(t) \leq \epsilon\}.\]
\[\tmix \triangleq \tmix(1/4).\]
\[\tmix(\epsilon) \leq \lceil\log_2(1/\epsilon)\rceil \tmix.\]
\[d(t) \leq \bar{d}(t) \leq 2^{-\lfloor t/\tmix \rfloor}.\]

Therefore, we can obtain the following relation between entries of $Z$ and $\tmix$.
\begin{align*}
Z_{jk} &= \sum_{t=0}^{\infty} \left(P_{jk}^t - \pi_k\right)
\leq \sum_{t=0}^{\infty} \left|P_{jk}^t - \pi_k\right|
\leq \sum_{t=0}^{\infty} \left\|P^t(j, \cdot) - \pi\right\|_{TV} \\
&\leq \sum_{t=0}^{\infty} d(t)
\leq \sum_{t=0}^{\infty} 2^{-\lfloor t/\tmix \rfloor}
\leq 2 \sum_{t=0}^{\infty} 2^{-t/\tmix}
= 2/(1 - 2^{-1/\tmix}) \approx 2 \ln(2) \tmix
\end{align*}

Therefore, $Z_{jk} = O(\tmix)$ for any $j,k \in \Sigma$. Our analysis and bounds will be given as a function of entries in the fundamental matrix $Z$, however observe that a bound on the mixing time also provides a bound on the maximum entry of $Z$. For countably infinite state space Markov chains, we have an equivalent notion of mixing time, which we discuss in Section \ref{sec:exp_decay}.

\subsection{Limitations of Poorly Mixing Markov Chains}

\begin{figure}
     \centering
        \subfigure[M/M/1 Queue]{
           \label{figure:MM1}
           \vspace{0.5in}
           \includegraphics[width=0.4\textwidth]{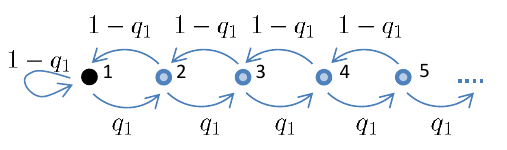}
        }
        \subfigure[Magnet Markov chain]{
           \label{figure:Magnet}
           \includegraphics[width=0.85\textwidth]{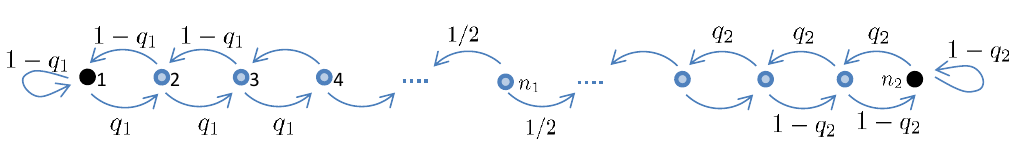}
        }
    \caption{We introduce two classes of Markov chains which have similar local random walk properties, and yet can have arbitrarily different stationary probabilities, illustrating the difficulty of approximating stationary probabilities using Monte Carlo type algorithms when the Markov chain can mix poorly.}
   \label{figure:graphs}
\end{figure}

We give an example to illustrate why the mixing properties affect the achievable estimation accuracy for a Monte Carlo method which samples random walks. Consider the Markov chains shown in Figures \ref{figure:MM1} and \ref{figure:Magnet}. The state space of the Markov chain in Figure \ref{figure:MM1} is the positive integers (therefore countably infinite). It models the length of a M/M/1 queue, where $q_1$ is the probability that an arrival occurs before a departure. This is also equivalent to a biased random walk on $\IntegersP$. In the Markov chain depicted by Figure \ref{figure:Magnet}, when $q_1$ and $q_2$ are less than one half, states 1 to $n_1 -1$ are attracted to state 1, and states $n_1$ to $n_2$ are attracted to state $n_2$. Since there are two opposite attracting states (1 and $n_2$), we call this Markov chain a ``Magnet''.

Consider the problem of estimating the stationary probability of state 1 in the Magnet Markov chain using sampled random walks. Due to the structure of the Magnet Markov chain, any random walk originating on the left will remain on the left with high probability. Therefore, the right portion of the Markov chain will only be explored with an exponentially small probability. The sample local random walks starting from state 1 will behave effectively the same for both the M/M/1 queue and the Magnet Markov chain, illustrating the difficulty of even distinguishing between the two Markov chains, much less to obtain an accurate estimate. This illustrates an unavoidable challenge for any Monte Carlo algorithm.

\section{Basic Algorithm and Main Results}

Recall that our algorithm is based on the characterization of stationary probability as given by Lemma \ref{eq:property}(b): $\pi_i = 1/\E_i[T_i]$. The \texttt{EstimateTruncatedRT}($i, N,\theta$) method, which forms the basic unit of our algorithm, estimates $\pi_i$ by collecting $N$ independent samples of the random variable $\min(T_i, \theta)$. Each sample is obtained by simulating the Markov chain starting from state $X_0 = i$ and stopping at the first time $t > 0$ that $t = \theta$ or $X_t = i$.  The sample average is used to approximate $\E_i[T_i]$. As the number of samples and $\theta$ go to infinity, the estimate will converge almost surely to $\pi_i$, due to the strong law of large numbers and positive recurrence of the Markov chain.

\vspace{1em} \fbox{
\begin{minipage}{0.8\textwidth}
\vspace{1em} 
\texttt{EstimateTruncatedRT}($i, N,\theta$): \\
\begin{enumerate}
\item Simulate $N$ independent realizations of the Markov chain with $X_0 = i$. For each sample $s \in \{1,2,\dots N\}$, let $t_s = \min\{t \geq 0 : t = \theta ~\text{ or }~ X_t = i\}$, distributed as $\min(T_i, \theta)$.
\item Return sample average $\hat{T}_i$, fraction truncated $\hat{p}_i$, and estimate $\hat{\pi}_i$
\[\hat{T}_i = \frac{1}{N} \sum_{s=1}^N t_s, ~~\hat{p} = \frac{\text{No. samples truncated}}{N}, ~~\hat{\pi}_i = \frac{1}{\hat{T}_i}\]
\end{enumerate}
\vspace{1em} 
\end{minipage}
} \vspace{1em}

It is not clear a priori what choice of $N$ and $\theta$ are sufficient to guarantee a good estimate while not costing too much computation. The \texttt{IteratedRefinement}($i, \epsilon, \alpha$) method iteratively improves the estimate by increasing $\theta$ and $N$. In each iteration, it doubles $\theta$ and increases $N$ according to the Chernoff's bound to ensure that with probability $1 - \alpha$, for all time $k$, $\hat{T}_i^{(k)} \in (1 \pm \epsilon) \E_i[\min(T_i,\theta^{(k)})]$. This allows us to use the estimate from the previous iteration to determine how many samples is sufficient for the current threshold $\theta^{(k)}$.

\vspace{1em} \fbox{
\begin{minipage}{0.8\textwidth}
\vspace{1em} 
\texttt{IteratedRefinement}($i, \epsilon, \alpha$): \\
\begin{enumerate}
\item $k = 1, \theta^{(1)} = 2, N^{(1)} = \left\lceil 6 (1+\epsilon) \ln(8 / \alpha) /\epsilon^2 \right\rceil$
\item $(\hat{T}_i^{(k)}, \hat{p}^{(k)}, \hat{\pi}_i^{(k)}) = $ \texttt{EstimateTruncatedRT}($i, N^{(k)},\theta^{(k)}$)
\item $\theta^{(k+1)} = 2 \theta^{(k)}, N^{(k+1)} \leftarrow \left\lceil 3 (1+\epsilon) \theta^{(k+1)} \ln(4 \theta^{(k+1)} / \alpha) / \hat{T}_i^{(k)} \epsilon^2 \right\rceil,$ increment $k$
\item Repeat from line 2
\end{enumerate}
\vspace{1em} 
\end{minipage}
} \vspace{1em}

The estimate $\hat{\pi}_i^{(k)}$ is larger than $\pi_i/(1+\epsilon)$ for all iterations $k$ with high probability. $\E_i[\hat{T}_i^{(k)}] = \E_i[\min(T_i, \theta^{(k)})]$ increases with each iteration; thus the expected estimate decreases in each iteration, converging to $\pi_i$ from above as $N^{(k)}$ and $\theta^{(k)}$ increase. In this section, for the sake of clarity, we present our results only for finite state space Markov chains, yet the extension of the analysis to countable state space will be discussed in Section \ref{sec:exp_decay} and presented in Theorems \ref{thm:countable_correctness} and \ref{thm:countable_convergence}. Theorem \ref{thm:fixed_t_conv_rate} provides error bounds which show the convergence rate of the estimator $\hat{\pi}_i^{(k)}$, and it also upper bounds the computation cost of the algorithm for the first $k$ iterations.

\begin{theorem} \label{thm:fixed_t_conv_rate}
For an irreducible finite state space Markov chain, for any $i \in \Sigma$, with probability greater than $1 - \alpha$, for all iterations $k$,

\begin{align*}
\left(1 - \epsilon - 4 \cdot 2^{-2^k/2 H_i} Z_{\max}(i) \right) \hat{\pi}_i^{(k)}
\leq \left(1 - \epsilon - 2 Z_{\max}(i) \Prob_i\left(T_i > \theta^{(k)}\right) \right) \hat{\pi}_i^{(k)}
\leq \pi_i \leq (1 + \epsilon) \hat{\pi}_i^{(k)},
\end{align*}
the number of random walk steps taken by the algorithm within the first $k$ iterations is bounded by
\[\tilde{O}\left( \frac{\ln(1/\alpha) 2^k}{\epsilon^2} \right).\]
\end{theorem}

Corollary \ref{cor:fixed_e_cost} directly follows from rearranging Theorem \ref{thm:fixed_t_conv_rate}, providing a bound on the cost of iterating the algorithm until the estimate is $\epsilon$-multiplicative close to the true value $\pi_i$.

\begin{corollary} \label{cor:fixed_e_cost}
For a finite state space Markov chain $\{X_t\}$, for any $i \in \Sigma$, with probability greater than $1 - \alpha$, the estimate $\hat{\pi}_i^{(k)}$ produced by the algorithm {\normalfont \texttt{IteratedRefinement}}($i, \epsilon/2, \alpha$) will satisfy $\pi_i \in (1 \pm \epsilon) \hat{\pi}_i^{(k)}$ for all
\[t \geq \log_2\left(2 H \log_2 \left( \frac{8 Z_{\max}(i)}{\epsilon}\right)\right) = O\left(\ln\left(\frac{\tmix}{\pi_i} \ln\left(\frac{\tmix}{\epsilon}\right)\right)\right).\]
The number of random walk steps simulated by the algorithm until $\pi_i \in (1 \pm \epsilon) \hat{\pi}_i^{(k)}$ is bounded by
\[\tilde{O}\left(\frac{H_i \ln(1 / \alpha)}{\epsilon^2} \ln \left(\frac{Z_{\max}(i)}{\epsilon}\right)\right) = \tilde{O}\left(\frac{\tmix \ln(1/\alpha)}{\pi_i \epsilon^2} \right).\]
\end{corollary}

The cost of our algorithm is comparable to standard Monte Carlo methods, as the length of each random walk to guarantee convergence to stationarity is $O(\tmix \ln(1/\epsilon \pi_i))$, and the number of samples to guarantee concentration by Chernoff's bound is $O(\ln(1/\alpha)/\pi_i \epsilon^2)$. Though this gives us an understanding of the convergence rate, we may not know $H_i, Z_{\max}(i)$, or $\tmix$ in the general case, and thus it does not provide practical guidance for how long to run the algorithm.

\subsection{Suggested Termination Criteria}
One intuitive termination criteria is to stop the algorithm when the fraction of samples truncated is less than some $\delta \in (0,1)$, since this indicates that the the bias produced by the truncation is small. Theorem \ref{thm:term_cond_guarantees_1} provides bound on both the error and the computation cost when the algorithm is terminated at $\hat{p}^{(k)} < \delta$.
\begin{theorem} \label{thm:term_cond_guarantees_1}
With probability greater than $1 - \alpha$, for all $k$ such that $\hat{p}^{(k)} < \delta$,
\begin{align*}
\left|\frac{\hat{\pi}_i^{(k)} - \pi_i}{\hat{\pi}_i^{(k)}} \right| \leq \epsilon \left(1 + 2 Z_{\max}(i)/3\right) + 2 \delta Z_{\max}(i) \leq \epsilon + 4 \ln(2) \tmix (\delta + \epsilon/3) .
\end{align*}
With probability greater than $1 - \alpha$, the number of random walk steps used by the algorithm before satisfying $\hat{p}^{(k)} < \delta$ is bounded above by
\[\tilde{O}\left( \frac{H_i \ln(1/\alpha) \ln(1/\delta))}{\epsilon^2} \right)
= \tilde{O}\left( \frac{ \tmix \ln(1/\alpha) \ln(1/\delta)}{\pi_i \epsilon^2} \right).\]
\end{theorem}

Theorem \ref{thm:term_cond_guarantees_1} indicates that the error is a function of both $\epsilon$ and $\delta$. The number of samples $N^{(t)}$ in each iteration of \texttt{IteratedRefinement} is chosen such that with high probability we obtain an $\epsilon$ approximation of the mean $\E[\hat{T}_i^{(t)}]$. Therefore, even if we were to run the algorithm until $\hat{p}^{(k)} = 0$, we would still only be able to guarantee an accuracy of $O(\epsilon \tmix)$ when the algorithm terminates, since the number of samples $N^{(t)}$ is only sufficient to guarantee less than $\epsilon$ error for the sample average estimates. Therefore, in the remainder of the paper, we choose $\delta$ to be on the same order as $\epsilon$, specifically $\delta = 2 \epsilon / 3$.

If we compare the results of Theorem \ref{thm:term_cond_guarantees_1} with Corollary \ref{cor:fixed_e_cost}, we observe that though the number of random walk steps scales similarly, Theorem \ref{thm:term_cond_guarantees_1} only guarantees an $O(\epsilon \tmix)$ multiplicative error bound, whereas Corollary \ref{cor:fixed_e_cost} reaches an $\epsilon$ close estimate. The difference is due to the fact that Corollary \ref{cor:fixed_e_cost} assumes we are able to determine when $\pi_i \in (1 \pm \epsilon) \hat{\pi}_i^{(k)}$, while the termination condition analyzed for Theorem \ref{thm:term_cond_guarantees_1} must rely only on measured quantities. The algorithm must determine how many samples and how far to truncate the random walks without knowledge of $\tmix$, and by using $\hat{p}$ as an estimate for $\Prob_i\left(T_i > \theta\right)$.

The example of distinguishing between the M/M/1 queue and the Magnet Markov chain in Figure \ref{figure:graphs} suggests why guaranteeing $\epsilon$ error without further knowledge of the mixing properties is impossible. Since the behavior in terms of sampled random walks from state 1 looks nearly identical for both Markov chains, an algorithm which knows nothing about the mixing properties will perform the same on both Markov chains, which actually have different stationary probabilities. In Makov chains which mix poorly such as the Magnet Markov chain, there may be states which look like they have a high stationary probability within the local neighborhood yet may not actually have large stationary probability globally.

Next we proceed to show that in a setting where we do not need an $O(\epsilon \tmix)$-close estimate for states with stationary probability less than some $\Delta \in (0,1)$, we can in fact provide a termination condition that upper bounds the computation time by $\tilde{O}(\ln(1/\alpha)/\epsilon^3 \Delta)$ independently of $\tmix$, while still maintaining the same error bound for states with stationary probability larger than $\Delta$.

\vspace{1em} \fbox{
\begin{minipage}{0.8\textwidth}
\vspace{1em} 
\texttt{TerminationCriteria}($\hat{\pi}_i, \hat{p}, \Delta, \epsilon$):\\

\hspace*{1em}
Stop when either ~~ {\bf (a)} $\hat{\pi}_i < \Delta/(1 + \epsilon)$ ~~ or ~~ {\bf (b)} $\hat{p} < 2 \epsilon / 3$.
\vspace{1em} 
\end{minipage}
} \vspace{1em}

The termination condition (a) is chosen by the fact that $(1 + \epsilon) \hat{\pi}_i$ is an upper bound on $\pi$ with high probability, thus when condition (a) is satisfied, we can safely conclude that $\pi_i < \Delta$ with high probability. The termination condition (b) is chosen according to the fact that we can upper bound the error as a function of $\Prob_i(T_i > \theta^{(k)})$. Therefore, we can use the fraction of samples truncated $\hat{p}$ to estimate this quantity, since $\hat{p}$ is a binomial random variable with mean $\Prob_i(T_i > \theta^{(k)})$. Therefore, when condition (b) is satisfied, we can safely conclude that $\Prob_i(T_i > \theta^{(k)}) < \epsilon$ with high probability, thus implying that the percentage error of estimate $\hat{\pi}_i^{(k)}$ is upper bounded by $O(\epsilon Z_{\max}(i))$.

\vspace{1em} \fbox{
\begin{minipage}{0.8\textwidth}
\vspace{1em} 
\texttt{BasicAlgorithm}($i, \epsilon, \alpha, \Delta$): \\

Run \texttt{IteratedRefinement}($i, \epsilon, \alpha$) until \texttt{TerminationCriteria}($\hat{\pi}_i^{(k)}, \hat{p}^{(k)}, \Delta, \epsilon$) is satisfied, at which point the algorithm outputs the final estimate $\hat{\pi}_i^{(k)}$.

\vspace{1em} 
\end{minipage}
} \vspace{1em}

For the remainder of the paper, we assume that this \texttt{TerminationCriteria} is used. It is easy to verify and does not require prior knowledge of the Markov chain to implement, as it only depends on the chosen parameters $\Delta$ and $\epsilon$. We highlight and discuss the benefits and limitations of using the termination criteria suggested. 

\begin{theorem} \label{thm:term_cond_guarantees_2}
With probability greater than $1 - \alpha$, the following three statements hold:
\begin{enumerate}[(a)]
\item If $\hat{\pi}_i^{(k)} < \Delta/(1 + \epsilon)$ for any $k$, then $\pi_i < \Delta$ with high probability.
\item For all $k$ such that $\hat{p}^{(k)} < 2 \epsilon / 3$,
\begin{align*}
\left|\frac{\hat{\pi}_i^{(k)} - \pi_i}{\hat{\pi}_i^{(k)}} \right| \leq \epsilon \left(2 Z_{\max}(i) + 1\right) \leq \epsilon (4 \ln(2) \tmix + 1) .
\end{align*}
\item The number of random walk steps used by the algorithm before satisfying either $\hat{\pi}_i^{(k)} < \Delta / (1 + \epsilon)$ or $\hat{p}^{(k)} < 2 \epsilon / 3$ is bounded above by
\[\tilde{O}\left( \frac{\ln(\frac{1}{\alpha})}{\epsilon^2} \min\left(H_i, \frac{1}{\epsilon \Delta}\right)\right) = \tilde{O}\left( \frac{\ln(\frac{1}{\alpha})}{\epsilon^2} \min\left(\frac{\tmix}{\pi_i}, \frac{1}{\epsilon \Delta}\right)\right).\]
\end{enumerate}
\end{theorem}

For ``important'' states $i$ such that $\pi_i > \Delta$, Theorem \ref{thm:term_cond_guarantees_2}(a) states that with high probability $\pi_i^{(t)} \geq \Delta /(1 + \epsilon)$ for all $t$, and thus the algorithm will terminate at criteria (b) $\hat{p}^{(t)} < 2\epsilon/\Delta$, which guarantees then an $O(\epsilon \tmix)$ multiplicative bound on the estimate error. Observe that the computation cost of the algorithm is upper bounded by $\tilde{O}(1/\epsilon^3 \Delta)$, which only depends on the algorithm parameters and is independent of the particular properties of the Markov chain. The cost is also bounded by $\tilde{O}(\tmix/\pi_i \epsilon^2 \Delta)$, which indicates that when the mixing time is smaller, the algorithm also terminates earlier for the same chosen algorithm parameters.

In some settings we may want to choose the parameters $\Delta$ and $\epsilon$ as a function of the Markov chain, whether as a function of the state space or the mixing properties. In settings where we have limited knowledge of the size of the state space or the mixing properties of the Markov chain, the algorithm can still be implemented as a heuristic. Since the estimates are an upper bound with high probability, we can observe and track the progress of the algorithm in each iteration as the estimate converges to the solution from above.

\subsection{Bias correction} \label{sec:bias_alg}

The algorithm presented above has a systematic bias due to the truncation. We present a second estimator which corrects for the bias under some conditions. In fact, in our basic simulations, we show that it corrects for the bias even for states with small stationary probabilty, which the original algorithm performs poorly on, since it terminates at $\hat{\pi}_i < \Delta/(1+\epsilon)$. Thus the surprising aspect of the bias corrected estimate is that it can obtain a good estimate at the same cost. This bias corrected estimate is based upon the characterization of stationary probability given in Lemma \ref{eq:property}(a), since the average visits to state $i$ along the sampled paths is given by $(1 - \hat{p})$.

\vspace{1em} \fbox{
\begin{minipage}{0.8\textwidth}
\vspace{1em} 
\texttt{BiasCorrectedAlgorithm}($i, \epsilon, \alpha, \Delta$): \\

Run \texttt{IteratedRefinement}($i, \epsilon, \alpha$) until \texttt{TerminationCriteria}($\hat{\pi}_i^{(k)}, \hat{p}^{(k)}, \Delta, \epsilon$) is satisfied, at which point the algorithm outputs the final estimate 
\[\tilde{\pi}_i = (1-\hat{p}) / \hat{T}_i.\]
\vspace{.1em}
\end{minipage}
} \vspace{1em}

While $\hat{\pi}_i$ is an upper bound of $\pi_i$ with high probability due to its use of truncation, $\tilde{\pi}_i$ is neither guaranteed to be an upper or lower bound of $\pi_i$. Theorem \ref{thm:bias} provides error bounds for the bias corrected estimator $\tilde{\pi}_i$.

\begin{theorem} \label{thm:bias}
For an irreducible finite state space Markov chain, for any $i \in \Sigma$, with probability greater than $1 - \alpha$, for all iterations $k$ such that $\Prob_i(T_i > \theta^{(k)}) < 1/2$,
\begin{align*}
\left| \frac{\tilde{\pi}_i^{(k)} - \pi_i}{\tilde{\pi}_i^{(k)}} \right| 
&\leq \frac{4 (1+ \epsilon)}{1- \epsilon} 2^{-\theta^{(k)}/2 H_i} \max(2 Z_{\max}(i) - 1,1) + \frac{2 \epsilon}{1 - \epsilon}.
\end{align*}
\end{theorem}

Theorem \ref{thm:bias} shows that with high probability, the percentage error between $\tilde{\pi}_i^{(k)}$ and $\pi_i$ decays exponentially in $\theta^{(k)}$. The condition $\Prob_i\left(T_i > \theta^{(k)}\right) < 1/2$ can be easily verified with high probability since $\hat{p}$ concentrates around $\Prob_i\left(T_i > \theta^{(k)}\right)$. We require $\Prob_i(T_i > \theta^{(k)}) < 1/2$ in order to ensure that $(1 - \hat{p}^{(k)})$ concentrates within a $(1 \pm \epsilon)$ multiplicative interval around $(1 - \Prob_i(T_i > \theta^{(k)}))$. If $\Prob_i\left(T_i > \theta^{(k)}\right)$ is too large and close to 1, then a majority of the sample random walks are truncated, and we cannot guarantee good multiplicative concentration of $(1 - \hat{p})$. The equivalent extension of this theorem to countable state space Markov chains is presented in Theorem \ref{thm:countable_bias}. Although the improvement of the bias corrected estimator is not clear from the theoretical error bounds, we will show simulations of computing PageRank, in which $\tilde{\pi}_i$ is a significantly closer estimate of $\pi_i$ than $\hat{\pi}_i$, especially for states with small stationary probability.

\subsection{Extension to multiple states} \label{sec:multiple_alg}

We can also simultaneously learn about other states in the Markov chain through these random walks from $i$. We refer to state $i$ as the {\em anchor state}. We will extend our algorithm to obtain estimates for the stationary probability of states within a subset $J \subseteq \Sigma$, which is given as an input to the algorithm. We refer to $J$ as the set of {\em observer states}. We estimate the stationary probability of any state $j \in J$ using the characterization given in Lemma \ref{eq:property}(a). We  modify the algorithm to keep track of how many times each state in $J$ is visited along the sample paths. The estimate $\tilde{\pi}_j$ is the fraction of visits to state $j$ along the sampled paths. We replace the subroutine in step 2 of the \texttt{IterativeRefinement} function with \texttt{EstimatedTruncatedRT-Multi}($i, J , N^{(k)}, \theta^{(k)}$), and we use the same \texttt{TerminationCriteria} previously defined.

\vspace{1em} \fbox{
\begin{minipage}{0.8\textwidth}
\vspace{1em} 
\texttt{EstimateTruncatedRT-Multi}($i, J, N,\theta$): \\
\begin{enumerate}
\item Sample $N$ independent truncated return paths to $i$
\[s_k \sim \min(T_i, \theta) ~\text{ for }~ k \in \{1,2,\dots N\}\]
\item Compute sample average $\hat{T}_i$ and fraction truncated $\hat{p}_i$
\[\hat{T}_i = \frac{1}{N} \sum_{k=1}^N s_k, ~~\hat{p} = \frac{\text{No. samples truncated}}{N}\]
\item For each $j \in J$, let the estimate $\tilde{\pi}_j$ be computed as
\[f_k(j) = \sum_{r=1}^{\theta^{(k)}} \Indicator\{X_r = j\} \Indicator\{r \leq T_i\}, ~~\hat{F}_j^{(k)} = \frac{1}{N^{(k)}} \sum_{k=1}^{N^{(k)}} f_k(j), ~~\text{ and }~~ \tilde{\pi}_j^{(k)} = \frac{\hat{F}_j^{(k)}}{\hat{T}_i^{(k)}}.\]
\end{enumerate}
\vspace{1em} 
\end{minipage}
} \vspace{1em}

Since the \texttt{IterativeRefinement} method sets the parameters $N, \theta$ independent of the states $j$ and their estimates, the error bounds for $\pi_j$ will be looser. The number of samples is only enough to guarantee that $\tilde{\pi}_j^{(k)}$ is an additive approximation of $\E_i[\hat{F}_j^{(k)}]/\E_i[\min(T_i,\theta^{(k)})]$. In addition, the effect of truncation is no longer as clear, since the frequency of visits to state $j$ along a sample return path from state $i$ to state $i$ can be distributed non-uniformly along the sample path. Therefore, the estimate cannot be guaranteed to be either an upper or lower bound. Theorem \ref{thm:freq} bounds the error for the estimates $\tilde{\pi}_j^{(k)}$ for states $j \neq i$. Due to the looser additive concentration guarantees for $\hat{F}_i^{(k)}$, Theorem \ref{thm:freq} provides an additive error bound rather than a bound on the percentage error.

\begin{theorem} \label{thm:freq}
For an irreducible finite state space Markov chain, for any $i,j \in \Sigma$ such that $j \neq i$, with probability greater than $1 - \alpha$, for all iterations $k$,
\begin{align*}
\left|\tilde{\pi}_j^{(k)} - \pi_j\right| &\leq 2 (1 + \epsilon) \Prob_i(T_i > \theta^{(k)}) Z_{\max}(j) \hat{\pi}_i^{(k)} + \epsilon\tilde{\pi}_j^{(k)} + \epsilon, \\
&\leq 4 (1 + \epsilon) 2^{-\theta^{(k)}/2 H_i} Z_{\max}(j) \hat{\pi}_i^{(k)} + \epsilon\tilde{\pi}_j^{(k)} + \epsilon.
\end {align*}
\end{theorem}

Theorem \ref{thm:freq} indicates that the accuracy of the estimator $\hat{\pi}_i^{(k)}$ depends on both $H_i$ and $Z_{\max}(j)$, the mixing properties centered at states $i$ and $j$. In order for the error to be small, both the anchor state $i$ and the observer state $j$ must have reasonable mixing and connectivity properties within the Markov chain. It is not surprising that it depends on the mixing properties related to both states, as the sample random walks are centered at state $i$, and the estimator consists of observing visits to state $j$. While the other theorems presented in this paper have equivalent results for countable state space Markov chains, Theorem \ref{thm:freq} does not directly extend to a countably infinite state space Markov chain because state $j$ can be arbitrarily far away from state $i$ such that random walks beginning at state $i$ rarely hit state $j$ before returning to state $i$.

\subsection{Implementation of multiple state algorithm}

This algorithm is simple to implement and is easy to parallelize. It requires only $O(|J|)$ space to keep track of the visits to each state in $J$, and a constant amount of space to keep track of the state of the random walk sample, and running totals such as $\hat{p}^{(k)}$ and $\hat{T}_i^{(k)}$. For each random walk step, the computer only needs to fetch the local neighborhood of the current state, which is upper bounded by the maximum degree. Thus, at any given instance in time, the algorithm only needs to access a small neighborhood within the graph. Each sample is completely independent, thus the task can be distributed among independent machines. In the process of sampling these random paths, the sequence of states along the path does not need to be stored or processed upon.

Consider implementing this over a distributed network, where the graph consists of the processors and the communication links between them. Each random walk over this network can be implemented by a message passing protocol. The anchor state $i$ initiates the random walk by sending a message to one of its neighbors chosen uniformly at random. Any state which receives the message forwards the message to one of its neighbors chosen uniformly at random. As the message travels over each link, it increments its internal counter. If the message ever returns to the anchor state $i$, then the message is no longer forwarded, and its counter provides a sample from $\min(T_i, \theta)$. When the counter exceeds $\theta$, then the message stops at the current state. After waiting for $\theta$ time steps, the anchor state $i$ can compute the estimate of its stationary probability within this network, taking into consideration the messages which have returned to state $i$. In addition, each observer state $j \in J$ can keep track of the number of times any of the messages are forwarded to state $j$. At the end of the $\theta$ time steps, state $i$ can broadcast the total number of steps to all states $j \in J$ so that they can properly normalize to obtain final estimates for $\pi_j$.

\section{Concentration bounds}

In order to analyze our algorithm, we first need to show that the statistics obtained from the random samples, specifically the values of $\hat{p}^{(k)}, \hat{T}_i^{(k)},$ and $\hat{F}_j^{(k)}$, concentrate around their mean with high probability, based upon standard concentration results for sums of independent identically distributed random variables. These statistics are used in computing the estimates and determining the termination time of \texttt{IterativeRefinement}.

Recall that the iterations are not independent, since the number of samples at iteration $k$ depends on the estimate $\hat{T}_i^{(k-1)}$ at iteration $k-1$, which itself is a random variable. Therefore, in order to prove any result about iteration $k$, we must consider the distribution over values of $\hat{T}_i^{(k-1)}$ from the previous iteration. The following Lemmas \ref{lemma:error_bounds} to \ref{lemma:f_concentrate} use iterative conditioning to show concentration bounds that hold for all iterations $k$ simultaneously with probability greater than $1 - \alpha$. 
Lemma \ref{lemma:error_bounds} shows concentration of $\hat{T}_i^{(k)}$, which directly implies concentration of the estimate $\hat{\pi}_i$ as well as $N^{(k)}$.


\begin{lemma} \label{lemma:error_bounds}
For every $k \in \IntegersP$,
\[\Prob_i\left( \bigcap_{h=1}^k \left\{ \hat{T}_i^{(h)} \in (1 \pm \epsilon) \E_i\left[\hat{T}_i^{(h)}\right] \right\} \right) \geq 1 - \alpha.\]
\end{lemma}

\proof{Proof of Lemma \ref{lemma:error_bounds}.}
We will sketch the proof here and leave the details to the Appendix. Let $A_h$ denote the event $\left\{ \hat{T}_i^{(h)} \in (1 \pm \epsilon) \E_i\left[\hat{T}_i^{(h)}\right] \right\}$. As discussed earlier, $N^{(h)}$ is a random variable that depends on $\hat{T}_i^{(h-1)}$. However, conditioned on the event $A_{h-1}$, we can lower bound $N^{(h)}$ as a function of $\E_i[\hat{T}_i^{(h-1)}]$. Then we apply Chernoff's bound for independent identically distributed bounded random variables and use the fact that $\E_i[\hat{T}_i^{(h)}]$ is nondecreasing in $h$ to show that 
\[\Prob_i\left( A_h | A_{h-1} \right) \geq 1- \frac{\alpha}{2^{h+1}} ~~\text{ for all } h.\]
Since iteration $h$ is only dependent on the outcome of previous iterations through the variable $\hat{T}_i^{(h-1)}$, we know that $A_{h'}$ is independent from $A_h$ for $h' < h$ conditioned on $A_{h-1}$. Therefore,
\[\Prob_i\left(\bigcap_{h=1}^k A_h\right) = \Prob_i\left(A_1\right) \prod_{h=2}^k \Prob_i\left(A_h | A_{h-1}\right).\]
We combine these two insights to complete the proof.\hfill \halmos
\endproof

Lemmas \ref{lemma:p_add_concentrate} to \ref{lemma:f_concentrate} also use the multiplicative concentration of $\hat{T}_i^{(k)}$ in order to lower bound the number of samples in each iteration. Their proofs are similar to the proof sketch given for Lemma \ref{lemma:error_bounds}, except that we have two events per iteration to consider. Conditioning on the event that $\hat{p}^{(h-1)} \in \Prob_i(T_i > \theta^{(h-1)}) \pm \epsilon/3$ and $\hat{T}_i^{(h-1)} \in (1 \pm \epsilon) \E_i[\hat{T}_i^{(h-1)}]$, we compute the probability that $\hat{p}^{(h)} \in \Prob_i(T_i > \theta^{(h)})$ and $\hat{T}_i^{(h)} \in (1 \pm \epsilon) \E_i[\hat{T}_i^{(h)}]$ using Chernoff's bound and union bound. Lemma \ref{lemma:p_add_concentrate} shows an additive concentration of $\hat{p}^{(h)}$. It is used to prove that when the algorithm terminates at condition (b), with high probability, $\Prob_i(T_i > \theta^{(h)}) < \epsilon$, which is used to upper bound the estimation error.

\begin{lemma} \label{lemma:p_add_concentrate}
For every $k \in \IntegersP$,
\[\Prob_i\left( \bigcap_{h=1}^k \left\{ \hat{p}^{(h)} \in \Prob_i(T_i > \theta^{(h)}) \pm \frac{\epsilon}{3} \right\} \bigcap_{h=1}^k \left\{ \hat{T}_i^{(h)} \in (1 \pm \epsilon) \E_i\left[\hat{T}_i^{(h)}\right] \right\} \right) \geq 1 - \alpha.\]
\end{lemma}
 
Lemma \ref{lemma:p_concentrate} gives a multiplicative concentration result for $(1-\hat{p}^{(k)})$, which is used in the analysis of the estimate $\tilde{\pi}_i^{(k)}$.

\begin{lemma} \label{lemma:p_concentrate}
Let $k_0$ be such that $\Prob_i(T_i > \theta^{(k_0)}) < 1/2$. For every $k \geq k_0$,
\[\Prob_i\left( \bigcap_{h=k_0}^k \left\{ (1 - \hat{p}^{(h)}) \in (1 \pm \epsilon) (1 - \Prob_i(T_i > \theta^{(h)})) \right\} \bigcap_{h=1}^k \left\{ \hat{T}_i^{(h)} \in (1 \pm \epsilon) \E_i\left[\hat{T}_i^{(h)}\right] \right\} \right) \geq 1 - \alpha.\]
\end{lemma}

Lemma \ref{lemma:f_concentrate} is used in the analysis of the estimate $\tilde{\pi}_j^{(k)}$ for $j \neq i$. It guarantees that $\hat{F}_j^{(k)}$ is within an additive value of $\epsilon \E_i[\hat{T}_i^{(k)}]$ around its mean. This allows us to show that the ratio between $\hat{F}_j^{(k)}$ and $\hat{T}_i^{(k)}$ is within an additive $\epsilon$ error around the ratio of their respective means. We are not able to obtain a small multiplicative error bound on $\hat{F}_j^{(k)}$ because we do not use any information from state $j$ to choose the number of samples $N^{(k)}$. $\E_i[\hat{F}_j^{(k)}]$ can be arbitrarily small compared to $\E_i[\hat{T}_i^{(k)}]$, so we may not have enough samples to estimate $\E_i[\hat{F}_j^{(k)}]$ closely. 

\begin{lemma} \label{lemma:f_concentrate}
For every $t \in \IntegersP$,
\[\Prob_i\left( \bigcap_{h=1}^k \left\{ \hat{F}_j^{(h)} \in \E_i\left[\hat{F}_j^{(h)}\right] \pm \epsilon \E_i\left[\hat{T}_i^{(h)}\right]\right\} \bigcap_{h=1}^k \left\{ \hat{T}_i^{(h)} \in (1 \pm \epsilon) \E_i\left[\hat{T}_i^{(h)}\right] \right\} \right) \geq 1 - \alpha.\]
\end{lemma}

\section{Exponential Decay of Return Times} \label{sec:exp_decay}

In this section, we discuss the error that arises due to truncating the random walks at threshold $\theta$. We show that the tail of the distribution of the return times to state $i$ decays exponentially as a function of the truncation parameter $\theta$. This is the key property which underlies the error and cost analysis of the algorithm. Intuitively, it means that the distribution over return times is concentrated around its mean, since it cannot have large probability at values far away from the mean. For finite state space Markov chains, this result is easy to show using the strong Markov property, as outlined by \citet{Aldous02} Chapter 2 Section 4.3 .

\begin{lemma} \citep{Aldous02} \label{lemma:finite}
Let Markov chain $\{X_t\}$ be defined on finite state space $\Sigma$. For any $i \in \Sigma$ and $t \in \Integers_+$, 
\[\Prob_i( T_i > t ) \leq 2 \cdot 2^{- t / 2 H_i},\]
where $H_i = \max_{j \in \Sigma}\E_j[T_i].$
\end{lemma}

Lemma \ref{lemma:E_diff} shows that since $\Prob_i(T_i > t)$ decays exponentially in $t$, the bias due to truncation likewise decays exponentially as a function of $\theta^{(k)}$. 

\begin{lemma} \label{lemma:E_diff}
\begin{align*}
\E_i[T_i] - \E_i[\hat{T}_i^{(k)}] &= \sum_{t=\theta^{(k)}}^{\infty} \Prob_i(T_i > t).
\end{align*}
\end{lemma}

\proof{Proof of Lemma \ref{lemma:E_diff}.}
Since $T_i$ is a nonnegative random variable, and by the definition of $\hat{T}_i^{(k)}$,
\begin{align*}
\E_i[T_i] - \E_i[\hat{T}_i^{(k)}] &= \E_i[T_i] - \E_i[\min(T_i,\theta^{(k)})]\\
& = \sum_{t=0}^{\infty} \Prob_i(T_i > t) - \sum_{t=0}^{\theta^{(k)} - 1} \Prob_i(T_i > t) \\
&= \sum_{t=\theta^{(k)}}^{\infty} \Prob_i(T_i > t).
\end{align*}\hfill \halmos
\endproof

Lemma \ref{lemma:finite} depends on the finite size of the state space. In order to obtain the same result for countable state space Markov chains, we use Lyapunov analysis techniques to prove that the tail of the distribution of $T_i$ decays exponentially for any state $i$ in any countable state space Markov chain that satisfies Assumption \ref{lyapunovAssumption}.

\begin{assumption} \label{lyapunovAssumption}
The Markov chain $\{X_t\}$ is irreducible. There exists a Lyapunov function $V: \Sigma \to \RealsP$ and constants $\nu_{\max}, \gamma > 0$, and $b \geq 0$, that satisfy the following conditions:
\begin{enumerate}
\item The set $B = \{x \in \Sigma: V(x) \leq b\}$ is finite,
\item For all $x,y \in \Sigma$ such that $\Prob\big(X_{t+1} = j | X_t = i\big) > 0$,
$|V(j)-V(i)| \leq \nu_{\max}$,
\item For all $x \in \Sigma$ such that $V(x) > b$,
$\E\big[V(X_{t+1}) - V(X_t) | X_t=x\big] < -\gamma$.
\end{enumerate}
\end{assumption}

\begin{figure}
\centering
\includegraphics[height=1.75in]{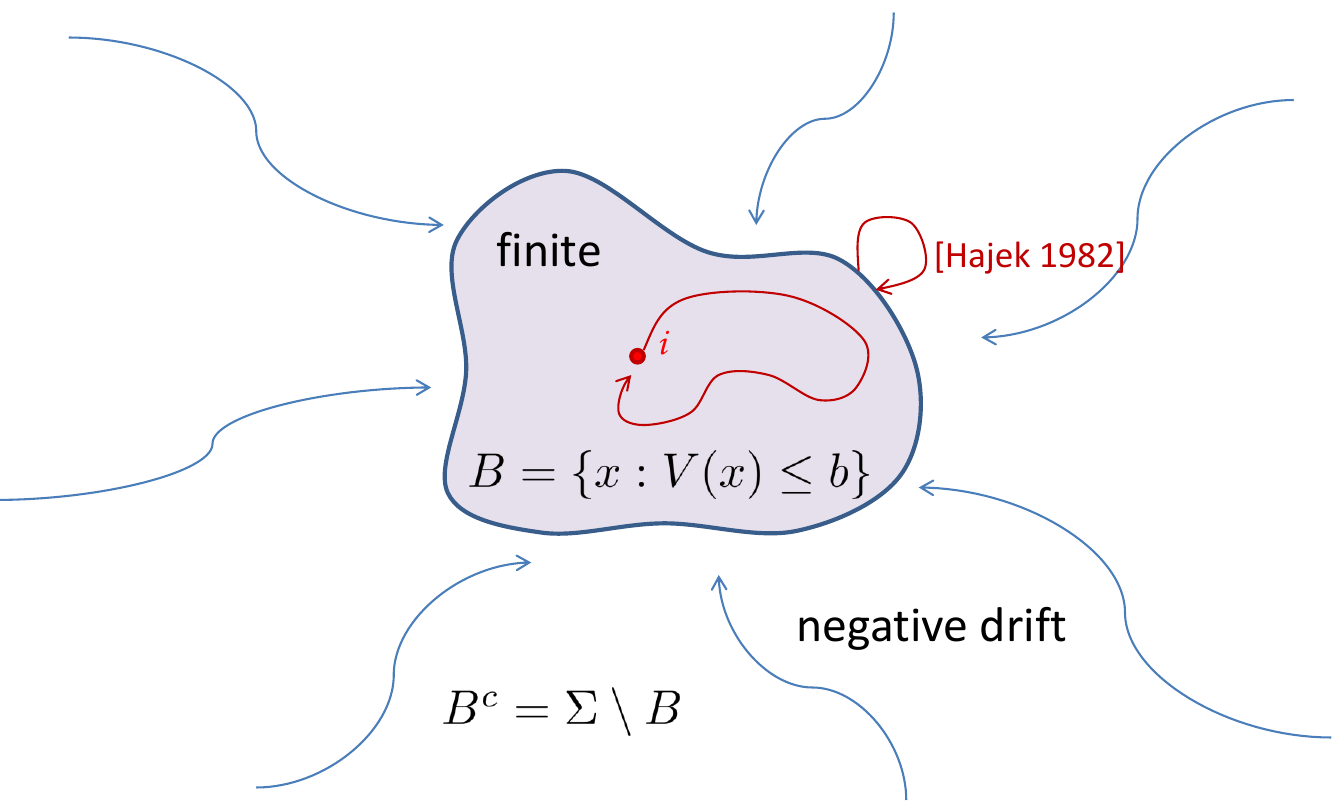}
\caption{This illustrates the implication of Assumption \ref{lyapunovAssumption}, which uses a Lyapunov function to decompose the state space into a finite region $B$ and a region with negative drift.} \label{decomposeSS}
\end{figure}

At first glance, this assumption may seem very restrictive. But in fact, this is quite reasonable: by the Foster-Lyapunov criteria (see Theorem \ref{foster} in Appendix), a countable state space Markov chain is positive recurrent if and only if there exists a Lyapunov function $V: \Sigma \to \RealsP$ that satisfies condition (1) and (3), as well as (2'): $\E[V(X_{t+1}) | X_{t} = x] < \infty$ for all $x \in \Sigma$. Assumption \ref{lyapunovAssumption} has (2), which is a restriction of the condition (2'). The implications of Assumption \ref{lyapunovAssumption} are visualized in Figure \ref{decomposeSS}. The existence of the Lyapunov function allows us to decompose the state space into sets $B$ and $B^c$ such that for all states $x \in B^c$, there is an expected decrease in the Lyapunov function in the next step or transition. Therefore, for all states in $B^c$, there is a negative drift towards set $B$. In addition, in any single step, the random walk cannot escape ``too far''.
The Lyapunov function helps to impose a natural ordering over the state space that allows us to prove properties of the Markov chain. There have been many results that use Lyapunov analysis to give bounds on the stationary probabilities, return times, and distribution of return times as a function of the Lyapunov function \citep{Hajek82, Bertsimas98}. Building upon results by Hajek, we prove the following lemma which establishes that return times have exponentially decaying tails even for countable-state space Markov chains, as long as they satisfy Assumption \ref{lyapunovAssumption}.
\begin{lemma} \label{lemma:concentration_RT}
Let $\{X_t\}$ be an irreducible Markov chain satisfying Assumption \ref{lyapunovAssumption}. For any $i \in B$ and for all $k \in \IntegersP$,
\[\Prob_i \left(T_i > k \right) \leq 4 \cdot 2^{- \frac{k}{R_i}},\]
where
\[R_i = O\left(\frac{H_i^B e^{2 \eta \nu_{\max}}}{(1-\rho)(e^{\eta \nu_{\max}} - \rho)}\right),\]
and $H_i^B$ is the maximal hitting time over the Markov chain with its state space restricted to the subset $B$. The scalars $\eta$ and $\rho$ are functions of $\gamma$ and $\nu_{\max}$ (see \eqref{eq:hajekConst} in Appendix \ref{app:lyapunov}).
\end{lemma}

Lemma \ref{lemma:concentration_RT} pertains to states $i \in B$ such that $V(i) \leq b$. This is not restrictive, since for any state $k$ of interest such that $V(k) = b' > b$, we can define a new Lyapunov function $V'(\cdot)$ such that $V'(k) = b$, and $V'(j) = V(j)$ for all $j \neq k$. Then we define $B' = \{ j \in \Sigma : V'(j) \leq b\} = B \cup \{k\}$ and $\nu_{\max}' = \nu_{\max} + b' - b$. By extension, Assumption \ref{lyapunovAssumption} holds for $V'(\cdot)$ with constants $\nu_{\max}', \gamma$, and $b'$. 

The quantity $R_i$ in Lemma \ref{lemma:concentration_RT} for countable state space Markov chains plays the same role as $H_i$ in Lemma \ref{lemma:finite} for finite state space Markov chains. Thus, equivalent theorems for the countable state space setting are obtained by using Lemma \ref{lemma:concentration_RT} rather than Lemma \ref{lemma:finite}. In the countable state space setting, $H_i$ and $Z_{\max}(i)$ no longer are well defined since the maximum over an infinite set may not exist. However, we recall that $Z_{\max}(i) = O(\pi_i H_i)$, and thus our analysis of the algorithm for finite state space Markov chains extend to countable state space Markov chains by substituting $R_i$ for $H_i$ and $\pi_i R_i$ for $Z_{\max}(i)$.

This Theorem leads to some interesting insights about the performance of our algorithm and computation of stationary probabilities over large finite Markov chains. In some sense, the bound in Lemma \ref{lemma:finite} is not very tight as the size of the state space grows, since it takes a maximum over all states. However, in fact, Lemma \ref{lemma:concentration_RT} indicates that it is perhaps the mixing properties of the local neighborhood that matters the most. In addition, Assumption 1 also lends insights into the properties that become significant for a large finite state space Markov chain. In some sense, if the large finite state space Markov chain mixes poorly, then it is kind of the notion of the Markov chain growing to a limiting countably infinite state space Markov chain which is no longer positive recurrent (i.e. becomes separate recurrence classes). In this setting, we argue that the true stationary distribution is no longer significant, and perhaps the significant quantity may be separate local stationary distributions over each community or subset.

\section{Analysis of Estimation Error} \label{sec:error_analysis}

In this section, we provide bounds on the estimates produced by the algorithm. The omitted proofs can be found in the Appendix. Recall that the estimate $\hat{\pi}_i^{(k)}$ concentrates around $1/\E_i[T_i^{(k)}]$, and $\pi_i = 1/\E_i[T_i]$. Therefore we begin by characterizing the difference between the truncated mean return time and the original mean return time.

\subsection{Precise Characterization of Expected Error via Local Mixing Times}

 Lemma \ref{lemma:exp_1} expresses the ratio between the $\E_i[T_i^{(k)}]$ and $\E_i[T_i]$ as a function of $\Prob_i\left(T_i > \theta^{(k)}\right)$ and the fundamental matrix $Z$. This lemma shows the error purely due to the truncation bias and not stochastic sampling error.

\begin{lemma} \label{lemma:exp_1}
For an irreducible, positive recurrent Markov chain $\{X_t\}$ with countable state space $\Sigma$ and transition probability matrix $P$, and for any $i \in \Sigma$ and $t \in \IntegersP$,
\begin{align} \label{eq:exp_1}
1 - \frac{\E_i[\hat{T}_i^{(k)}]}{\E_i[T_i]} &= \Prob_i\left(T_i > \theta^{(k)}\right) \Gamma_i(\theta^{(k)}),
\end{align}
where
\begin{align} \label{eq:local_mixing}
\Gamma_i(\theta) \triangleq \left(\sum_{q \in \Sigma \setminus \{i\}} \Prob_i\left(\left.X_{\theta} = q \right| T_i > \theta\right) (Z_{ii} - Z_{qi}) \right).
\end{align}
\end{lemma}

\proof{Proof of Lemma \ref{lemma:exp_1}.}
We divide the equation given in Lemma \ref{lemma:E_diff} by $\E_i[T_i]$. Then we apply Bayes' rule, the law of total probability, and the Markov property.
\begin{align}
&1 - \frac{\E_i[\hat{T}_i^{(k)}]}{\E_i[T_i]} = \frac{1}{\E_i[T_i]}\sum_{k=\theta^{(k)}}^{\infty} \Prob_i(T_i > k) \label{eq:exp_1_step_0}\\
& = \frac{\Prob_i\left(T_i > \theta^{(k)}\right)}{\E_i[T_i]} \sum_{k=\theta^{(k)}}^{\infty} \Prob_i\left(\left. T_i > k \right| T_i > \theta^{(k)}\right) \nonumber\\
& = \frac{\Prob_i\left(T_i > \theta^{(k)}\right)}{\E_i[T_i]} \sum_{k=\theta^{(k)}}^{\infty} \sum_{q \in \Sigma \setminus \{i\}} \Prob_i\left(\left.T_i > k \right| X_{\theta^{(k)}} = q, T_i > \theta^{(k)}\right) \Prob_i\left(\left. X_{\theta^{(k)}} = q\right| T_i > \theta^{(k)}\right) \nonumber\\
& = \Prob_i\left(T_i > \theta^{(k)}\right) \sum_{q \in \Sigma \setminus \{i\}} \Prob_i\left(\left.X_{\theta^{(k)}} = q\right| T_i > \theta^{(k)}\right) \frac{\E_q[T_i]}{\E_i[T_i]}. \label{eq:exp_1_step}
\end{align}
Finally, we use Lemma \ref{lemma:aldous_1} to complete the proof.\hfill \halmos
\endproof

In order to understand the expression $\Gamma_i(\theta)$, we observe that by the definition of $Z$,
\[Z_{ii} - Z_{qi} = \sum_{k=0}^{\infty} \left(P_{ii}^{(k)} - P_{qi}^{(k)}\right) \leq \sum_{k=0}^{\infty} \bar{d}(t)
\leq \sum_{k=0}^{\infty} 2^{-\lfloor t/\tmix \rfloor} \approx 2 \ln(2) \tmix.\]

Although $\Gamma_i(\theta)$ is globally upper bounded by $2 Z_{\max}$ and $2 \ln(2) \tmix$ for all $i$ and $\theta$, it is actually a convex combination of the quantities $(Z_{ii} - Z_{qi})$, where each term is weighted according to the probability that the random walk is at state $q$ after $\theta$ steps. Because the random walks begin at state $i$, we expect this distribution to more heavily weight states which are closer to $i$, in fact the support of this distribution is limited to states that are within a $\theta$ distance from $i$. Thus, $\Gamma_i(\theta)$ can be interpreted as a locally weighted variant of the mixing time, which we term the ``local mixing time'', measuring the time it takes for a random walk beginning at some distribution of states around $i$ to reach stationarity at state $i$, where the size of the local neighborhood considered depends on $\theta$. In scenarios where these ``local mixing times'' for different states differ from the global mixing time, then our algorithm which utilizes local random walk samples may provide tighter results than standard MCMC methods. This suggests an interesting mathematical inquiry of how to characterize local mixing times of Markov chains, and in what settings they may be homogenous as opposed to heterogenous.

This lemma gives us a key insight into the termination conditions of the algorithm. Recall that termination condition (b) is satisfied when $\hat{p}^{(k)} < 2 \epsilon / 3$. Since $\hat{p}^{(k)}$ concentrates around $\Prob_i\left(T_i > \theta^{(k)}\right)$, and $\hat{\pi}_i^{(k)}$ concentrates around $1/\E_i[\hat{T}_i^{(k)}]$, Lemma \ref{lemma:exp_1} indicates that when the algorithm stops at condition (b), the multiplicative error between $\hat{\pi}_i^{(k)}$ and $\pi_i$ is approximately $\epsilon \Gamma_i (\theta^{(k)}) \leq 2 \epsilon Z_{\max}(i)$. 

\subsection{Error of Basic Algorithm Estimates}

Theorem \ref{thm:main_general_b} states that with high probability, for any irreducible, positive recurrent Markov chain, the estimate produced by the basic algorithm is always an upper bound with high probability.

\begin{theorem} \label{thm:main_general_b}
For an irreducible, positive recurrent, countable state space Markov chain, and for any $i \in \Sigma$, with probability greater than $(1-\alpha)$, for all $k$,
\begin{align*} 
\hat{\pi}_i^{(k)} \geq \frac{\pi_i}{1 + \epsilon}.
\end{align*}
\end{theorem}


\proof{Proof of Theorem \ref{thm:main_general_b}.}
This result follows direcly from Lemma \ref{lemma:error_bounds}, which implies that $\hat{\pi}_i^{(k)}$ lies within $1 / (1 \pm \epsilon) \E_i[\hat{T}_i^{(k)}]$. Due to truncation, $\E_i[\hat{T}_i^{(k)}] \leq \E_i[T_i]$, thus the estimate is an upper bound of $\pi_i$ with high probability. \hfill \halmos
\endproof

Theorem \ref{thm:finite_correctness} upper bounds the the percentage error between $\hat{\pi}_i^{(k)}$ and $\pi_i$.

\begin{theorem} \label{thm:finite_correctness}
For an irreducible finite state space Markov chain, for any $i \in \Sigma$, with probability greater than $1 - \alpha$, for all iterations $k$,
\begin{align*}
\left|\frac{\hat{\pi}_i^{(k)} - \pi_i}{\hat{\pi}_i^{(k)}} \right| 
&\leq 2 (1 - \epsilon) \Prob_i(T_i > \theta^{(k)}) Z_{\max}(i) + \epsilon, \\
&\leq 4 (1 - \epsilon) 2^{-\theta^{(k)}/2 H_i} Z_{\max}(i) + \epsilon.
\end{align*}
\end{theorem}

Corollary \ref{cor:phat_error} directly follows from Theorem \ref{thm:finite_correctness} and Lemma \ref{lemma:p_add_concentrate}, allowing us to upper bound the error as a function of $\hat{p}^{(k)}$. This corollary motivates the choice of termination criteria, indicating that terminating when $\hat{p} \leq 2 \epsilon / 3$ results in an error bound of $\epsilon (2 Z_{\max}(i) + 1)$.

\begin{corollary} \label{cor:phat_error}
With probability greater than $1 - \alpha$, for all iterations $k$,
\begin{align*}
\left|\frac{\hat{\pi}_i^{(k)} - \pi_i}{\hat{\pi}_i^{(k)}} \right| 
&\leq \epsilon \left(2 Z_{\max}(i)/3 + 1\right) + 2 \hat{p}^{(k)} Z_{\max}(i) .
\end{align*}
\end{corollary}

We proceed to prove Theorem \ref{thm:finite_correctness} by combining Lemmas \ref{lemma:exp_1}, \ref{lemma:error_bounds},  \ref{lemma:p_add_concentrate}, and \ref{lemma:finite}.

\proof{Proof of Theorem \ref{thm:finite_correctness}.}
By Theorem \ref{thm:main_general_b}, it follows that $(\pi_i - \hat{\pi}_i^{(k)})/\hat{\pi}_i^{(k)} < \epsilon$ with high probability. By Lemma \ref{lemma:error_bounds} and Lemma \ref{lemma:exp_1}, it follows that with high probability,
\begin{align}
\frac{\hat{\pi}_i^{(k)}-\pi_i}{\hat{\pi}_i^{(k)}} &= 1 - \frac{\hat{T}_i^{(k)}}{\E_i[T_i]} \nonumber \\
&\leq 1 - \frac{(1-\epsilon)\E_i[\hat{T}_i^{(k)}]}{\E_i[T_i]} \label{eq:lemma_exp_1} \\
&= 1 - (1-\epsilon) (1 - \Prob_i(T_i > \theta^{(k)}) \Gamma_i(\theta^{(k)}) )  \nonumber \\
&= (1 - \epsilon) \Prob_i(T_i > \theta^{(k)}) \Gamma_i(\theta^{(k)}) + \epsilon. \nonumber \\
&= 2 (1 - \epsilon) \Prob_i(T_i > \theta^{(k)}) Z_{\max}(i) + \epsilon. \nonumber
\end{align}
When the algorithm terminates at condition (b), $\hat{p}^{(k)} < 2 \epsilon / 3$. By Lemma \ref{lemma:p_add_concentrate}, with probability greater than $1 - \alpha$, $\Prob_i(T_i > \theta^{(k)}) \leq \hat{p}^{(k)} + \frac{\epsilon}{3} \leq \epsilon$. Therefore, it follows that
\begin{align*}
\frac{\hat{\pi}_i^{(k)} - \pi_i}{\hat{\pi}_i^{(k)}}
&\leq \epsilon \left(2 Z_{\max}(i) + 1\right).
\end{align*}\hfill \halmos
\endproof
Theorem \ref{thm:finite_correctness} shows that the error bound decays exponentially in $\theta^{(k)}$, which doubles in each iteration. Thus, for every subsequent iteration $k$, the estimate $\hat{\pi}_i^{(k)}$ approaches $\pi_i$ exponentially fast. The key part of the proof relies on the fact that the distribution of the return time $T_i$ has an exponentially decaying tail, ensuring that the return time $T_i$ concentrates around its mean $\E_i[T_i]$. Theorem \ref{thm:countable_correctness} presents the error bounds for countable state space Markov chains, relying upon the exponentially decaying tail proved in Lemma \ref{lemma:concentration_RT}.

\begin{theorem} \label{thm:countable_correctness}
For a Markov chain satisfying Assumption \ref{lyapunovAssumption}, for any $i \in B$, with probability greater than $1 - \alpha$, for all iterations $k$,
\begin{align*}
\left| \frac{\hat{\pi}_i^{(k)} - \pi_i}{\hat{\pi}_i^{(k)}} \right| &\leq 4 (1 - \epsilon) \left(\frac{2^{-\theta^{(k)}/R_i}}{1-2^{-1/R_i}}\right) \pi_i + \epsilon, \\
&\approx 4 ( 1 - \epsilon) \ln(2) \pi_i R_i 2^{-\theta^{(k)}/R_i} + \epsilon.
\end {align*}
\end{theorem}

\proof{Proof of Theorem \ref{thm:countable_correctness}.}
This proof follows a similar proof of Theorem \ref{thm:finite_correctness}. Substitute Lemma \ref{lemma:concentration_RT} into \eqref{eq:exp_1_step_0} to show that
\begin{align}
1 - \frac{\E_i[\hat{T}_i^{(k)}]}{\E_i[T_i]} = \frac{1}{\E_i[T_i]}\sum_{k=\theta^{(k)}}^{\infty} \Prob_i(T_i > k) \leq \pi_i \left(\frac{4 \cdot 2^{-\theta^{(k)} / R_i}}{1 - 2^{-1 /R_i}}\right). \label{eq:countable_correctness}
\end{align}
Substitute \eqref{eq:countable_correctness} into \eqref{eq:lemma_exp_1} to complete the proof.\hfill \halmos
\endproof

\subsection{Error of Bias Corrected Estimate}

Lemma \ref{lemma:exp_2} gives an expression for how the bias corrected estimate differs from the true value if we had the exact expected return time as well as the probability of truncation. By comparing Lemma \ref{lemma:exp_2} with Lemma \ref{lemma:exp_1}, we gain some intuition of the difference in the expected error for the original estimator $\hat{\pi}_i$ and the bias corrected estimator $\tilde{\pi}_i$.
Recall that $\tilde{\pi}_i^{(k)}$ is equal to $(1 - \hat{p}^{(k)})/\hat{T}_i^{(k)}$. Lemma \ref{lemma:exp_2} gives an expression for the additive difference between $(1 - \Prob_i(T_i > \theta^{(k)}))/\E_i[T_i^{(k)}]$ and $E_i[T_i]$.

\begin{lemma} \label{lemma:exp_2}
For an irreducible, positive recurrent Markov chain $\{X_t\}$ with countable state space $\Sigma$ and transition probability matrix $P$, and for any $i \in \Sigma$ and $t \in \IntegersP$,
\begin{align} \label{eq:exp_2}
\frac{(1-\Prob_i(T_i > \theta^{(k)}))}{\E_i[\hat{T}_i^{(k)}]} - \pi_i &= \frac{\Prob_i(T_i > \theta^{(k)})}{\E_i[\hat{T}_i^{(k)}]} \left(\Gamma_i(\theta^{(k)}) - 1\right),
\end{align}
where
\begin{align}
\Gamma_i(\theta) \triangleq \left(\sum_{q \in \Sigma \setminus \{i\}} \Prob_i\left(\left.X_{\theta} = q \right| T_i > \theta\right) (Z_{ii} - Z_{qi}) \right).
\end{align}
\end{lemma}

\proof{Proof of Lemma \ref{lemma:exp_2}.}
This lemma follows directly from Lemma \ref{lemma:exp_1}. \hfill \halmos
\endproof

In comparing Lemma \ref{lemma:exp_1} and \ref{lemma:exp_2}, we see that the additive estimation error for both estimators are almost the same except for $(\Gamma_i(\theta^{(k)}) - 1)$ in Lemma \ref{lemma:exp_2} as opposed to $\Gamma_i(\theta^{(k)})$ in Lemma \ref{lemma:exp_1}. Therefore, when $\Gamma_i(\theta^{(k)})$ is small, we expect $\tilde{\pi}_i$ to be a better estimate than $\hat{\pi}_i$; however, when $\Gamma_i(\theta^{(k)})$ is large, then the two estimates will have approximately the same error.  Theorem \ref{thm:countable_bias} presents an equivalent bound for the error of $\tilde{\pi}_i$ when the algorithm is implemented on a countable state space Markov chain.

\begin{theorem} \label{thm:countable_bias}
For a Markov chain satisfying Assumption \ref{lyapunovAssumption}, for any $i \in B$, with probability greater than $1 - \alpha$, for all iterations $k$ such that $\Prob_i(T_i > \theta^{(k)}) < 1/2$,
\begin{align*}
\left| \frac{\tilde{\pi}_i^{(k)} - \pi_i}{\tilde{\pi}_i^{(k)}} \right| 
&\leq \frac{8 (1+ \epsilon)}{1- \epsilon} 2^{-\theta^{(k)}/R_i} \max \left(\frac{\pi_i}{1-2^{-1/R_i}},1\right) + \frac{2 \epsilon}{1 - \epsilon}, \\
&\approx \frac{8 (1+ \epsilon)}{1- \epsilon} 2^{-\theta^{(k)}/R_i} \max \left(\ln(2) \pi_i R_i,1\right) + \frac{2 \epsilon}{1 - \epsilon}.
\end{align*}
\end{theorem}

\subsection{Error of Estimates for observer states $j$}

As the number of samples increases, the sample mean converges to the true epxected value, and thus the error bound stated in Lemme \ref{lemma:exp_3} shows the bias of the estimates $\tilde{\pi}_j^{(k)}$.

\begin{lemma} \label{lemma:exp_3}
For an irreducible, positive recurrent Markov chain $\{X_t\}$ with countable state space $\Sigma$ and transition probability matrix $P$, and for any $i,j \in \Sigma$, and $t \in \IntegersP$,
\begin{align} \label{eq:exp_3}
\frac{\E_i[\hat{F}_j^{(k)}]}{\E_i[\hat{T}_i^{(k)}]} - \pi_j &= \frac{\Prob_i\left(T_i > \theta^{(k)}\right)}{\E_i[\hat{T}_i^{(k)}]} \left(\sum_{q \in \Sigma \setminus \{i\}} \Prob_i\left(\left.X_{\theta^{(k)}} = q \right| T_i > \theta^{(k)}\right) (Z_{ij} - Z_{qj}) \right)
\end {align}
\end{lemma}

Compare Lemmas \ref{lemma:exp_1} and \ref{lemma:exp_3}. Although they look similar, observe that if we bound $(Z_{ij} - Z_{qj})$ by $2 Z_{\max}(j)$, it becomes clear that Lemma \ref{lemma:exp_3} depends on the Markov chain mixing properties with respect to both state $i$ through $\Prob_i(T_i > \theta^{(k)})$, and state $j$ through $Z_{\max}(j)$.

\subsection{Tightness of Analysis}

In this section, we discuss the tightness of our analysis. Lemmas \ref{lemma:exp_1}, \ref{lemma:exp_2}, and \ref{lemma:exp_3} give exact expressions of the estimation error that arises from the truncation of the sample random walks. For a specific Markov chain, Theorems \ref{thm:finite_correctness}, \ref{thm:bias}, and \ref{thm:freq} could be loose due to two approximations. First, $2 Z_{\max}(i)$ could be a loose upper bound upon $\Gamma_i(\theta^{(k)})$. Second, Lemma \ref{lemma:finite} could be loose due to its use of the Markov inequality. Since $\hat{\pi}_i$ is greater than $\pi_i$ with high probability, Theorem \ref{thm:finite_correctness} is only useful when the upper bound is less than 1. We will show that for a specific family of graphs, namely clique graphs, our bound scales correctly as a function of $\theta^{(k)}$, $H_i$, and $Z_{\max}(i)$.

Consider a family of clique graphs $G_n$ indexed by $n \in \IntegersP$, such that $G_n$ is the clique graph over $n$ vertices. For $G_n$, we can directly compute the hitting time $H_i = n$, the truncation probability $\Prob_i(T_i > k)\approx e^{-(\theta^{(t)}-1)/(n-1)}$, and values of the fundamental matrix $Z$, specifically $Z_{ii} - Z_{ji} = \pi_i \E_j[T_i] = \frac{n-1}{n}$ and $Z_{\max}(i) = \frac{n^2-n+1}{n^2}$. By substituting these into Lemma \ref{lemma:exp_1}, it follows that the expected error is approximately given by
\begin{align*}
1 - \frac{\E_i[\hat{T}_i^{(k)}]}{\E_i[T_i]} &= e^{-(\theta^{(k)}-1)/(n-1)} \left(\frac{n-1}{n}\right).
\end{align*}
By substituting these into Theorem \ref{thm:finite_correctness}, we show that with probability at least $1 - \alpha$,
\begin{align*}
\left|1 - \frac{\pi_i}{\hat{\pi}_i^{(k)}} \right| &\leq 4 (1 - \epsilon) e^{-\theta^{(k)}\ln(2)/2n} \left(\frac{n^2-n+1}{n^2}\right) + \epsilon.
\end{align*}
While Theorem \ref{thm:finite_correctness} gives that the percentage error is upper bounded by $O(e^{-\theta^{(k)} \ln(2) /2 H_i} Z_{\max}(i))$, by Lemma \ref{lemma:exp_1} the percentage error of our algorithm on the clique graph is no better than $\Omega(e^{-\theta^{(k)}/H_i} Z_{\max}(i))$. If we used these bounds to determine a threshold $\theta^{(k)}$ large enough to guarantee that the multiplicative error is bounded by $\epsilon$, the threshold computed via Theorem \ref{thm:finite_correctness} would only be a constant factor of $2/\ln(2)$ larger than the threshold computed via Lemma \ref{lemma:exp_1}. Our algorithm leverages the property that there is some concentration of measure, or ``locality'', over the state space. It is the worst when there is no concentration of measure, and the random walks spread over the space quickly and take a long time to return, such as a random walk over the clique graph. For Markov chains that have strong concentration of measure, such as a biased random walk on the positive integers, the techniques given in Section \ref{sec:exp_decay} for analyzing countable state space Markov chains using Lyapunov functions will obtain tighter bounds, as compared to using the hitting time $H_i$ and $Z_{\max}(i)$, since these quantities are computed as a worst case over all states, even if the random walk only stays within a local region around $i$.

\section{Cost of Computation} \label{sec:comp_cost}

In this section, we compute bounds on the computation cost of the algorithm. We first prove that the total number of random walk steps taken by the algorithm within the first $k$ iterations scales with $2^k$, which we recall is equivalent to $\theta^{(k)}$ by design. 

\begin{lemma} \label{lemma:stepsTaken}
With probability greater than $(1 - \alpha)$, the total number of random walk steps taken by the algorithm within the first $k$ iterations is bounded by
\[\tilde{O}\left( \frac{\ln(1/\alpha) 2^k}{\epsilon^2} \right).\]
\end{lemma}

\proof{Proof of Lemma \ref{lemma:stepsTaken}.}
The total number of random walk steps (i.e., oracle calls) used in the algorithm over all $k$ iterations is equal to $\sum_{h=1}^k N^{(h)} \hat{T}_i^{(h)}$. We condition on the event that $\hat{T}_i^{(h)}$ is within a $(1 \pm \epsilon)$ multiplicative interval around its mean for all $h \in 1, \dots k$, which occurs with probability greater than $(1 - \alpha)$ by Lemma \ref{lemma:error_bounds}. Because $\theta^{(h)}$ doubles in each iteration, $\E_i[\hat{T}_i^{(h)}] \leq 2 \E_i[\hat{T}_i^{(h-1)}]$. By combining these facts with the definition of $N^{(h)}$, we obtain an upper bound as a function of $\theta^{(h)}, \alpha, \epsilon,$ and $\E_i[\hat{T}_i^{(h)}]$. We suppress the insignificant factors $(1 + \epsilon)$ and $(1 - \epsilon)$. Since $N^{(h)} \hat{T}_i^{(h)}$ grows super exponentially, the largest term of the summation dominates.
\begin{align*}
\sum_{h=1}^k N^{(h)} \hat{T}_i^{(h)} &= \sum_{h=1}^k O\left(\frac{2^h \ln(2^h / \alpha)}{\epsilon^2}\right) \\
&= O\left(\frac{2^k \ln(2^k / \alpha)}{\epsilon^2}\right) = \tilde{O}\left(\frac{2^k \ln(1/\alpha)}{\epsilon^2}\right).
\end{align*}\hfill \halmos
\endproof

The next two theorems provide upper bounds for the number of iterations until \texttt{TerminationCriteria} is satisfied. Theorem \ref{thm:main_general_a} asserts that with high probability, the algorithm terminates in finite time as a function of the parameters of the algorithm, independent from the size of the Markov chain state space. It is proved by showing that if $\theta^{(k)} > 3 (1 + \epsilon) / 2 \epsilon \Delta$, then either termination condition (a) or (b) must be satisfied. 

\begin{theorem} \label{thm:main_general_a}
For an irreducible, positive recurrent, countable state space Markov chain, and for any $i \in \Sigma$, with probability 1, the total number of iterations $k$ before the algorithm satisfies either $\hat{\pi}_i^{(k)} < \Delta/(1 + \epsilon)$ or $\hat{p}^{(k)} < 2 \epsilon / 3$ is bounded above by
\[\log_2\left(\frac{3 (1 + \epsilon)}{2 \epsilon \Delta}\right).\]
With probability greater than $1 - \alpha$, the computation time, or the total number of random walk steps (i.e. oracle calls) used by the algorithm is bounded above by
\[\tilde{O}\left( \frac{\ln(\frac{1}{\alpha})}{\epsilon^3 \Delta} \right).\]
\end{theorem} 

\proof{Proof of Theorem \ref{thm:main_general_a}.}
By definition, $\hat{T}_i^{(k)} \geq \hat{p}^{(k)} \theta^{(k)}$, which implies that $\hat{p}^{(k)} \leq 1/\hat{\pi}_i^{(k)}\theta^{(k)}$. When $\theta^{(k)} \geq 3 (1 + \epsilon) / 2 \epsilon \Delta$, if termination condition (a) has not been satisfied, then $\hat{\pi}_i^{(k)} \geq \Delta/(1 + \epsilon)$. This implies that $\hat{p}^{(k)} \leq 1/\hat{\pi}_i^{(k)}\theta^{(k)} \leq 2 \epsilon/3$, satisfying termination condition (b). This provides an upper bound for the number of iterations before the algorithm terminates, and we can substitute this into Lemma \ref{lemma:stepsTaken} to complete the proof.\hfill \halmos
\endproof

As can be seen through the proof, Theorem \ref{thm:main_general_a} does not utilize any information or properties of the Markov chain. Theorems \ref{thm:finite_convergence_a} and \ref{thm:finite_convergence_a} use the exponential tail bounds in Lemma \ref{lemma:finite} to prove that as a function of the mixing preoperties, if $\theta^{(k)}$ is large enough, the fraction of truncated samples is small, and the error is bounded, such that either one of the termination conditions are satisfied.

\begin{theorem} \label{thm:finite_convergence_a}
For an irreducible finite state space Markov chain, for any state $i \in \Sigma$ such that $\pi_i < (1 - \epsilon)\Delta / (1 + \epsilon)$, with probability greater than $1 - \alpha$,
\begin{align}
\min\{t : \hat{\pi}_i^{(k)} < \Delta / (1 + \epsilon)\} \leq \log_2\left(2 H_i \log_2\left( 4 Z_{\max}(i) \left(1 - \frac{(1+\epsilon) \pi_i}{(1 - \epsilon)\Delta}\right)^{-1}\right)\right).\label{eq:t_max_a}
\end{align}
Thus the total number of random walk steps (i.e. oracle calls) used by the algorithm before satisfying $\hat{\pi}_i^{(k)} < \Delta /(1 + \epsilon)$ is bounded above by
\[\tilde{O}\left( \frac{H_i \ln(1/\alpha)}{\epsilon^2} \ln\left(Z_{\max}(i) \left(1 - \frac{\pi_i}{\Delta}\right)^{-1}\right)\right).\]
\end{theorem}

\proof{Proof of Theorem \ref{thm:finite_convergence_a}.}
For $k$ larger than the expression given in \eqref{eq:t_max_a}, we compute an upper bound on $\hat{\pi}_i^{(k)}$ by substituting into Theorem \ref{thm:finite_correctness}. It follows that $\hat{\pi}_i^{(k)} < \Delta / (1 + \epsilon)$ with probability greater than $1 - \alpha$. \hfill \halmos
\endproof

\begin{theorem} \label{thm:finite_convergence_b} 
For an irreducible finite state space Markov chain, for all $i \in \Sigma$, with probability greater than $1 - \alpha$,
\begin{align}
\min\{t : \hat{p}^{(k)} < \delta\} \leq \log_2\left(2 H_i \log_2\left(6/(3 \delta - \epsilon)\right)\right).\label{eq:t_max_b}
\end{align}
The total number of random steps (i.e. oracle calls) used by the algorithm before satisfying $\hat{p}^{(k)} < \delta$ is bounded above by
\[\tilde{O}\left( \frac{H_i \ln(1/\alpha) \ln(6/(3 \delta - \epsilon))}{\epsilon^2} \right).\]
\end{theorem}

\proof{Proof of Theorem \ref{thm:finite_convergence_b}.}
For $k$ larger than the expression specified in \eqref{eq:t_max_b}, we show that $\Prob_i\left(T_i > \theta^{(k)}\right) < \delta - \epsilon/3$ by substituting into Lemma \ref{lemma:finite}. We use Lemma \ref{lemma:p_add_concentrate} to show that $\hat{p}^{(k)} < \delta$ with probability greater than $1 - \alpha$.\hfill \halmos
\endproof

For states $i$ in a Markov chain such that the maximal hitting time is small, the bounds given in Theorem \ref{thm:finite_convergence_b} will be smaller than the general bound given in Theorem \ref{thm:main_general_a}. Given a state $i$ such that $\pi_i < (1 - \epsilon) \Delta/(1 + \epsilon)$, our tightest bound is given by the minimum over the expressions from Theorems \ref{thm:main_general_a}, \ref{thm:finite_convergence_a}, and \ref{thm:finite_convergence_b}. For a state $i$ such that $\pi_i > \Delta$, our tightest bound is given by the minimum between Theorem \ref{thm:main_general_a} and \ref{thm:finite_convergence_b}.

Theorem \ref{thm:countable_convergence} presents the equivalent result for countable state space Markov chains.

\begin{theorem} \label{thm:countable_convergence} For a Markov chain satisfying Assumption \ref{lyapunovAssumption}, \\
{\bf(a)} For any state $i \in B$ such that $\pi_i < (1 - \epsilon) \Delta / (1 + \epsilon) $, with probability greater than $1 - \alpha$, the total number of steps used by the algorithm before satisfying $\hat{\pi}_i^{(k)}$ is bounded above by
\[\tilde{O}\left( \frac{R_i \ln(\frac{1}{\alpha})}{\epsilon^2} \ln\left(\pi_i R_i \left(1 - \frac{\pi_i}{\Delta}\right)^{-1}\right)\right).\]
{\bf (b)} For all states $i \in B$, with probability greater than $1 - \alpha$, the total number of steps used by the algorithm before satisfying $\hat{p}^{(k)} < 2 \epsilon / 3$ is bounded above by
\[\tilde{O}\left( \frac{R_i \ln(\frac{1}{\alpha})}{\epsilon^2} \right).\]
\end{theorem}

\proof{Proof of Theorem \ref{thm:countable_convergence}.}
The proof is exactly the same as the proof of Theorems \ref{thm:finite_convergence_a} and \ref{thm:finite_convergence_b}, except that we use Theorem \ref{thm:countable_correctness} instead of Theorem \ref{thm:finite_correctness}, and Lemma \ref{lemma:concentration_RT} instead of Lemma \ref{lemma:finite}.\hfill \halmos
\endproof

\section{Examples and Simulations} \label{section:simulations}

We present the results of applying our algorithm to concrete examples of Markov chains. The examples illustrate the wide applicability of our algorithm for estimating stationary probabilities.

\begin{example}[\pagerank]
In analyzing the web graph, \pagerank~is a frequently used measure to compute the importance of webpages. We are given a scalar parameter $\beta$ and an underlying directed graph over $n$ nodes, described by the adjacency matrix $A$ (i.e., $A_{ij} = 1$ if $(i,j) \in E$ and 0 otherwise). The transition probability matrix of the \pagerank~random walk is given by
\begin{align}
P = \frac{\beta}{n} \bOne \cdot \bOne^T + (1-\beta) D^{-1} A, \label{eq:PageRank_simulation}
\end{align}
where $D$ denotes the diagonal matrix whose diagonal entries are the out-degrees of the nodes. The state space is equivalent to the set of nodes in the graph, and it follows that
\[P_{rs} = \Prob(X_{t+1} = s | X_t = r) = \beta\left(\frac{1}{n}\right) + (1-\beta)\left(\frac{A_{rs}}{\sum_{v} A_{rv}}\right).\]
Thus, in every step, there is a $\beta$ probability of jumping uniformly randomly to any other node in the graph. In our simulation, $n = 100$, 
$\beta = 0.15$, and the underlying graph is generated according to the configuration model with a power law degree distribution: $\Prob(d) \propto d^{-1.5}$. We choose $\beta = 0.15$ to match the value used in the original definition of PageRank by Brin and Page. The exponent 1.5 was chosen so that the distribution is easy to plot and view for a scale of 100 nodes. We computed that $Z_{\max} \approx 3.5$. 
\end{example}

\begin{example}[Queueing System]
In queuing theory, Markov chains are commonly used to model the length of the queue of jobs waiting to be processed by a server, which evolves over time as jobs arrive and are processed. For illustrative purposes, we chose the M/M/1 queue, equivalent to a random walk on $\IntegersP$. The state space $\IntegersP$ is countably infinite. Assume we have a single server where the jobs arrive according to a Poisson process with parameter $\lambda$, and the processing time for a single job is distributed exponentially with parameter $\mu$. The queue length can be modeled with the random walk shown in Figure \ref{figure:MM1}, where $q_1$ is the probability that a new job arrives before the current job is finished processing, given by $\lambda/(\lambda + \mu)$. For the purposes of our simulation, we choose $q_1 = 0.3$, and estimate the stationary probabilities for the queue to have length $i$ for $i \in \{1,2,3, \dots 50\}$. The parameter $q_1 = 0.3$ was chosen so that the distribution is easy to plot and view for the scale of 50 states, and $\lambda$ and $\mu$ can be any values such that $\lambda/(\lambda + \mu) = 0.3$, for example $\lambda = 0.3$ and $\mu = 0.7$.
\end{example}

\begin{example}[Magnet Graph]
This example illustrates a Markov chain with poor mixing properties. The Markov chain is depicted in Figure \ref{figure:Magnet}, and can be described as a random walk over a finite section of the integers such that there are two attracting states, labeled in Figure \ref{figure:Magnet} as states 1 and $n_2$. We assume that $q_1, q_2 < 1/2$, such that for all states left of state $n_1$, the random walk will drift towards state 1 with probablity $1 - q_1$ in each step, and for all states right of state $n_1$, the random walk will drift towards state $n_2$ with probability $1 - q_2$ in each step. Due to this bipolar attraction, a random walk that begins on the left will tend to stay on the left, and similarly, a random walk that begins on the right will tend to stay on the right. For our simulations, we chose $q_1 = q_2 = 0.3$, $n_1 = 25,$ and $n_2 = 50$. We computed that $Z_{\max} \approx 1.4 \times 10^9$.
\end{example}

We show the results of applying our algorithm to estimate the stationary probabilities in these three different Markov chains, using algorithm parameters $\Delta = 0.02$, $\epsilon = 0.15$, and $\alpha = 0.2$. The three Markov chains have different mixing properties, chosen to illustrate the performance of our algorithm on Markov chains with different values of  $H_i$ and $Z_{\max}(i)$. Let $t_{\max}$ denote the final iteration at which the algorithm terminates. In the following figures and discussion, we observe the accuracy of the estimates, the computation cost, truncation threshold, and fraction of samples truncated as a function of the estimation threshold $\Delta$, the stationary probability of the anchor state, and the properties of the Markov chain.

\subsection{Basic Estimate vs. Bias Corrected Estimates}

\begin{figure}[h]
     \centering
        \subfigure[PageRank Estimates.]{
           \label{figure:PageRank_estimate}
           \includegraphics[width=0.48\textwidth]{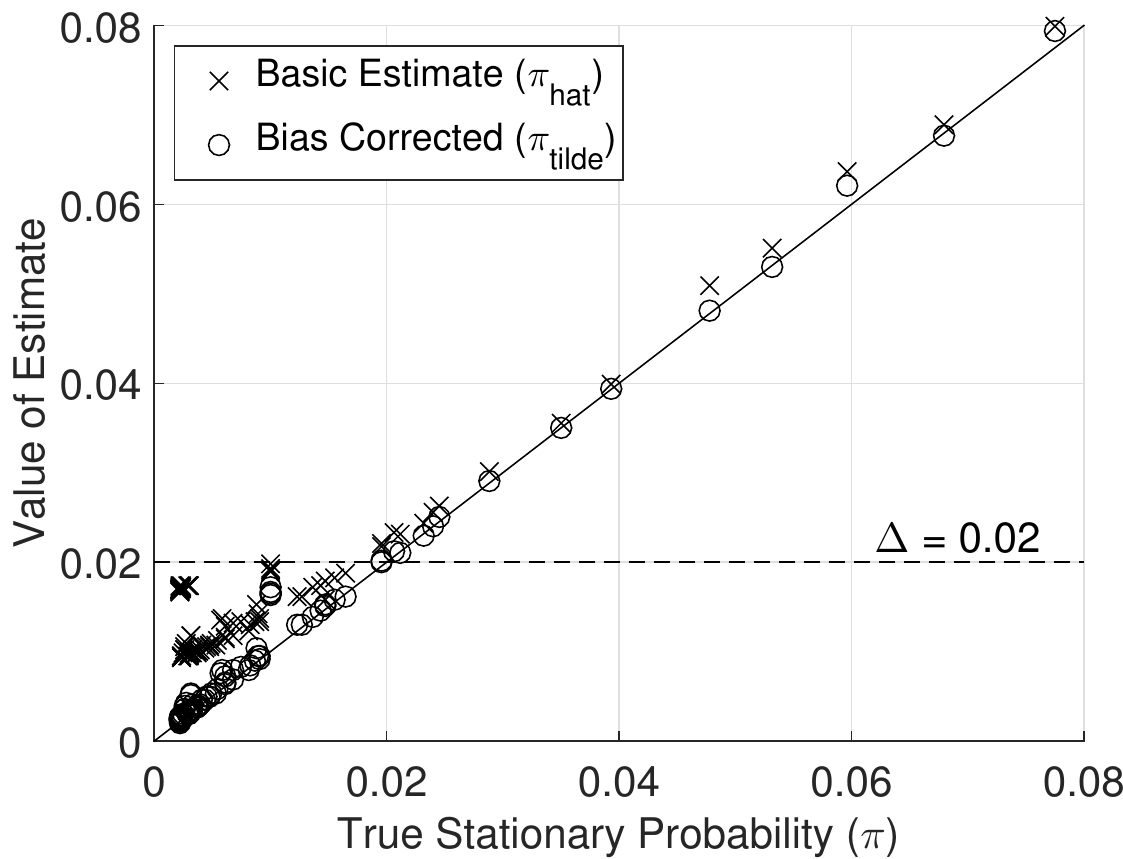}
        }
				\subfigure[PageRank Error.]{
           \label{figure:PageRank_error}
           \includegraphics[width=0.45\textwidth]{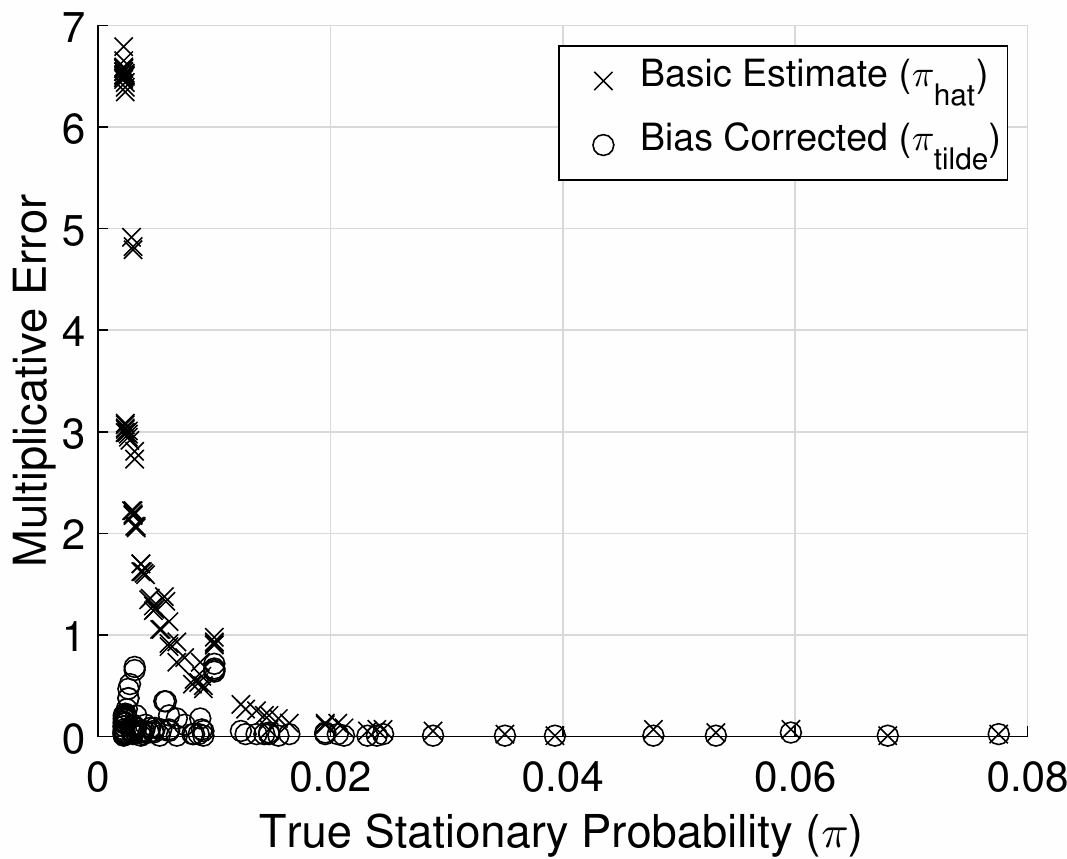}
        } 
    \caption{These plots show the estimates as well as the multiplicative error resulting from applying both the basic algorithm and the bias corrected algorithm on the PageRank Markov chain.}
   \label{figure:PageRank}
\end{figure}

We applied both the basic and bias corrected algorithms to estimate each state in the PageRank Markov chain. Figure \ref{figure:PageRank_estimate} plots both estimates $\hat{\pi}_i$ and $\tilde{\pi}_i$ as a function of the true stationary probability $\pi_i$ for all states $i$. Figure \ref{figure:PageRank_error} plots the multiplicative error given by $|\hat{\pi}_i - \pi_i|/\pi_i$ and $|\tilde{\pi}_i - \pi_i|/\pi_i$. We observe that the results validate Theorem \ref{thm:term_cond_guarantees_2}, which proves that for all states $i$ such that $\pi_i > \Delta$, the multiplicative error is bounded approximately by $2 \epsilon Z_{\max}$, while for states $i$ such that $\pi_i \leq \Delta$, we only guarantee $\hat{\pi}_i \leq \Delta$. In fact, we verified that for most pairs of states $(i,q)$ in the PageRank Markov chain, $Z_{ii} - Z_{qi} \approx 1$, and thus $\Gamma_i(\theta) \approx 1$. Lemma \ref{lemma:exp_1} then predicts that the error should be bounded by $\epsilon \Gamma_i \approx 0.15$, which we verify holds in our simulations. 

Figure \ref{figure:PageRank_estimate} clearly shows a significant reduction in the multiplicative error of the bias corrected estimate $\tilde{\pi}_i$. This is analyzed in Lemmas \ref{lemma:exp_1} and \ref{lemma:exp_2}, which prove that the additive error of the expected estimates will be a factor of $(\Gamma_i - 1)/\Gamma_i$ smaller for the bias corrected estimate as opposed to the basic estimate. Again, we point out that this is a surprising gain due to the fact that we are still using the same termination criteria with early truncation and that the bias corrected estimate does not use any more samples than the basic estimate.

\subsection{Markov chains with different mixing properties}

\begin{figure}[h]
     \centering
        \subfigure[MM1 Estimates.]{
           \label{figure:MM1_estimate}
           \includegraphics[width=0.45\textwidth]{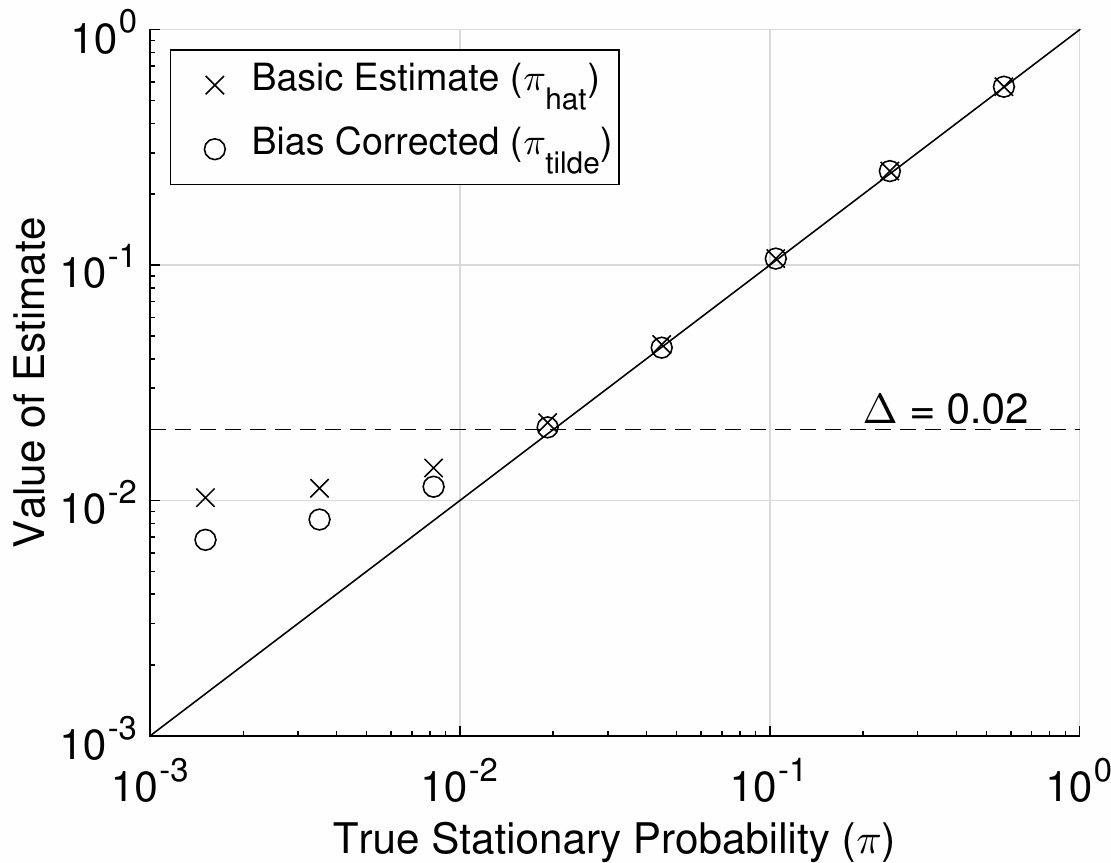}
        }
				\subfigure[Magnet Estimates.]{
           \label{figure:Magnet_estimate}
           \includegraphics[width=0.45\textwidth]{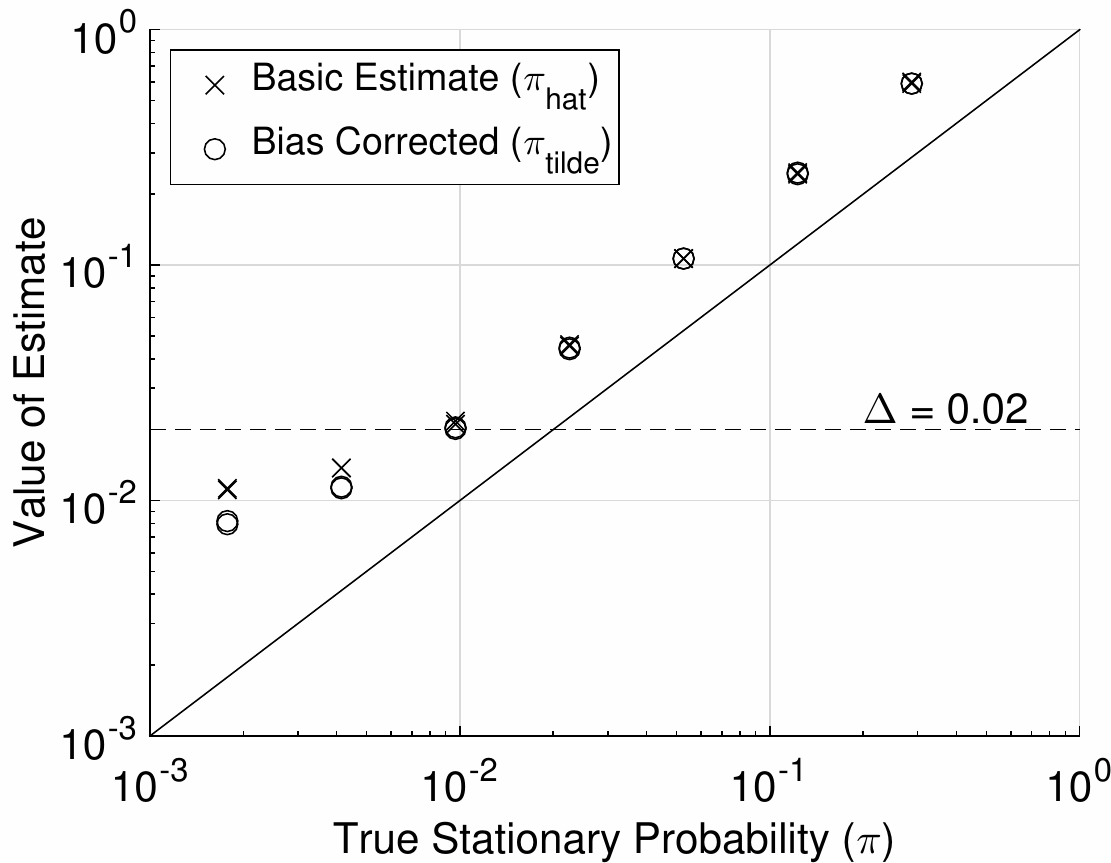}
        } 
    \caption{These plots show the estimates obtained by applying both the basic algorithm and the bias corrected algorithm on the MM1 and the Magnet Markov chains.}
   \label{figure:MM1_Magnet}
\end{figure}

In order to gain an understanding of the behavior of the algorithm as a function of the mixing properties of the Markov chain, we applied both the basic and bias corrected algorithms to estimate the first 50 states of the the M/M/1 Markov chain, and all states of the Magnet Markov chain, since these two chains are locally similar, yet have different global mixing properties. Figure \ref{figure:MM1_Magnet} plots both estimates $\hat{\pi}_i$ and $\tilde{\pi}_i$ as a function of the true stationary probability $\pi_i$ for both Markov chains. Since the stationary probabilities decay exponentially, the figures are plotted on a log-log scale. For the M/M/1 queue, we observe a similar pattern in Figure \ref{figure:MM1_estimate} as we did for PageRank, in which the states $i$ with $\pi_i > \Delta$ are approximated closely, and states $i$ such that $\pi_i \leq \Delta$ are thresholded, i.e. $\hat{\pi}_i \leq \Delta$. 

In constrast, Figure \ref{figure:Magnet_estimate} shows the result for the Magnet Markov chain, which mixes very slowly. The algorithm overestimates the stationary probabilities by almost two times the true value, which is depicted in the figure by the estimates being noticeably above the diagonal. This is due to the fact that the random samples have close to zero probability of sampling from the opposite half of the graph. Therefore the estimates are computed without being able to observe the opposite half of the graph. As the challenge is due to the poor mixing properties of the graph, both $\hat{\pi}_i$ and $\tilde{\pi}_i$ are poor estimates. In the figure, it is difficult to distinguish the two estimates because they are nearly the same and thus superimposed upon each other. We compute the fundamental matrix $Z$ for this Markov chain, and find that for most pairs $i,j \in \Sigma$, $|Z_{ij}|$ is on the order of $10^9$.

Standard methods such as power iteration or MCMC will also perform poorly on this graph, as it would take an incredibly large amount of time for the random walk to fully mix across the middle border. The final outputs of both power iteration and MCMC are very sensitive to the initial vector, since with high probability each random walk will stay on the half of the graph in which it was initialized. The estimates are neither guaranteed to be upper or lower bounds upon the true stationary probability. An advantage of our algorithm even in settings with badly mixing Markov chains, is that $\hat{\pi}_i$ is always guaranteed to be an upper bound for $\pi_i$ with high probability.

\subsection{Computation cost as a function of stationary probability}

\begin{figure}[h]
     \centering
        \subfigure[$\hat{p}^{(t_{\max})}$ vs. $\pi_i$]{
           \label{figure:PageRank_p}
           \includegraphics[width=0.325\textwidth]{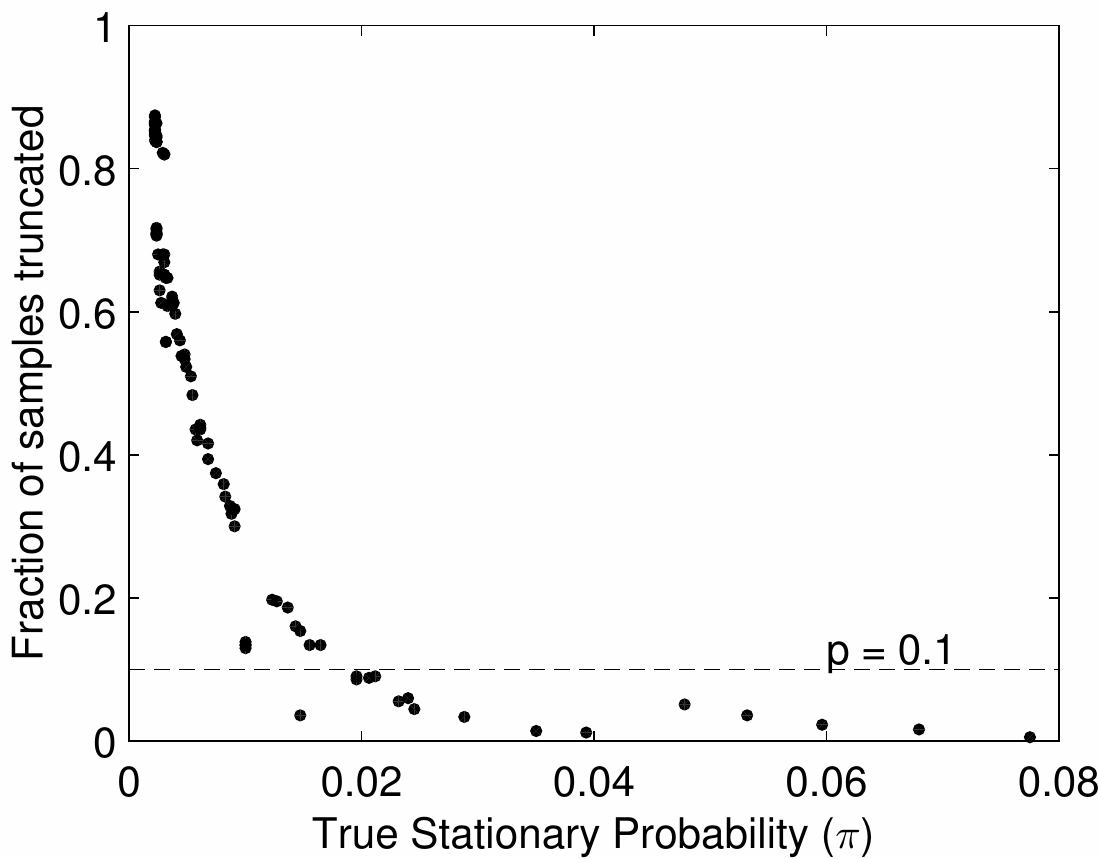}
        }
				\subfigure[$\theta^{(t_{\max})}$ vs. $\pi_i$]{
           \label{figure:PageRank_theta}
           \includegraphics[width=0.3\textwidth]{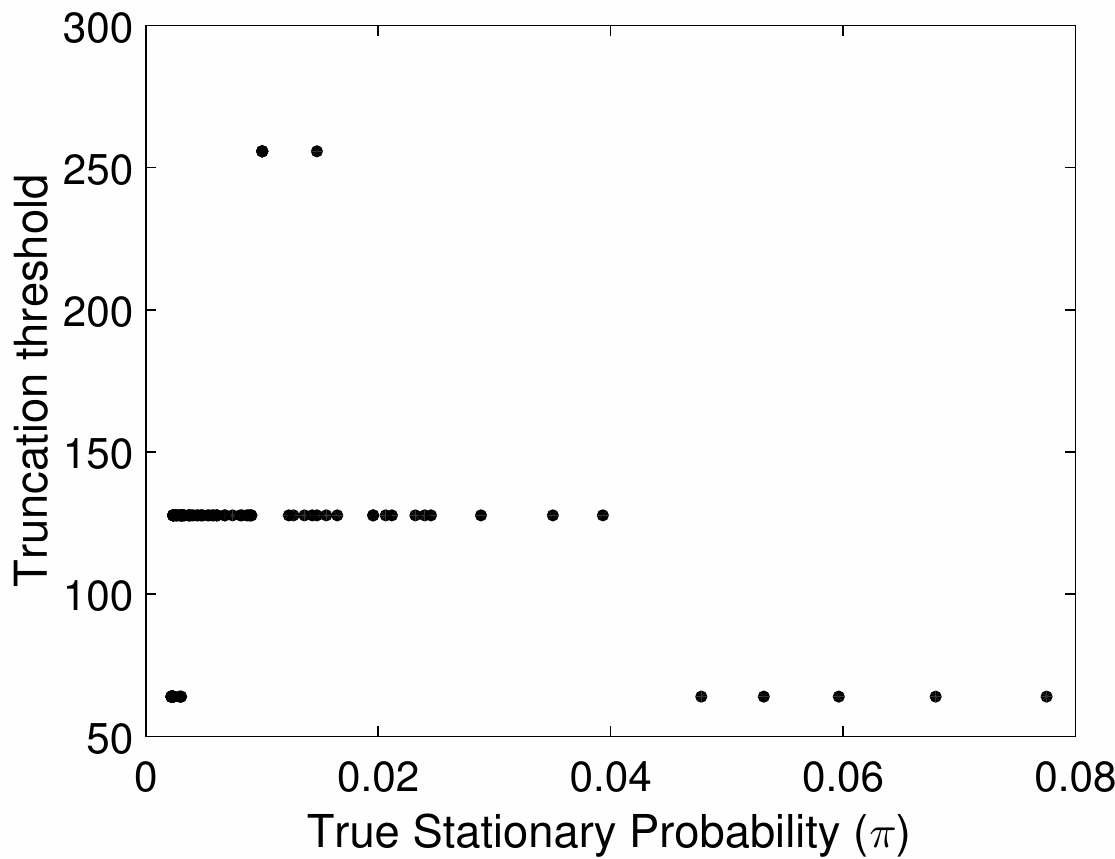}
        }
				\subfigure[Total Steps vs. $\pi_i$]{
           \label{figure:PageRank_NT}
           \includegraphics[width=0.29\textwidth]{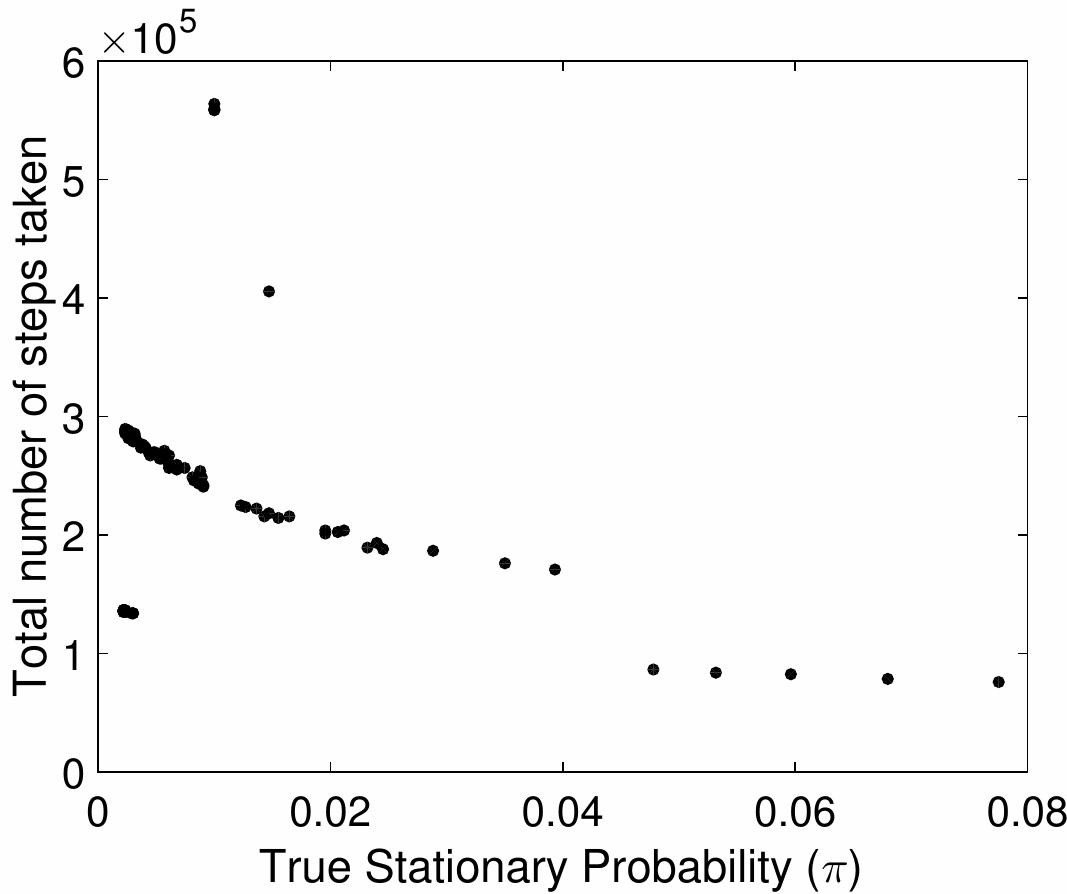}
        } 
    \caption{This shows the value of three variables from the final iteration $t_{\max}$ of the algorithm applied to the PageRank Markov chain: (a) fraction of samples truncated = $\hat{p}^{(t_{\max})}$; (b) truncation threshold = $\theta^{(t_{\max})}$; (c) total number of random walk steps taken = $N^{(t_{\max})} \cdot \hat{T}_i^{(t_{\max})}$.}
   \label{figure:ComputationCost}
\end{figure}

Figure \ref{figure:ComputationCost} plots the quantities $\hat{p}^{(t_{\max})}$, $\theta^{(t_{\max})}$, and $N^{(t_{\max})} \cdot \hat{T}_i^{(t_{\max})}$ for the execution of our algorithm on the PageRank Markov chain, as a function of the stationary probability of the target state. Recall that the algorithm terminates when either $\hat{\pi}_i^{(t)} < \Delta / (1 + \epsilon)$ or $\hat{p}^{(t)} < 2 \epsilon/3$. In our setting, we chose $\epsilon = 0.15$, such that the algorithm terminates when $\hat{p}^{(t)} < 0.1$. 

In Figure \ref{figure:PageRank_p}, we notice that all states $i$ such that $\pi_i > \Delta$ terminated at the condition $\hat{p}^{(t)} < 2 \epsilon / 3$, as follows from Theorem \ref{thm:term_cond_guarantees_2}. The fraction of samples truncated increases as $\pi_i$ decreases. For states with small stationary probability, the algorithm terminates with a large fraction of samples truncated, even as large as 0.8.

Similarly, the truncation threshold and the total computation time also initially increase as $\pi_i$ decreases, but then decreases again for very small stationary probability states. This illustrates the effect of the truncation and termination conditions. For states with large stationary probability the expected return time $\E[T_i]$ is small, leading to lower truncation threshold and number of steps taken. For states with very small stationary probability, although $\E_i[T_i]$ is large, the algorithm terminates quickly at $\hat{\pi}_i < \Delta/(1+ \epsilon)$, thus also leading to a lower truncation threshold and total number of steps taken. This figure hints at the the computational savings of our algorithm due to the design of truncation and termination conditions. The algorithm can quickly determines that a state has small stationary probability without wasting extra time to obtain unnecessary precision.

\subsection{Algorithm performance as a function of $\Delta$}

\begin{figure}[h]
     \centering
     		\subfigure[$\hat{\pi}_i$ vs. $\Delta$]{
           \label{figure:PageRank_delta_pi}
           \includegraphics[width=0.45\textwidth]{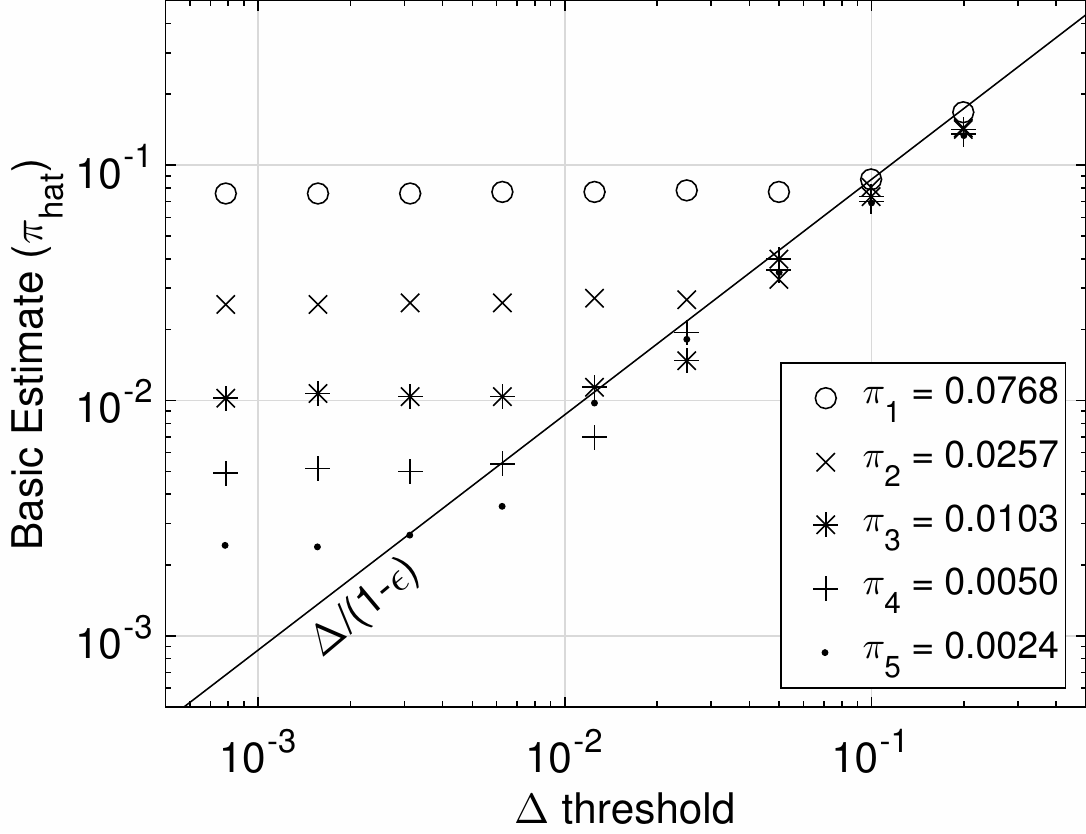}
        }
        \subfigure[Total Steps vs. $\Delta$]{
           \label{figure:PageRank_delta_NT}
           \includegraphics[width=0.45\textwidth]{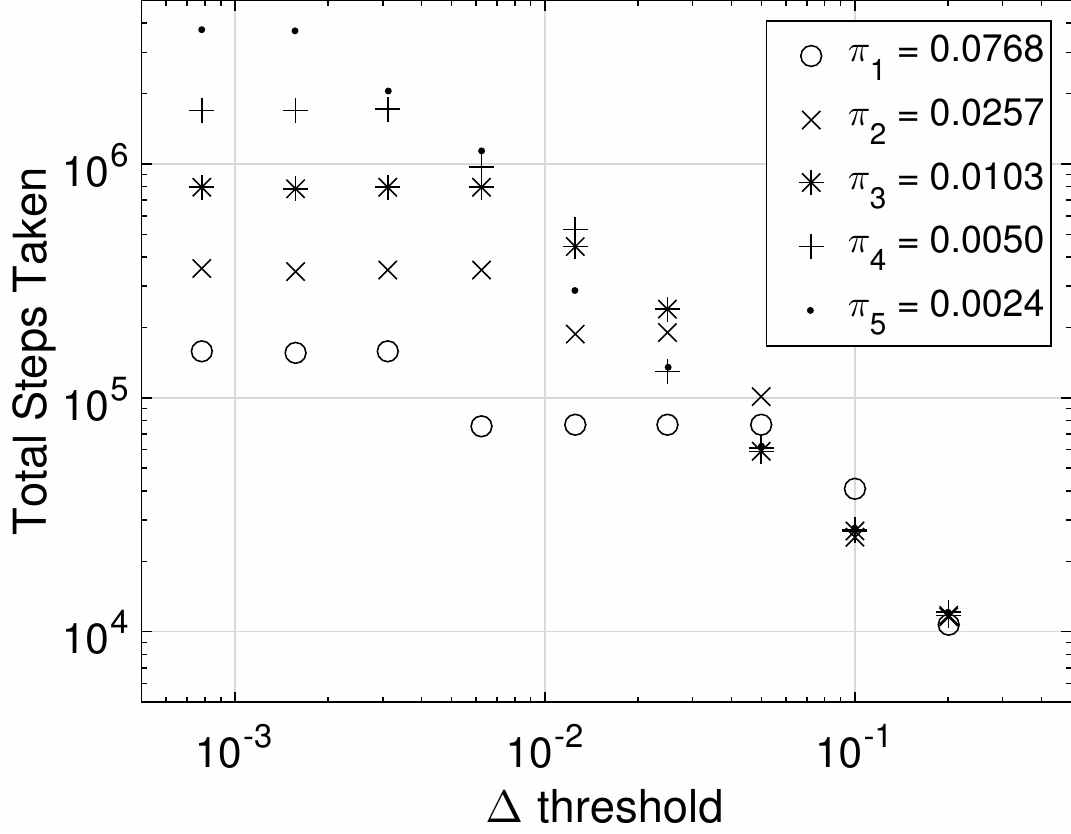}
        }    
    \caption{These figures show the dependence of our algorithm as a function of the parameter $\Delta$. We plot the basic algorithm estimate $\hat{\pi}_i$ and the total number of steps taken by the algorithm when applied to the PageRank Markov chain. The figures are shown on a log-log scale.}
   \label{figure:PageRank_delta}
\end{figure}

Figure \ref{figure:PageRank_delta} shows the results of our algorithm as a function of the parameter $\Delta$, when applied to the PageRank Markov chain. The figures are shown on a log-log scale. Recall that parameter $\Delta$ only affects the termination conditions of the algorithm. We show results from separate executions of the algorithm to estimate five different states in the Markov chain with varying stationary probabilities. Figure \ref{figure:PageRank_delta_pi} plots the basic algorithm estimates for the stationary probability, along with a diagonal line indicating the termination criteria corresponding to $\hat{\pi}_i < \Delta /(1 + \epsilon)$. When $\Delta > \hat{\pi}_i$, due to the termination conditions, the estimate produced is approximately $\Delta /(1 + \epsilon)$. When $\Delta \leq \hat{\pi}_i$, then we see that $\hat{\pi}_i$ concentrates around $\pi_i$ and no longer decreases with $\Delta$.

Figure \ref{figure:PageRank_delta_NT} plots the total steps taken in the last iteration of the algorithm, which we recall is provably of the same order as the total random walk steps over all iterations. Figure \ref{figure:PageRank_delta_NT} confirms that the computation time of the algorithm is upper bounded by $O(1/\Delta)$, which is linear when plotted on log-log scale. When $\Delta > \pi_i$, the computation time behaves as $\Theta(1/\Delta)$. When $\Delta \leq \pi_i$, the computation time levels off and grows slower than $O(1/\Delta)$.

\subsection{Algorithm Results for Multiple State Algorithm}

\begin{figure}[h]
     \centering
     		\subfigure[PageRank Markov chain]{
           \label{figure:PageRank_multiNode}
           \includegraphics[width=0.31\textwidth]{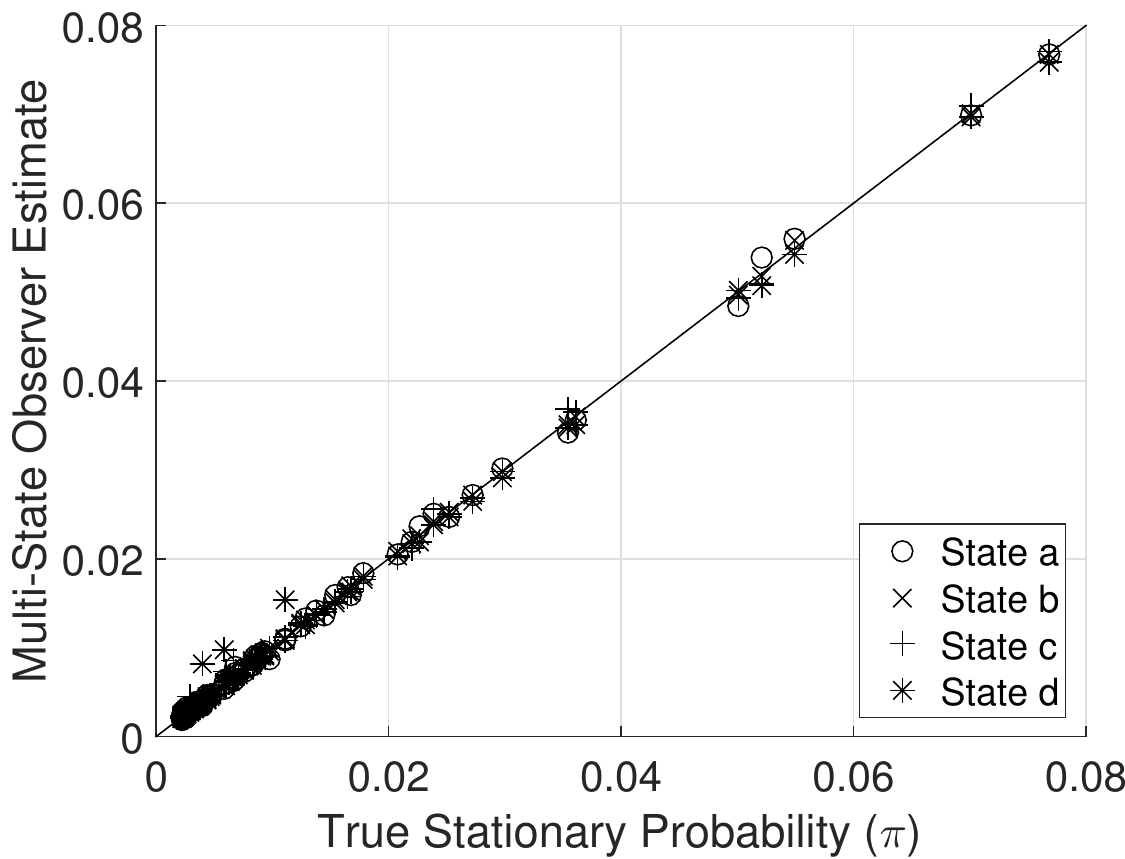}
        }
        \subfigure[Magnet Markov chain]{
           \label{figure:Magnet_multiNode}
           \includegraphics[width=0.3\textwidth]{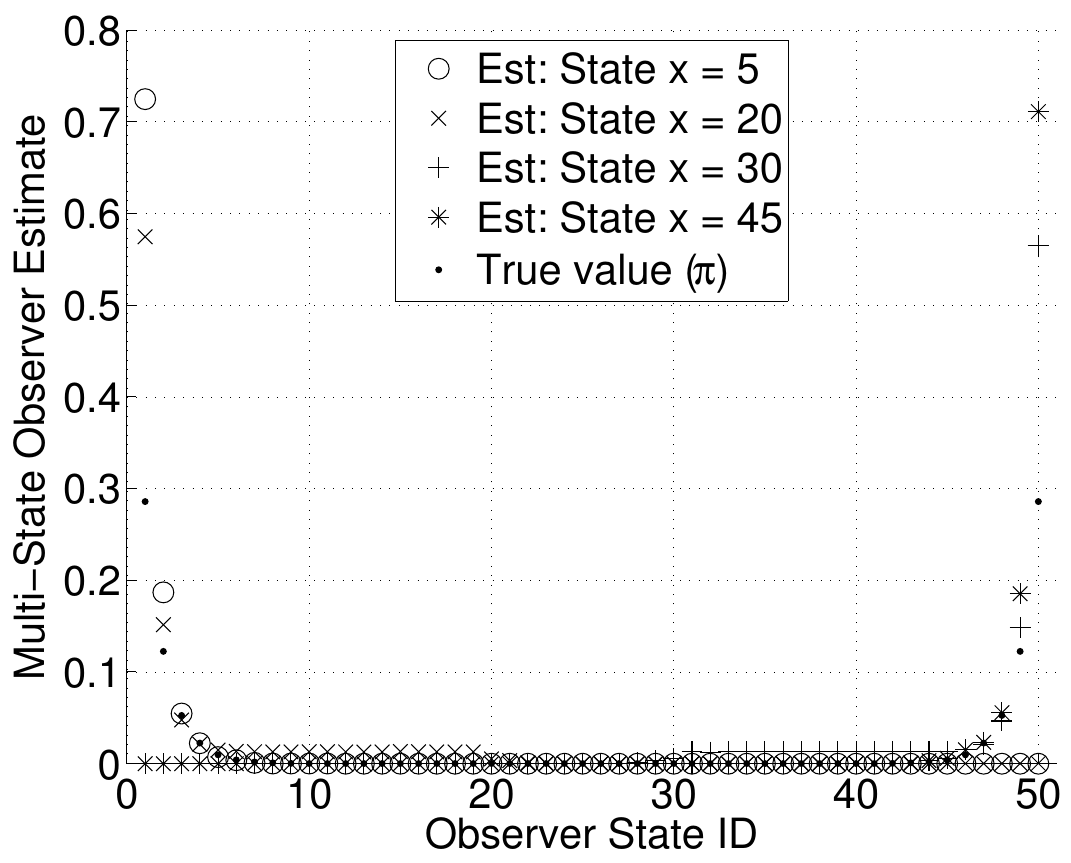}
        }
        \subfigure[Magnet Markov chain - log scale]{
           \label{figure:Magnet_multiNode_log}
           \includegraphics[width=0.3\textwidth]{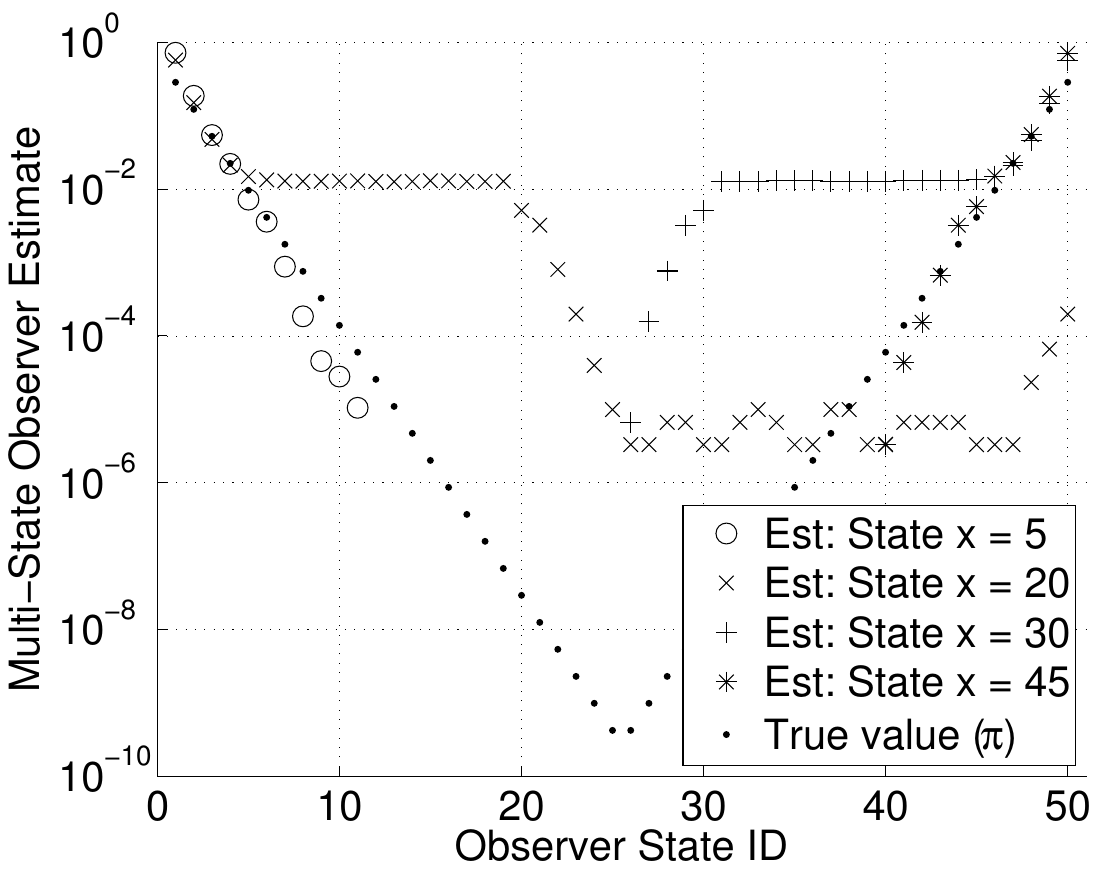}
        }
    \caption{These figures show the results of the multiple state extension of the algorithm. We chose 4 different states in the state space to fix as the anchor state. Then we used the multiple state extension to approximate the full stationary distribution by keeping track of frequency of visits along the sample paths. We show the results from both the PageRank and Magnet Markov chains. Due to the exponential decay of stationary probabilities in the Magnet Markov chain, we display the results plotted with a linear scale and log scale.}
   \label{figure:MultiState}
\end{figure}

In this section, we show that in simulations the multiple state extension of our algorithm performs quite well on the PageRank Markov chain, regardless of which state is chosen as the anchor state. The algorithm has interesting behavior on the Magnet Markov chain, varying according to the anchor state. We chose 4 different states in the state space with varying stationary probability as the anchor state. We apply the multiple state extension of the algorithm to estimate stationary probability for the observer states, which involves keeping track of the frequency of visits to the observer states along the return paths to the anchor state. 

Figure \ref{figure:PageRank_multiNode} shows that for all choices of the anchor state, the estimates were close to the true stationary probability, almost fully coinciding with the diagonal. Note that these estimates are computed with the same sample sequences collected from the basic algorithm, and thus the truncation and termination are a function of the anchor state.

When we apply the algorithm to the Magnet Markov chain, we observe that the results are highly dependent on the anchor state chosen, due to the poor mixing properties of the Markov chain. Figure \ref{figure:Magnet_multiNode} shows that for all the choices of anchor states, the algorithm over estimated states in the same half of the state space, and underestimated states in the opposite half of the state space. In Figure \ref{figure:Magnet_multiNode_log}, the estimates are plotted on a log-scale in order to more finely observe the behavior in the middle section of the state space which has exponentially small stationary probabilities. We observe that the random walks sampled from anchor states 5, 30, and 45 do not cross over to the other half of the Markov chain, and thus completely ignore the other half of the state space. Some of the random walks beginning from anchor state 20 did cross over to the other half of the state space, however it was still not significant enough to properly adjust the estimates. The significance of overestimating one side and underestimating the other side will depend on the choice of $q_1$ and $q_2$. This is a graph on which any Monte Carlo algorithm will perform poorly on, since the random walks originating in one half will not be aware of the other half of the state space.

\section{Discussion: Understanding Properties of \pagerank}

Equation \eqref{eq:PageRank_simulation} shows that the \pagerank~random walk is a convex combination between a directed graph given by the adjacency matrix $A$, and a complete graph modeled by the uniform random jumps between any two states. When $\beta = 0$, then the Markov chain is specified by $A$, which can be fully general when the edges in the graph are weighted. By tuning the parameter $\beta$, we can control whether the \pagerank~Markov chain is closer to the underlying directed graph or to the complete graph. The existing algorithms and analysis for computing \pagerank~locally only utilize the property that with probability $\beta$, the transition is chosen from the complete graph \citep{JehWidom03, Fogaras05, Avrachenkov07, Bahmani10, Borgs12}. When $\beta$ is large, the Markov chain mixes quickly due to the complete graph, and these algorithms also performs efficiently; however, when $\beta$ is small, regardless of the properties of the underlying graph, the algorithm scales with $1/\beta$, even though the Markov chain might still mix quickly depending on the underlying graph. Specifically, we have observed that our algorithm finds a natural tradeoff to balance precision of estimates with the computation cost, differentiating between high and low stationary probability states. In this section we explore the properties of the PageRank Markov chain when $\beta$ is a parameter that can be tuned to modify the Markov chain to be closer to the classic PageRank Markov chain with $\beta = 0.15$, or any general random walk defined by matrix $A$. These properties will lead to implications and tradeoffs between using existing methods as opposed to our algorithm to estimate PageRank as a function of $\beta$.

In the following simulations, we restrict ourselves to the personalized \pagerank~setting. For a fixed central state $x$ and scalar $\beta$, we consider the Markov chain with transition probability matrix $\tilde{P}$:
\begin{align}
\tilde{P} = \beta \bOne \cdot e_x^T + (1-\beta) D^{-1} A, \label{eq:ppr}
\end{align}
where $A$ is the adjacency matrix of the underlying graph, generated according to the configuration model with a power law degree distribution, and $D$ is a diagonal matrix of the out-degrees. In every step, there is a probability $\beta$ of returning to the state $x$, and probability $1 - \beta$ of choosing a neighbor at random (according to the configuration model generated graph). We choose several values for $\beta$ and four different states in the graph to be the central state $x$. For each combination of $x$ and $\beta$, we sample ten long random walks starting at state $x$ having five million steps each. After every ten thousand steps, we compute the current fraction of visits to state $x$ out of the total steps taken thus far. This quantity converges to $\pi_x$ as the length goes to infinity. 

We can split the random walk sequence into adjacent contiguous subsequences denoted by when the random walk jumps back to state $x$ with probability $\beta$. These events denote renewal times of the process, and due to the Markov property these sequences are independent and identically distributed, conditioned on their previous and ending state being $x$. The length of each subsequence is distributed as a geometric random variable with parameter $\beta$. In fact, this insight forms the foundation for the existing \pagerank~algorithms by \citet{Fogaras05} and \citet{Avrachenkov07}, which collects geometric length random walk samples on the underlying graph given by matrix $A$, beginning at state $x$. These independently sampled sequences can be stitched together to form a long random walk which has the same distribution as if the Markov chain was simulated sequentially. Our work can be seen as extending this approach for general Markov Chains by using the return visits to state $x$ to denote the renewal time of the process, which includes the $\beta$ jumps in the case of Personalized PageRank. While the frequency of $\beta$ jumps is always distributed as a Bernoulli process, the frequency of visits to state $x$ depends on both $\beta$ and the neighborhood of state $x$.

\begin{figure}[h]
     \centering
     		\subfigure[Number of steps until 0.01 error vs. $1/\beta$]{
           \label{figure:PageRank_beta_stepsReq}
           \includegraphics[width=0.45\textwidth]{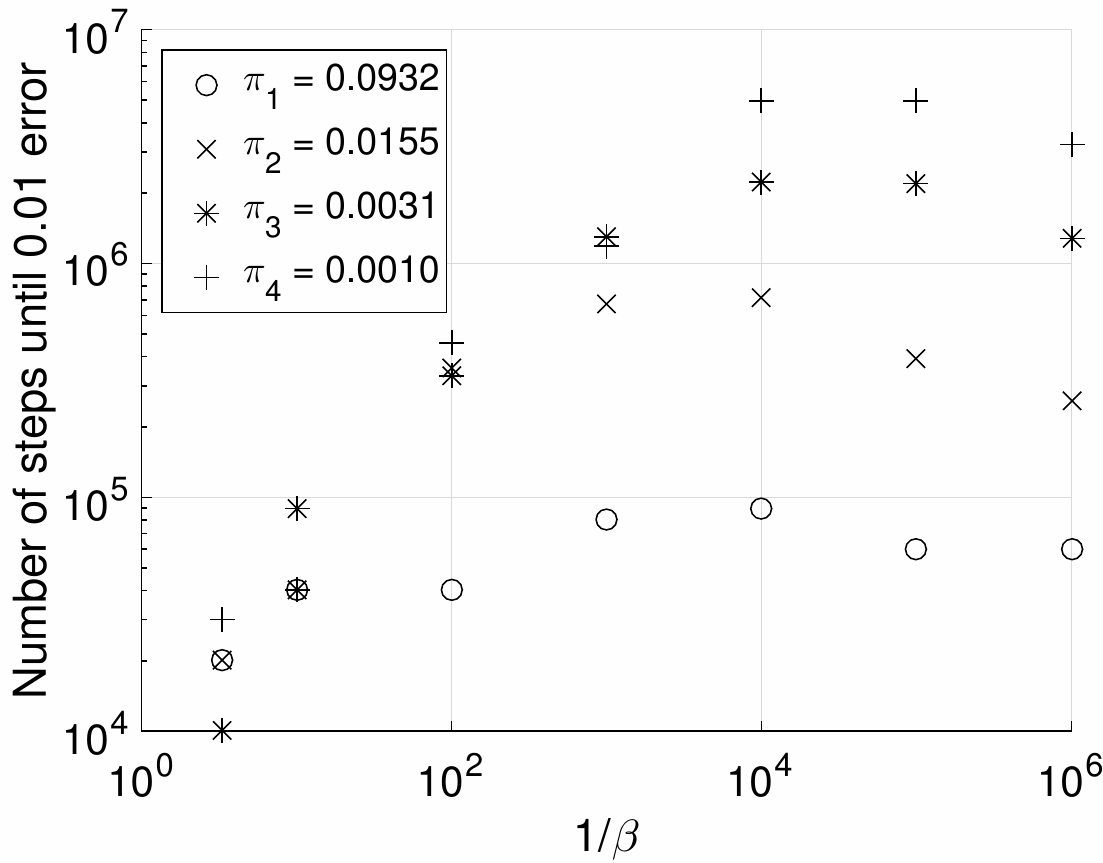}
        }
        \subfigure[Expected return time vs. $1/\beta$]{
           \label{figure:PageRank_beta_ET}
           \includegraphics[width=0.45\textwidth]{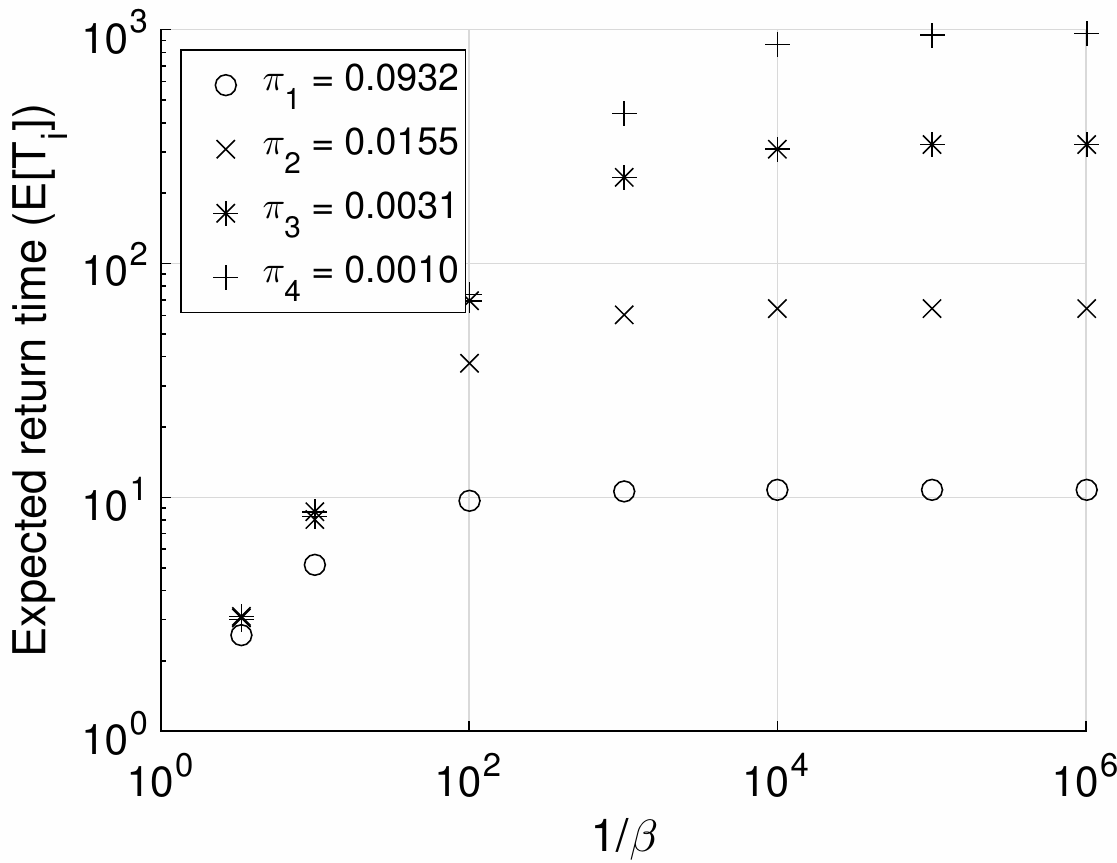}
        }
    \caption{These figures show the number of random walk steps needed to accurately estimate the stationary probability of a state, as well as the expected return time to a state in a PageRank Markov chain, for four different states, and as a function of the parameter $\beta$.}
   \label{figure:PageRank_beta}
\end{figure}

In Figure \ref{figure:PageRank_beta_ET}, we plot the expected length of a sample in Fogaras' algorithm, which is given by $1/\beta$, as opposed to the expected length of a sample in our algorithm, given by $\E[T_i]$. When $\beta$ is large ($1/\beta < 100$), we notice that $\E[T_i] \approx 1/\beta$ uniformly for all states, which is expected since the Markov chain will be dominated by the $\beta$ jumps. When $\beta$ is very small, the expected return times differentiate between states, reflecting the different local connectivity properties around each state. 

Next we investigate the rates at which the sample frequency of visits to a state in the Markov chain converges to the true stationary probability. This helps us to quantify the amount of necessary computation, or simulated random walk steps, until we have sufficient information to obtain a good estimate for the stationary probability of a state. For each combination of $x$ and $\beta$ and for each long random walk sample, we compute the multiplicative error between the stationary probability, and the fraction of visits to state $x$ along the path. We average the error across the 10 samples, and compute the smallest number of steps such that the average error is less than 0.01. This is shown in Figure \ref{figure:PageRank_beta_stepsReq} as a function of $\beta$ for the four different chosen states $x$. 

When $\beta$ is large, the $\beta$ jump back to $x$ dominates the random walk, causing the random walk to behave similarly for different chosen states $x$. In this setting $\beta$ can be used to determine the required number of samples to achieve good accuracy. When $\beta$ is small, the random walk approaches that of the underlying graph, thus the choice of state $x$ greatly affects the computation. The number of steps to achieve below 0.01 error does not increase much as $\beta$ decreases for states $x$ that have large $\pi_x$ in the underlying graph (such as state 1). However, for states such that $\pi_x$ is small in the underlying graph (such as state 4), the required number of steps increases significantly as $\beta$ decreases. We observe a relationship between $\E_x[T_x]$ and the total amount of computation steps needed for 0.01 accuracy. This clearly highlights that in the setting when $\beta$ is very small, it becomes critical to consider information from the underlying graph in determining the number of steps to sample. The existing algorithm only uses the parameter $\beta$ to govern the number of steps taken for the algorithm, where the length of a sample random walk scales with $1 / \beta$. However, for small $\beta$, this is unnecessarily expensive, as some states (such as state 1) do not in fact require the number of sample steps to increase with $1 / \beta$. Our proposed algorithm relies on the expected return time to state $x$ as the length of each sample path, which adjusts for states that have varying stationary probabilities in the underlying graph.

\begin{APPENDICES}

\section{Formal Proofs for Theorems presented in Section 3}

\proof{Proof of Theorem \ref{thm:fixed_t_conv_rate}.}
The error bounds of Theorem \ref{thm:fixed_t_conv_rate} directly follow from Theorems \ref{thm:main_general_b} and \ref{thm:finite_correctness} in Section \ref{sec:error_analysis}. The analysis of computation cost is directly proved by Lemma \ref{lemma:stepsTaken}. 
\hfill \halmos \\
\endproof

\proof{Proof of Theorem \ref{thm:term_cond_guarantees_1}.}
The error bound of Theorem \ref{thm:term_cond_guarantees_1} directly follows from Corollary \ref{cor:phat_error}, and using the property that $Z_{\max} \leq 2 \ln(2) \tmix$. The bound on the computation cost in Theorem \ref{thm:term_cond_guarantees_1} follows directly from Theorem \ref{thm:finite_convergence_b}, and the property that $H_i = O(Z_{\max}(i)/\pi_i) = O(t_{\tmix}/\pi_i)$.
\hfill \halmos \\
\endproof

\proof{Proof of Theorem \ref{thm:term_cond_guarantees_2}.}
Theorem \ref{thm:term_cond_guarantees_2}(a) follows from Theorem \ref{thm:main_general_b}. Theorem \ref{thm:term_cond_guarantees_2}(b) follows from plugging in $2 \epsilon / 3$ for $\delta$ in Theorem \ref{thm:term_cond_guarantees_1}. Theorem \ref{thm:term_cond_guarantees_2}(c) follows from taking the minimum of bounds given in Theorems \ref{thm:main_general_a} and \ref{thm:finite_convergence_b}, again substituting $2 \epsilon / 3$ for $\delta$ and using the property that $H_i = O(t_{\tmix}/\pi_i)$. \hfill \halmos \\
\endproof

\proof{Proof of Theorem \ref{thm:bias}.}
By Lemma \ref{lemma:p_concentrate}, with probability greater than $1 - \alpha$, $\hat{T}_i^{(k)} \in (1 \pm \epsilon) \E_i[\hat{T}_i^{(k)}]$ for all $k$, and $(1 - \hat{p}^{(k)}) \in (1 \pm \epsilon) (1 - \Prob_i(T_i > \theta^{(k)}))$ for all $k$ such that $\Prob_i(T_i > \theta^{(k)}) > 1/2$. Therefore, with probability greater than $1 - \alpha$, for every $k$ such that $\Prob_i(T_i > \theta^{(k)}) > 1/2$,
\begin{align}
&\left| \frac{\tilde{\pi}_i^{(k)} - \pi_i}{\tilde{\pi}_i^{(k)}} \right| = \left|1 - \frac{\hat{T}_i^{(k)}}{(1-\hat{p}^{(k)}) \E_i[T_i]} \right| \nonumber\\
&\leq \max\left(1 - \frac{(1 - \epsilon)\E_i[\hat{T}_i^{(k)}]}{(1 + \epsilon)(1-\Prob_i(T_i > \theta^{(k)})) \E_i[T_i]},\frac{(1 + \epsilon)\E_i[\hat{T}_i^{(k)}]}{(1 - \epsilon)(1-\Prob_i(T_i > \theta^{(k)})) \E_i[T_i]} - 1\right) \nonumber\\
&\leq \max\left(\frac{1 - \epsilon}{1 + \epsilon}\left(1 - \frac{\E_i[\hat{T}_i^{(k)}]}{(1-\Prob_i(T_i > \theta^{(k)})) \E_i[T_i]}\right),\frac{1 + \epsilon}{1 - \epsilon}\left(\frac{\E_i[\hat{T}_i^{(k)}]}{(1-\Prob_i(T_i > \theta^{(k)})) \E_i[T_i]} - 1\right)\right) + \frac{2 \epsilon}{1 - \epsilon} \nonumber \\
&\leq \frac{1 + \epsilon}{1 - \epsilon} \left|1 - \frac{\E_i[\hat{T}_i^{(k)}]}{(1-\Prob_i(T_i > \theta^{(k)})) \E_i[T_i]}\right| + \frac{2 \epsilon}{1 - \epsilon}. \label{eq:lemma_exp_2}
\end{align}

By upper bounding $(Z_{ii} - Z_{qi})$ by $2 Z_{\max}(i)$ for all $q$, it follows that $\Gamma_i(\theta^{(k)}) < 2 Z_{\max}(i)$. Thus by rearranging \eqref{eq:exp_2} from Lemma \ref{lemma:exp_2}, it follows that
\begin{align} \label{eq:ez2}
\left|1 - \frac{\E_i[\hat{T}_i^{(k)}]}{(1-\Prob_i(T_i > \theta^{(k)})) \E_i[T_i]}\right| &\leq \max(2 Z_{\max}(i) - 1, 1) \left(\frac{\Prob_i\left(T_i^{(k)} > \theta^{(k)}\right)}{1 - \Prob_i\left(T_i^{(k)} > \theta^{(k)}\right)}\right).
\end{align}

Substitute \eqref{eq:ez2} into \eqref{eq:lemma_exp_2}, and apply Lemma \ref{lemma:finite} to complete the proof.
\hfill \halmos \\
\endproof

\proof{Proof of Theorem \ref{thm:freq}.}
By Lemma \ref{lemma:f_concentrate}, with probability greater than $1 - \alpha$, $\hat{F}_j^{(k)} \in \E_i[\hat{F}_j^{(k)}] \pm \epsilon \E_i[\hat{T}_i^{(k)}]$ and $\hat{T}_i^{(k)} \in (1 \pm \epsilon) \E_i[\hat{T}_i^{(k)}]$ for all $k$. Therefore with probability greater than $1 - \alpha$,
\begin{align*}
\frac{(1 - \epsilon) \hat{F}_j^{(k)}}{\hat{T}_i^{(k)}} &\leq \frac{\E_i[\hat{F}_j^{(k)}] + \epsilon \E_i[\hat{T}_i^{(k)}]}{\E_i[\hat{T}_i^{(k)}]}.
\end{align*}
Therefore,
\begin{align*}
\hat{\pi}_j^{(k)} &\leq \frac{\E_i[\hat{F}_j^{(k)}]}{\E_i[\hat{T}_i^{(k)}]} + \epsilon + \epsilon \hat{\pi}_j^{(k)}.
\end{align*}
Similarly,
\begin{align*}
\frac{(1 + \epsilon) \hat{F}_j^{(k)}}{\hat{T}_i^{(k)}} &\geq \frac{\E_i[\hat{F}_j^{(k)}] - \epsilon \E_i[\hat{T}_i^{(k)}]}{\E_i[\hat{T}_i^{(k)}]} \\
\hat{\pi}_j^{(k)} &\geq \frac{\E_i[\hat{F}_j^{(k)}]}{\E_i[\hat{T}_i^{(k)}]} - \epsilon - \epsilon \hat{\pi}_j^{(k)}.
\end{align*}
Therefore, with probability greater than $1 - \alpha$,
\begin{align*}
\left|\tilde{\pi}_j^{(k)} - \pi_j\right| &\leq \max \left(\frac{\E_i[\hat{F}_j^{(k)}]}{\E_i[\hat{T}_i^{(k)}]} + \epsilon(1 + \hat{\pi}_j^{(k)}) - \pi_j,
\pi_j - \frac{\E_i[\hat{F}_j^{(k)}]}{\E_i[\hat{T}_i^{(k)}]} + \epsilon (1 + \hat{\pi}_j^{(k)})\right) \\
&= \left|\frac{\E_i[\hat{F}_j^{(k)}]}{\E_i[\hat{T}_i^{(k)}]} - \pi_j\right| + \epsilon (1 + \tilde{\pi}_j^{(k)}).
\end{align*}
Use Lemma \ref{lemma:exp_3} to show that with probability greater than $1 - \alpha$,
\begin{align*}
\left|\tilde{\pi}_j^{(k)} - \pi_j\right| &\leq \left|\frac{\Prob_i\left(T_i > \theta^{(k)}\right)}{\E_i[\hat{T}_i^{(k)}]} \left(\sum_{k \in \Sigma \setminus \{i\}} \Prob_i\left(\left. X_{\theta^{(k)}} = k \right| T_i > \theta^{(k)}\right) Z_{ij}  - Z_{kj}\right)\right| + \epsilon (1 + \tilde{\pi}_j^{(k)}) \\
&\leq \frac{\Prob_i\left(T_i > \theta^{(k)}\right)}{\E_i[\hat{T}_i^{(k)}]} 2 Z_{\max}(j) + \epsilon (1 + \tilde{\pi}_j^{(k)}).
\end{align*}
Since $\hat{T}_i^{(k)} \in (1 \pm \epsilon) \E_i[\hat{T}_i^{(k)}]$, it follows that
\begin{align*}
\left|\tilde{\pi}_j^{(k)} - \pi_j\right| &\leq (1 + \epsilon) \Prob_i\left(T_i > \theta^{(k)}\right) 2 Z_{\max}(j) \hat{\pi}_i^{(k)} + \epsilon (1 + \tilde{\pi}_j^{(k)}).
\end{align*}
Apply Lemma \ref{lemma:finite} to complete the proof.\hfill \halmos
\endproof

\section{Additional Proofs of Lemmas presented in Section 4}

The proofs of Lemmas \ref{lemma:error_bounds} to \ref{lemma:f_concentrate} use the following fact:
\begin{fact} \label{fact:algebra}
If $\sum_{h=1}^k x_h \leq 1$, then $\prod_{h=1}^k (1 - x_h) \geq 1 - \sum_{h=1}^k x_h$.
\end{fact}

\proof{Proof of Lemma \ref{lemma:error_bounds}.}
Let $A_h$ denote the event $\left\{ \hat{T}_i^{(h)} \in (1 \pm \epsilon) \E_i\left[\hat{T}_i^{(h)}\right] \right\}$. Since $N^{(h)}$ is a random variable due to its dependence on $\hat{T}_i^{(h-1)}$, the distribution of $\hat{T}_i^{(h)}$ depends on the value of $\hat{T}_i^{(h-1)}$. Conditioned on the event $A_{h-1}$,
\begin{align}
N^{(h)} &= \left\lceil \frac{3 (1+\epsilon) \theta^{(h)} \ln(4 \theta^{(h)} / \alpha)}{\hat{T}_i^{(h-1)} \epsilon^2} \right\rceil \nonumber \\
&\geq \frac{3 (1+\epsilon) \theta^{(h)} \ln(4 \theta^{(h)} / \alpha)}{(1+\epsilon) \E_i[\hat{T}_i^{(h-1)}] \epsilon^2} = \frac{3 \theta^{(h)} \ln(4 \theta^{(h)} / \alpha)}{\E_i[\hat{T}_i^{(h-1)}] \epsilon^2}. \label{eq:N_t}
\end{align}
Then we apply Chernoff's bound for independent identically distributed bounded random variables (see Theorem \ref{chernoff_bounded} in Appendix), substitute in for $N^{(h)}$, and use the facts that $\E_i[\hat{T}_i^{(h)}] \geq \E_i[\hat{T}_i^{(h-1)}]$ and $\theta^{(h)} = 2^h$ for all $h$, to show that
\begin{align}
\Prob_i\left( \lnot A_h | A_{h-1} \right) &\leq 2 \exp\left(-\frac{\epsilon^2 N^{(h)} \E_i[\hat{T}_i^{(h)}]}{3 \theta^{(h)}}\right) \nonumber \\
&\leq 2 \exp\left(-\frac{\E_i[\hat{T}_i^{(h)}]\ln(4 \theta^{(h)} / \alpha)}{\E_i[\hat{T}_i^{(h-1)}]}\right) \nonumber \\
&\leq \frac{\alpha}{2 \theta^{(h)}} \leq \frac{\alpha}{2^{h+1}}. \label{eq:prob_not_A}
\end{align}
It can be verified that $\Prob_i(\lnot A_1)$ is similarly upper bounded by $\alpha/4$ using the definition of $N^{(1)}$. Therefore, by Bayes rule and by the fact that $\hat{T}_i^{(h')}$ is independent from $\hat{T}_i^{(h)}$ conditioned on $\hat{T}_i^{(h-1)}$ for all $h' < h$, we show that
\begin{align}
\Prob_i\left(\bigcap_{h=1}^k A_h\right) = \Prob_i\left(A_1\right) \prod_{h=2}^k \Prob_i\left(A_h | A_{h-1}\right) \geq \prod_{h=1}^k \left(1- \frac{\alpha}{2^{h+1}}\right). \label{eq:lemm_err_bd_1}
\end{align}
By applying Fact \ref{fact:algebra}, it follows that
\begin{align}\label{eq:error_bounds_2}
\Prob_i\left(\bigcap_{h=1}^k A_h\right) \geq 1-\sum_{h=1}^k \frac{\alpha}{2^{h+1}} = 1 - \frac{\alpha}{2} (1 - 2^{-k}) \geq 1 - \alpha.
\end{align}\hfill \halmos
\endproof

\proof{Proof of Lemma \ref{lemma:p_add_concentrate}.}
Let $A_h$ denote the event $\left\{ \hat{T}_i^{(h)} \in (1 \pm \epsilon) \E_i\left[\hat{T}_i^{(h)}\right] \right\}$. Let $B_h$ denote the event \\
$\left\{ \hat{p}^{(h)} \in \Prob_i(T_i > \theta^{(h)}) \pm \frac{\epsilon}{3}\right\}.$
$\hat{T}_i^{(h)}$ is independent from $\hat{p}_i^{(h-1)}$ conditioned on $\hat{T}_i^{(h-1)}$. Therefore, it follows from \eqref{eq:prob_not_A} that
\[\Prob_i(\lnot A_h | A_{h-1} \cap B_{h-1}) \leq \frac{\alpha}{2^{h+1}}.\]
By applying Hoeffding's Inequality for Bernoulli random variables to $\hat{p}^{(h)}$ and substituting \eqref{eq:N_t}, we show that
\begin{align*}
\Prob_i\left( \lnot B_h | A_{h-1} \cap B_{h-1} \right) &\leq 2 \exp\left(-\frac{2}{9} \epsilon^2 N^{(h)}\right) \\
&\leq 2 \exp\left(-\frac{2 \theta^{(h)} \ln(4 \theta^{(h)}/\alpha)}{3 \E_i[\hat{T}_i^{(h-1)}]} \right).
\end{align*}
Since $\E_i[\hat{T}_i^{(h-1)}] \leq \theta^{(h-1)} = \theta^{(h)} / 2$,
 \begin{align*}
\Prob_i\left( \lnot B_h | A_{h-1} \cap B_{h-1} \right) &\leq 2 \exp\left(-\frac{4}{3} \ln(4 \theta^{(h)}/\alpha) \right) \leq \frac{\alpha}{2 \theta^{(h)}} \leq \frac{\alpha}{2^{t+1}}.
\end{align*}
By applying the union bound, it follows that 
\begin{align}
\Prob_i\left( A_h \cap B_h | A_{h-1} \cap B_{h-1}\right) 
&\geq 1 - \frac{\alpha}{2^t}.
\end{align}
We can easily verify using the same techniques that 
\[\Prob_i(A_1 \cap B_1 ) \geq 1 - \frac{\alpha}{2}.\]
Therefore,
\[\Prob_i\left(\bigcap_{h=1}^k (B_h \cap A_h) \right) \geq \prod_{h=1}^{k} \left(1- \frac{\alpha}{2^{h}}\right).\]
We use Fact \ref{fact:algebra} to show that
\[\Prob_i\left(\bigcap_{h=1}^k (B_h \cap A_h) \right) \geq 1-\sum_{h=1}^k \frac{\alpha}{2^{h}} = 1 - \alpha (1 - 2^{-k}) \geq 1 - \alpha.\]\hfill \halmos
\endproof

\proof{Proof of Lemma \ref{lemma:p_concentrate}.}
Recall that $k_0$ is defined such that $\Prob_i(T_i > \theta^{(k_0)}) < \frac12$. Let $A_h$ denote the event \\
$\left\{ \hat{T}_i^{(h)} \in (1 \pm \epsilon) \E_i\left[\hat{T}_i^{(h)}\right] \right\}$. Let $B_h$ denote the event
$\left\{ (1 - \hat{p}^{(h)}) \in (1 \pm \epsilon) (1 - \Prob_i(T_i > \theta^{(h)})) \right\}$.
\begin{align} \label{eq:lemm_p_1}
\Prob_i\left(\bigcap_{h=k_0}^k B_h \bigcap_{h=1}^k A_h \right) = \Prob_i\left(\left.\bigcap_{h=k_0}^k (B_h \cap A_h) \right| \bigcap_{h=1}^{k_0-1} A_h \right) \Prob_i\left(\bigcap_{h=1}^{k_0 - 1} A_h \right).
\end{align}
$\hat{T}_i^{(h)}$ and $\hat{p}^{(h)}$ are not independent from $\hat{T}_i^{(h-1)}$ and $\hat{p}^{(h-1)}$ due to the dependence of $N^{(h)}$ as a random variable upon $\hat{T}_i^{(h-1)}$. However, $\hat{T}_i^{(h)}$ is independent from $\hat{p}_i^{(h-1)}$ conditioned on $\hat{T}_i^{(h-1)}$. Therefore, it follows from \eqref{eq:prob_not_A} that
\begin{align}
\Prob_i(\lnot A_h | A_{h-1} \cap B_{h-1}) \leq \frac{\alpha}{2^{h+1}}. \label{eq:not_A_p1}
\end{align}
By using Chernoff's bound for Bernoulli random variables and substituting \eqref{eq:N_t}, we show that
\begin{align}
\Prob_i(\lnot B_h | A_{h-1} \cap B_{h-1}) &\leq 2 \exp\left(-\frac{\epsilon^2 N^{(h)} (1 - \Prob_i(T_i > \theta^{(h)}))}{3}\right) \nonumber \\
&\leq 2 \exp\left(-\frac{\theta^{(h)} \ln(4 \theta^{(h)} / \alpha) (1-\Prob_i(T_i > \theta^{(h)}))}{\E_i[\hat{T}_i^{(h-1)}]}\right). \label{eq:prob_not_B_p1}
\end{align}
By definition, $\E_i[\hat{T}_i^{(h-1)}] \leq \theta^{(h-1)} = \theta^{(h)} / 2$. Since we are given that $\Prob_i(T_i > \theta^{(h)}) \leq \frac12$, then
\begin{align}
\frac{\theta^{(h)} (1-\Prob_i(T_i > \theta^{(h)}))}{\E_i[\hat{T}_i^{(h-1)}]} \geq 1. \label{eq:prob_not_B_p1_2}
\end{align}
By substituting \eqref{eq:prob_not_B_p1_2} into \eqref{eq:prob_not_B_p1}, it follows that
\begin{align}
\Prob_i(\lnot B_h | A_{h-1} \cap B_{h-1}) \leq \frac{2\alpha}{ 4 \theta^{(h)}} \leq \frac{\alpha}{2^{h+1}}.\label{eq:not_B_p1}
\end{align}
We combine \eqref{eq:not_A_p1} and \eqref{eq:not_B_p1}, and apply the union bound to show that
\begin{align}
\Prob_i\left( A_h \cap B_h | A_{h-1} \cap B_{h-1}\right) 
&\geq 1 - \frac{\alpha}{2^t}. \label{eq:lemm_p_2}
\end{align}
We can easily verify using the same techniques that 
\[\Prob_i\left(A_{k_0} \cap B_{k_0} \left| \bigcap_{h=1}^{k_0-1} A_k \right.\right) \geq 1 - \frac{\alpha}{2^{k_0}}.\]
We use Bayes' rule and substitute \eqref{eq:lemm_err_bd_1} and \eqref{eq:lemm_p_2} into \eqref{eq:lemm_p_1} to show that
\[\Prob_i\left(\bigcap_{h=k_0}^k B_h \bigcap_{h=1}^k A_h \right) \geq \prod_{h=k_0}^{k} \left(1- \frac{\alpha}{2^{h}}\right) \prod_{h=1}^{k_0 - 1} \left(1- \frac{\alpha}{2^{h+1}}\right) .\]
We apply Fact \ref{fact:algebra} to show that
\[\Prob_i\left(\bigcap_{h=k_0}^k B_h \bigcap_{h=1}^k A_h \right) \geq 1-\sum_{h=1}^k \frac{\alpha}{2^{h}} + \sum_{h=1}^{k_0} \frac{\alpha}{2^{h+1}} = 1 - \alpha (1 - 2^{-k}) + \sum_{h=1}^{k_0} \frac{\alpha}{2^{h+1}} \geq 1 - \alpha.\]\hfill \halmos
\endproof

\proof{Proof of Lemma \ref{lemma:f_concentrate}.}
Let $A_h$ denote the event $\left\{ \hat{T}_i^{(h)} \in (1 \pm \epsilon) \E_i\left[\hat{T}_i^{(h)}\right] \right\}$. Let $B_h$ denote the event \\
$\left\{ \hat{F}_j^{(h)} \in \E_i\left[\hat{F}_j^{(h)}\right] \pm \epsilon \E_i\left[\hat{T}_i^{(h)}\right] \right\}.$
$\hat{T}_i^{(h)}$ is independent from $\hat{F}_j^{(h-1)}$ conditioned on $\hat{T}_i^{(h-1)}$. Therefore, it follows from \eqref{eq:prob_not_A} that
\[\Prob_i(\lnot A_h | A_{h-1} \cap B_{h-1}) \leq \frac{\alpha}{2^{h+1}}.\]
Recall from the definition of $\hat{F}_j^{(h)}$ that $\hat{F}_j^{(h)} \in [0,\theta^{(h)}]$. Therefore, by applying Chernoff's bound for independent identically distributed bounded random variables, and substituting \eqref{eq:N_t}, we show that
\begin{align}
\Prob_i\left( \lnot B_h | A_{h-1} \cap B_{h-1} \right) &\leq 2 \exp\left(-\left(\frac{\epsilon \E_i[\hat{T}_i^{(h)}]}{\E_i[\hat{F}_j^{(h)}]}\right)^2 \frac{N^{(h)} \E_i[\hat{F}_j^{(h)}]}{3 \theta^{(h)}}\right) \nonumber \\
&\leq 2 \exp\left(-\frac{\E_i[\hat{T}_i^{(h)}]^2\ln(4 \theta^{(h)} / \alpha)}{\E_i[\hat{F}_j^{(h)}] \E_i[\hat{T}_i^{(h-1)}]}\right). \label{eq:prob_not_B_f}
\end{align}
Since $\E_i[\hat{T}_i^{(h)}] \geq \E_i[\hat{T}_i^{(h-1)}]$ and $\E_i[\hat{T}_i^{(h)}] \geq \E_i[\hat{F}_j^{(h)}]$, then
\[\frac{\E_i[\hat{T}_i^{(h)}]^2}{\E_i[\hat{F}_j^{(h)}] \E_i[\hat{T}_i^{(h-1)}]} \geq 1.\]
By substituting into \eqref{eq:prob_not_B_f}, it follows that
\begin{align*}
\Prob_i\left( \lnot B_h | A_{h-1} \cap B_{h-1} \right) &\leq 2 \frac{\alpha}{4 \theta^{(h)}} \leq \frac{\alpha}{2^{h+1}}.
\end{align*}
By applying the union bound, we show that
\begin{align}
\Prob_i\left( A_h \cap B_h | A_{h-1} \cap B_{h-1}\right) 
&\geq 1 - \frac{\alpha}{2^h}.
\end{align}
We can easily verify using the same techniques that 
\[\Prob_i(A_1 \cap B_1 ) \geq 1 - \frac{\alpha}{2}.\]
Therefore,
\[\Prob_i\left(\bigcap_{h=1}^k (B_h \cap A_h) \right) \geq \prod_{h=1}^{k} \left(1- \frac{\alpha}{2^{h}}\right).\]
We use the Fact \ref{fact:algebra} to show that
\[\Prob_i\left(\bigcap_{h=1}^k (B_h \cap A_h) \right) \geq 1-\sum_{h=1}^k \frac{\alpha}{2^{h}} = 1 - \alpha (1 - 2^{-k}) \geq 1 - \alpha.\]\hfill \halmos
\endproof

\section{Additional Proofs for Lemmas presented in Section 5}

\proof{Proof of Lemma \ref{lemma:finite}.}
This proof is adapted from a proof found in Chapter 2 Section 4.3 of \citet{Aldous02}. It is included here for completeness.
\begin{align*}
\Prob_i(T_i > m \alpha) &= \prod_{s=0}^{m-1}\Prob_i(T_i > (s+1) \alpha ~|~ T_i > s \alpha) \\
&= \prod_{s=0}^{m-1} \left( \sum_{j \in \Sigma} \Prob_i(T_i > (s+1) \alpha ~|~ X_{s \alpha} = j, T_i > s \alpha) ~\Prob_i(X_{s \alpha} = j ~|~ T_i > s \alpha) \right).
\end{align*}
By the Markov property,
\begin{align*}
\Prob_i(T_i > m \alpha) &= \prod_{s=0}^{m-1} \left( \sum_{j \in \Sigma} \Prob_j(T_i > \alpha) ~\Prob_i(X_{s \alpha} = j ~|~ T_i > s \alpha) \right).
\end{align*}
By Markov's inequality,
\begin{align*}
\Prob_i(T_i > m \alpha) &\leq \prod_{s=0}^{m-1} \left( \sum_{j \in \Sigma} \left(\frac{\E_j[T_i]}{\alpha}\right) \Prob_i(X_{s \alpha} = j ~|~ T_i > s \alpha)\right) \\
&\leq \prod_{s=0}^{m-1} \left(\frac{\max_{j \in \Sigma} \E_j[T_i]}{\alpha} \right) = \left(\frac{H_i}{\alpha}\right)^m.
\end{align*}
By choosing $\alpha = 2 H_i$ and $m = \lfloor k / 2 H_i \rfloor$, it follows that
\[\Prob_i(T_i > k) \leq \Prob_i\left(T_i > \left\lfloor \frac{k}{2 H_i}\right\rfloor \cdot 2 H_i\right) \leq 2^{-\lfloor k / 2 H_i \rfloor} \leq 2 \cdot 2^{-k / 2 H_i}.\]\hfill \halmos
\endproof

To prove that the return times concentrate for countably infinite state space Markov chains, we use Lyapunov function analysis introduced by Foster (1953). We establish that indeed return times have exponentially decaying tail even for countable-state space Markov chain as long as they satisfy Assumption \ref{lyapunovAssumption}.

\medskip
\noindent{\bf Useful notation.} We introduce formal notation for observing $\{X_t\}_{t \geq 0}$ over the subset $B$. Let $\{Y_t\}_{t \geq 0}$ be a Markov chain with state space $B$. $\{Y_t\}$ is a subsequence of $\{X_t\}$ constructed in the following way. Define the subsequence $\{S_k\}$ of $\IntegersP$ as:
\[S_0 \triangleq 0, S_k \triangleq \min\{t > S_{k-1} : X_t \in B\},\]
and define $Y_t \triangleq X_{S_t}$ such that
\[P_Y(x, y) = \Prob\left(X_{S_1} = y | X_0 = x\right) ~~\text{ for any}~ x,y \in B.\]
Let $Q_t \triangleq S_{t+1} - S_t$, the length of the path between $X_{S_t}$ and $X_{S_{t+1}}$. \\
Let $T_i^B \triangleq \inf\{t \geq 1 ~|~ Y_t = i\}$, the return time to $i$ for the chain $\{Y_t\}$. \\
Let $H_i^B \triangleq \max_{j \in B}\E_j\left[T_i^B\right]$, the maximal expected hitting time to state $i$ for the chain $\{Y_t\}$. We can use these variables to express the return time to state $i$ by
\begin{align} \label{eq:expression_T}
T_i = S_{T_i^B} = \sum_{k=0}^{T_i^B-1} Q_k.
\end{align}

\begin{lemma} \label{lemma:negdrift}
Let $\{X_t\}$ be a countable state space Markov chain satisfying Assumption \ref{lyapunovAssumption}. Let $\{Y_t\}$ be defined above as the Markov chain restricted to $B$, and let $Q_k$ be the length between visits to $B$. For any $W \in \IntegersP$, $Z \geq W$, and $i \in B$,
\[\Prob_i\left(\sum_{k=0}^{W-1} Q_k > Z\right) \leq \exp\left( \frac{0.8(1 - \rho)}{e^{\eta \nu_{\max}}} \left(\frac{1.25 W e^{2 \eta \nu_{\max}}}{(1-\rho)(e^{\eta \nu_{\max}} - \rho)} + W - Z \right) \right).\]
The constants $\gamma$ and $\nu_{\max}$ are given by the Assumption \ref{lyapunovAssumption}, and the scalars $\eta$ and $\rho$ are functions of $\gamma$ and $\nu_{\max}$, as defined in \eqref{eq:hajekConst} in Appendix \ref{app:lyapunov}.
\end{lemma}

\proof{Proof of Lemma \ref{lemma:negdrift}.}
By the law of iterated expectation,
\begin{align*}
&\E_i\left[\prod_{k=0}^{W-1} \exp\left(\lambda Q_k\right) \right]
= \E_i\left[ \exp\left(\lambda Q_0 \right) \E_i\left[ \left. \prod_{k=1}^{W-1} \exp\left(\lambda Q_k\right) \right| Q_0 \right] \right]
\end{align*}
Conditioned on $Y_1 = j$, $Y_0$ and $Q_0$ are independent of $\{Q_k\}_{k > 0}$, because $Y_1 = X_{Q_0}$. Thus by the strong Markov property,
\[\E_i\left[ \left. \prod_{k=1}^{W-1} \exp\left(\lambda Q_k\right) \right| Q_0 \right] \leq \max_{j \in B} \E\left[ \left. \prod_{k=1}^{W-1} \exp\left(\lambda Q_k\right) \right|Y_1=j \right],\]
so that 
\[\E_i\left[\prod_{k=0}^{W-1} \exp\left(\lambda Q_k\right) \right] \leq \E_i [\exp(\lambda Q_0)] \max_{j \in B} \E\left[ \left. \prod_{k=1}^{W-1} \exp\left(\lambda Q_k\right) \right|Y_1=j \right].\]
We iteratively apply conditioning to show that
\[\E_i\left[ \prod_{k=0}^{W-1} \exp\left(\lambda Q_k\right) \right] \leq \E_i [\exp(\lambda Q_0)] \prod_{k=1}^{W-1} \left( \max_{j \in B} \E\left[\left. \exp\left(\lambda Q_k\right) \right|Y_k=j \right] \right).\]
We can upper bound $Q_k$ by assuming that it always goes on an excursion from $B$, such that
\[Q_k \leq 1 + (\text{length of an excursion into } B^c).\]
We invoke Hajek's result of Theorem \ref{hajek}, with $V(x) < b + \nu_{\max}$ to bound the exponential moments of the excursion. For any $i \in B$,
\[\E_i\left[\prod_{k=0}^{W-1} \exp\left(\lambda Q_k\right) \right] \leq \left( e^\lambda \left(e^{\eta \nu_{\text{max}}} \left(\frac{e^\lambda - 1}{1-\rho e^\lambda}\right) + 1\right) \right)^W,\]
where $\eta$ and $\rho$ are functions of $\gamma$ and $\nu_{\max}$, as defined in \eqref{eq:hajekConst} in Appendix \ref{app:lyapunov}. They satisfy the conditions given in \citet{Hajek82}. For $\lambda < \min\left(0.43, \frac{0.8 (1-\rho)}{\rho}\right),$
\[e^{\lambda} <1 + 1.25 \lambda ~~\text{ and }~~ 1 - \rho e^{\lambda} > 1 - \rho - 1.25 \rho \lambda.\]
By substituting in these approximations and using $1 + x < e^x$, we obtain
\begin{align*}
\left( e^\lambda \left(e^{\eta \nu_{\text{max}}} \left(\frac{e^\lambda - 1}{1-\rho e^\lambda}\right) + 1\right) \right)^W &< \left( e^\lambda \left(e^{\eta \nu_{\text{max}}} \left(\frac{1.25 \lambda}{1-\rho - 1.25 \rho \lambda}\right) + 1\right) \right)^W \\
&< \exp\left( \lambda W \right) \exp \left(\frac{1.25 \lambda W e^{\eta \nu_{\max}}}{1 - \rho - 1.25 \rho \lambda}\right) \\
&< \exp\left( \lambda W \left(\frac{1.25 e^{\eta \nu_{\max}}}{1 - \rho - 1.25 \rho \lambda} + 1 \right) \right).
\end{align*}
By Markov's inequality,
\begin{align*}
\Prob_i\left(\sum_{k=0}^{W-1} Q_k > Z\right) &\leq \frac{\E_i \left[ \exp \left (\lambda \sum_{k=0}^{W-1} Q_k \right) \right]}{\exp(\lambda Z)} \\
&\leq \exp\left( \lambda W \left(\frac{1.25 e^{\eta \nu_{\max}}}{1 - \rho - 1.25 \rho \lambda} + 1 \right) - \lambda Z\right).
\end{align*}
Choose $\lambda = \frac{0.8 (1 - \rho)}{e^{\eta \nu_{\max}}}$. We can verify that for our choice of $\eta$ and $\rho$ according to \eqref{eq:hajekConst}, $\lambda < \max\left(0.43,\frac{0.8 (1-\rho)}{\rho}\right)$ always holds. Therefore, we complete the proof by substituting in for $\lambda$,
\[\Prob_i\left(\sum_{k=0}^{W-1} Q_k > Z\right) \leq \exp\left( \frac{0.8(1 - \rho)}{e^{\eta \nu_{\max}}} \left(\frac{1.25 W e^{2 \eta \nu_{\max}}}{(1-\rho)(e^{\eta \nu_{\max}} - \rho)} + W - Z \right) \right).\]\hfill \halmos
\endproof

\proof{Proof of Lemma \ref{lemma:concentration_RT}.}
By \eqref{eq:expression_T}, for any constants $W,Z \in \IntegersP$,
\[\{T_i^B \leq W\} \bigcap \left\{\sum_{k=0}^{W-1} Q_k \leq Z\right\} \implies \{T_i \leq Z\}.\]
We use the union bound on the contrapositive statement to obtain the inequalities
\begin{align}
\Prob_i(T_i > Z) &\leq \Prob_i\left(\{T_i^B > W\} \cup \left\{\sum_{k=0}^{W-1} Q_k > Z\right\}\right) \nonumber \\
&\leq \Prob_i(T_i^B > W) + \Prob_i\left(\sum_{k=0}^{W-1} Q_k > Z\right). \label{eq:union_bound_app}
\end{align}
Choose $W = 2 H_i^B \left(2 + \frac{k}{R_i}\right)$ and
\[Z = 4 H_i^B \left(\frac{1.25 e^{2 \eta \nu_{\max}}}{(1-\rho)(e^{\eta \nu_{\max}} - \rho)} + 1 \right) + \frac{\ln(2) e^{\eta \nu_{\max}}}{0.8(1-\rho)} + k = 2 R_i - \frac{\ln(2) e^{\eta \nu_{\max}}}{0.8(1-\rho)} + k.\]
The next inequality follows from substituting these expressions for $W$ and $Z$ into \eqref{eq:union_bound_app}, and applying Lemmas \ref{lemma:finite} and \ref{lemma:negdrift}:
\[\Prob_i \left(T_i > 2 R_i - \frac{\ln(2) e^{\eta \nu_{\max}}}{0.8(1-\rho)} + k \right) \leq 2^{- \frac{k}{R_i}},\]
\[\Prob_i \left(T_i > 2 R_i + k \right) \leq \Prob_i \left(T_i > 2 R_i - \frac{\ln(2) e^{\eta \nu_{\max}}}{0.8(1-\rho)} + k \right) \leq 2^{- \frac{k}{R_i}},\]
\[\Prob_i \left(T_i > k \right) \leq 2^{- \frac{k - 2 R_i}{R_i}} \leq 4 \cdot 2^{-\frac{k}{R_i}}.\]\hfill \halmos
\endproof

\section{Additional Proofs for Results in Section 6}

\proof{Proof of Theorem \ref{thm:countable_bias}.}
This proof follows a similar proof of Theorem \ref{thm:bias}. By dividing \eqref{eq:exp_1_step_0} by $(1 - \Prob_i(T_i > \theta^{(k)}))$, it follows that
\begin{align*}
\left|1 - \frac{\E_i[\hat{T}_i^{(k)}]}{(1 - \Prob_i(T_i > \theta^{(k)}))\E_i[T_i]}\right| &= \left|\frac{1}{(1 - \Prob_i(T_i > \theta^{(k)}))} \left(\pi_i \sum_{k=\theta^{(k)}}^{\infty} \Prob_i(T_i > k) - \Prob_i(T_i > \theta^{(k)})\right)\right| \\
&\leq \frac{1}{(1 - \Prob_i(T_i > \theta^{(k)}))} \max\left(\pi_i \sum_{k=\theta^{(k)}}^{\infty} \Prob_i(T_i > k), \Prob_i(T_i > \theta^{(k)})\right).
\end{align*}

Then we apply Lemma \ref{lemma:concentration_RT} and use the fact that $\Prob_i(T_i > \theta^{(k)}) < \frac12$ to show that
\begin{align}
\left|1 - \frac{\E_i[\hat{T}_i^{(k)}]}{(1 - \Prob_i(T_i > \theta^{(k)}))\E_i[T_i]}\right| 
&\leq 2 \max\left(\pi_i \left(\frac{4 \cdot 2^{-\theta^{(k)} / R_i}}{1 - 2^{-1 /R_i}}\right), 4 \cdot 2^{-\theta^{(k)} / R_i}\right) \nonumber \\
&= 8 \cdot 2^{-\theta^{(k)} / R_i} \max\left(\frac{\pi_i}{1 - 2^{-1 /R_i}}, 1\right). \label{eq:countable_bias}
\end{align}

Substitute \eqref{eq:countable_bias} into \eqref{eq:lemma_exp_2} to complete the proof.
\hfill \halmos \\
\endproof

In order to analyze the distribution over the number of visits to state $j$ on a return path to state $i$, we will use the following Lemma as stated by \citet{Aldous02} in Chapter 2 Section 2.2 Lemma 9.
\begin{lemma} \citep{Aldous02} \label{lemma:aldous_2}
For distinct $i,k \in \Sigma$, the expected number of visits to a state $j$ beginning from state $k$ before visiting $i$ is equal to
\[\E_k\left[\sum_{t=1}^{\infty} \Indicator\{X_t=j\} \Indicator\{t \leq T_i\}\right] = \pi_j (E_k[T_i] + E_i[T_j] - E_k[T_j]).\]
\end{lemma}

\proof{Proof of Lemma \ref{lemma:exp_3}.}
By definition,
\begin{align*}
&\E_i[F_j] - \E_i[\hat{F}_j^{(k)}] \\
&= \E_i\left[\sum_{t=1}^{\infty} \Indicator\{X_t = j\} \Indicator\{t \leq T_i\}\right] - \E_i\left[\sum_{t=1}^{\theta^{(k)}} \Indicator\{X_t = j\} \Indicator\{t \leq T_i\} \right] \\
&= \Prob_i\left(T_i > \theta^{(k)}\right) \E_i\left[ \left. \sum_{t=\theta^{(k)}}^{\infty} \Indicator\{X_t = j\} \Indicator\{t \leq T_i\} \right| T_i > \theta^{(k)}\right] \\
&= \Prob_i\left(T_i > \theta^{(k)}\right) \sum_{q \in \Sigma \setminus \{i\}} \Prob_i\left(\left.X_{\theta^{(k)}} = q \right| T_i > \theta^{(k)}\right) \E_i\left[ \left.\sum_{t=\theta^{(k)} + 1}^{\infty} \Indicator\{X_t=j\} \Indicator\{t \leq T_i\} \right| X_{\theta^{(k)}} = q, T_i > \theta^{(k)}\right]  \\
&= \Prob_i\left(T_i > \theta^{(k)}\right) \sum_{q \in \Sigma \setminus \{i\}} \Prob_i\left(\left.X_{\theta^{(k)}} = q \right| T_i > \theta^{(k)}\right) \E_q\left[\sum_{t=1}^{\infty} \Indicator\{X_t=j\} \Indicator\{t \leq T_i\}\right]. 
\end{align*}

We divide by $\E_i[F_j]$ and use Lemma \ref{lemma:aldous_2} and Lemma \ref{eq:property}(a) to show that
\begin{align}
&1- \frac{\E_i[\hat{F}_j^{(k)}]}{\E_i[F_j]} \nonumber \\
&= \frac{\Prob_i\left(T_i > \theta^{(k)}\right)}{\E_i[F_j]} \sum_{q \in \Sigma \setminus \{i\}} \Prob_i\left(\left. X_{\theta^{(k)}} = q \right| T_i > \theta^{(k)}\right) \pi_j (E_q[T_i] + E_i[T_j] - E_q[T_j]) \nonumber \\
&=\Prob_i\left(T_i > \theta^{(k)}\right) \sum_{q \in \Sigma \setminus \{i\}} \Prob_i\left(\left. X_{\theta^{(k)}} = q \right| T_i > \theta^{(k)}\right) \frac{(E_q[T_i] + E_i[T_j] - E_q[T_j])}{\E_i[T_i]} \label{eq:F_diff}.
\end{align}

By multiplying \eqref{eq:exp_1_step} 
by $\E_i[\hat{T}_i^{(k)}]$, it follows that
\begin{align}
1 - \frac{\E_i[\hat{T}_i^{(k)}]}{\E_i[T_i]} = \Prob_i\left(T_i > \theta^{(k)}\right) \sum_{q \in \Sigma \setminus \{i\}} \Prob_i\left(\left.X_{\theta^{(k)}} = q\right| T_i > \theta^{(k)}\right) \frac{\E_q[T_i]}{\E_i[T_i]}. \label{eq:}
\end{align}

We use \eqref{eq:F_diff} and \eqref{eq:})= to show that

\begin{align*}
\frac{\E_i[\hat{F}_j^{(k)}]}{\E_i[\hat{T}_i^{(k)}]} - \pi_j &= \frac{\E_i[F_j]}{\E_i[\hat{T}_i^{(k)}]} \left(\frac{\E_i[\hat{F}_j^{(k)}]}{\E_i[F_j]} - \frac{\E_i[\hat{T}_i^{(k)}]}{\E_i[T_i]}\right) \\
&= \frac{\E_i[F_j]}{\E_i[\hat{T}_i^{(k)}]} \Prob_i\left(T_i > \theta^{(k)}\right) \sum_{q \in \Sigma \setminus \{i\}} \Prob_i\left(\left. X_{\theta^{(k)}} = q \right| T_i > \theta^{(k)}\right) \frac{(E_q[T_j] - E_i[T_j])}{\E_i[T_i]}  \\
&= \frac{\Prob_i\left(T_i > \theta^{(k)}\right)}{\E_i[\hat{T}_i^{(k)}]} \sum_{q \in \Sigma \setminus \{i\}} \Prob_i\left(\left. X_{\theta^{(k)}} = q \right| T_i > \theta^{(k)}\right) \pi_j (E_q[T_j] - E_i[T_j]).
\end{align*}

By Lemma \ref{lemma:aldous_1},

\begin{align*}
\frac{\E_i[\hat{F}_j^{(k)}]}{\E_i[\hat{T}_i]} - \pi_j &= \frac{\Prob_i\left(T_i > \theta^{(k)}\right)}{\E_i[\hat{T}_i^{(k)}]} \left(\sum_{q \in \Sigma \setminus \{i\}} \Prob_i\left(\left. X_{\theta^{(k)}} = q \right| T_i > \theta^{(k)}\right) (Z_{ij}  - Z_{qj})\right).
\end{align*}\hfill \halmos
\endproof

\section{Chernoff Bounds}

\begin{theorem} [{\normalfont {\em Chernoff's Multiplicative Bound for Binomials}}] \label{chernoff_binomial} ~\\
Let $\{X_1, X_2, X_3, \dots X_N\}$ be a sequence of independent identically distributed Bernoulli random variables, such that for all $i$, $X_i = 1$ with probability $p$ and $X_i = 0$ otherwise. Then for any $\epsilon > 0$,
\begin{align*}
\Prob\left( \left| \frac{1}{N} \sum_{i=1}^N X_i - \E[X] \right| \geq \epsilon \E[X] \right)
&\leq 2 e^{-\frac{\epsilon^2 N p}{3}}.
\end{align*}
\end{theorem}

\begin{theorem} [{\normalfont {\em Chernoff's Multiplicative Bound for Bounded Variables}}] \label{chernoff_bounded} ~\\
Let $\{X_1, X_2, X_3, \dots X_N\}$ be a sequence of independent identically distributed strictly bounded nonnegative random variables, such that $X_i \sim X$ for all $i$, and $X \in [0,\theta]$. Then for any $\epsilon > 0$,
\begin{align*}
\Prob\left( \left| \frac{1}{N} \sum_{i=1}^N X_i - \E[X] \right| \geq \epsilon \E[X] \right) 
&\leq 2 e^{-\frac{\epsilon^2 N \E[X]}{3 \theta}}.
\end{align*}
\end{theorem}

\section{Lyapunov Function Analysis} \label{app:lyapunov}

\begin{theorem} \citep{Foster53} \label{foster}
Let $\{X_t\}$ be a discrete time, irreducible Markov chain on countable state space $\Sigma$ with transition probability matrix $P$. $\{X_t\}$ is positive recurrent if and only if there exists a Lyapunov function $V: \Sigma \to \RealsP$, $\gamma > 0$ and $b \geq 0$, such that 
\begin{enumerate}
\item For all $x \in \Sigma$,
\[\E\left[V(X_{t+1})| X_t=x \right] \leq \infty,\]
\item For all $x \in \Sigma$ such that $V(x) > b$,
\[\E\left[V(X_{t+1}) - V(X_t) | X_t=x \right] \leq - \gamma.\]
\end{enumerate}
\end{theorem}

In words, given a positive recurrent Markov chain, there exists a Lyapunov function $V:\Sigma \rightarrow \RealsP$ and a decomposition of the state space into $B = \{x \in \Sigma: V(x) \leq b\}$ and $B^c = \{x \in \Sigma: V(x) > b\}$ such that there is a uniform negative drift in $B^c$ towards $B$ and $|B|$ is finite.

For any irreducible, Markov chain, the following function is a valid Lyapunov function for $\gamma = 1$ and $b = 0.5$: Choose any state $i \in \Sigma$, and fix this as the ``central state''. Define the function $V:\Sigma \rightarrow \RealsP$ such that $V(i) = 0$ and for all $j \in \Sigma \setminus \{i\}$, $V(j) = \E_j[T_i]$. By definition, $B = \{i\}$. For all $x \in \Sigma$, by positive recurrence, $\E\left[V(X_{t+1})| X_t=x \right] \leq \infty$. Similarly, for all $x$ such that $V(x) > b$, 
\[V(x) = \E_x[T_i] = 1 + \sum_{y \in \Sigma} P_{xy} E_y[T_i] = 1 + \E[V(X_{t+1})|X_t = x].\]
Therefore, for all $x \in B^c$,
\[\E\left[V(X_{t+1}) - V(X_t) | X_t=x \right] = -1 \leq -\gamma.\]

\begin{theorem} \citep{Hajek82} \label{hajek}
Let $\{X_t\}$ be an irreducible, positive recurrent Markov chain on a countable state space $\Sigma$ with transition probability matrix $P$. Assume that there exists a Lyapunov function $V: \Sigma \to \RealsP$ and values $\nu_{\max}, \gamma > 0$, and $b \geq 0$ satisfying Assumption \ref{lyapunovAssumption}. Let the random variable $\tau_B = \inf\{t: X_t \in B\}$. Then for any $x$ such that $V(x) > b$, and for any choice of constants $\omega > 0$, $\eta$, $\rho$, and $\lambda$ satisfying
\[0 < \eta \leq \min\left(\omega, \frac{\gamma \omega^2}{e^{\omega \nu_{\max}} - (1 + \omega \nu_{\max})}\right),\]
\[\rho = 1 - \gamma \eta + \frac{\left(e^{\omega \nu_{\max}} - (1 + \omega \nu_{\max})\right)\eta^2}{\omega^2},\]
\[\text{and } 0 < \lambda < \ln(\frac{1}{\rho}),\]
the following two inequalities hold:
\[ \Prob[\tau_B > k | X_0=x] \leq e^{\eta(V(x) - b)} \rho^k,\]
\[\text{and } \E[e^{\lambda \tau_B} | X_0=x] \leq e^{\eta(V(x) - b)} \left(\frac{e^{\lambda}-1}{1-\rho e^{\lambda}}\right) + 1.\]
\end{theorem}

\noindent
A concrete set of constants that satisfy the conditions above are
\begin{align} \label{eq:hajekConst}
\omega = \frac{1}{\nu_{\max}}, \eta = \frac{\gamma}{2(e-2)\nu_{\max}^2}, \text{ and } \rho = 1 - \frac{\gamma^2}{4(e-2)\nu_{\max}^2}.
\end{align}

\end{APPENDICES}


\ACKNOWLEDGMENT{This work is supported in parts by ARO under MURI awards 58153-MA-MUR and W911NF-11-1-0036, and grant 56549-NS, and by NSF under grant CIF 1217043 and a Graduate Fellowship.}

\bibliographystyle{ormsv080}
\bibliography{bibliography}

\end{document}